\documentclass[preprint,3p,number]{elsarticle}

\usepackage{graphicx,lipsum}
\usepackage{algorithm,algpseudocode}
\usepackage[dvipsnames]{xcolor}
\usepackage{figlatex,wrapfig}
\usepackage{listings,amssymb,mathtools}
\usepackage{stmaryrd}
\usepackage{mathrsfs}
\usepackage{array,multirow}
\usepackage{pifont}
 \usepackage[font=small]{caption}
\usepackage{wrapfig}
\usepackage{textcomp}
\usepackage{xspace}
\usepackage{trfrac}
\usepackage{amsmath}
\usepackage{afterpage}
\usepackage{tcolorbox}
\usepackage[T1]{fontenc}
\usepackage[utf8]{inputenc} 
\usepackage{tikz}
\usepackage{caption}  
\usepackage{float}
\usepackage{esvect}
\usepackage{verbatim}
\usepackage{stmaryrd}
\usepackage{turnstile}
\usepackage{soul}
\usetikzlibrary{decorations.pathreplacing}
\usepackage{lineno,hyperref}

\definecolor{BrickRed}{RGB}{203,65,84}
\definecolor{eclipseBlue}{RGB}{42,0.0,255}
\definecolor{eclipseGreen}{RGB}{63,127,95}
\definecolor{eclipsePurple}{RGB}{127,0,85}
\definecolor{myred}{RGB}{100,0,0}
\definecolor{mygrey}{RGB}{211,211,211}
\definecolor{mygrey2}{RGB}{112,128,124}

\hypersetup{
	colorlinks = true,
	citecolor = {violet},
	linkcolor = {purple},
	urlcolor  = {blue}
}

\newtheorem{theorem}{Theorem}
\newtheorem{lemma}[theorem]{Lemma}

\newtheorem{definition}{Definition}[section]
\newtheorem{example}{Example}[section]

\DeclareSymbolFont{symbolsC}{U}{txsyc}{m}{n}
\DeclareMathSymbol{\Searrow}{\mathrel}{symbolsC}{117}

\newcommand{\mathhl}[2]{\colorbox{#1}{$\displaystyle #2$}}

\newcommand{\lustre}{{\sc{Lustre}}}
\newcommand{\nlustre}{{\sc{NLustre}}}

\newcommand{\tbf}[1]{\textbf{#1}}
\newcommand{\ttt}[1]{\texttt{#1}}
\newcommand{\tti}[1]{\textit{#1}}

\newcommand{\type}[1]{\textcolor{eclipseBlue}{#1}}

\newcommand{\stlub}{\sqcup}

\newcommand{\rel}{\xspace \sqsubseteq }
\newcommand{\strel}{\xspace \type{\sqsubseteq} }
\newcommand{\lub}{\xspace \sqcup}

\newcommand{\for}[2]{#2\mapsto #1}

\newcommand{\val}[1]{\textcolor{mygrey2}{ #1 } }
\newcommand{\when}[2]{#1~\texttt{when}~#2} 
 
\newcommand{\binop}[2]{ #1 \oplus #2}
\newcommand{\unop}[1]{ \diamond ~#1}
\newcommand{\ite}[3]{\ttt{if } #1\ttt{ then } #2 \ttt{ else } #3}

\newcommand{\lmerge}[3]{\ttt{merge}~#1~#2~#3}
\newcommand{\ck}[2]{#1~\ttt{on}~#2}

\newcommand{\on}[3]{#1 ~\ttt{on}~ #2=\val{#3}}
\newcommand{\onF}[3]{#1 ~\ttt{on}~ #2=\val{\neg #3}}
\newcommand{\fby}[2]{#1~\texttt{fby}~#2}
\newcommand{\lfby}[2]{#1~\texttt{fby}_l~#2}
\newcommand{\nlfby}[2]{#1~\texttt{fby}_{nl}~#2}
\newcommand{\eqn}[3]{#1 =_{#3} #2}
\newcommand{\nodecall}[2]{#1 (\vv{#2})} 
\newcommand{\Node}{
  \ttt{node}~ f(\vv{x^{ck}}) ~ \texttt{returns} ~\vv{y^{ck}} \\
 {}&~\ttt{var}~ \vv{z} ~ \ttt{let}~ \vv{eq} ~\ttt{tel}
}

\newcommand{\norm}[1]{\lfloor #1 \rfloor}

\newcommand{\node}{
 \Bigg\{ \begin{array}{c}
      \textsf{name} =~ \ttt{f};~ \textsf{in}=~\vv{x}; ~\textsf{var}=~\vv{z};\\
 \textsf{out}=~\vv{y};~ \textsf{eqs}=~\vv{eq}\\
 \end{array}   \Bigg\}
}

\newcommand{\securitySignature}[5]{\stackrel{Node}{\vdash} \texttt{\textcolor{eclipsePurple}{Node} #1 } \textcolor{eclipseBlue}{#2^{#4}} \xrightarrow{#5} \textcolor{eclipseBlue}{#3}}

\newcommand{\vars}[2]{(#1_{1},\dots#1_{#2})}

\newcommand{\set}[1]{\{ #1\}}

\makeatletter
\newcommand{\leqnos}{\tagsleft@true\let\veqno\@@leqno}
\newcommand{\reqnos}{\tagsleft@false\let\veqno\@@eqno}
\reqnos
\makeatother

\newcommand{\mjdg}[2]{ \trfrac[]{ #1 }{  #2 } }
\newcommand{\namedJdg}[3]{ \trfrac[~(\footnotesize{#3})]{ #1 }{  #2 } }

\newcommand{\mexprPred}[3]{ #1 \textcolor{BrickRed}{\stackrel{e}{\vdash}} #2: \textcolor{eclipseBlue}{#3} }

\newcommand{\mcexprPred}[3]{ #1 \textcolor{OliveGreen}{\stackrel{ce}{\vdash}}  #2: \textcolor{eclipseBlue}{#3} }

\newcommand{\mclkPred}[3]{ #1 \textcolor{NavyBlue}{\stackrel{ck}{\vdash}}  #2: \textcolor{eclipseBlue}{#3} }

\newcommand{\manyexprPred}[3]{ #1 \textcolor{Black}{\stackrel{?}{\vdash}} #2: \textcolor{eclipseBlue}{#3} }

\newcommand{\meqnPred}[3]{  #1\stackrel{ eqn }{\vdash}  #2\ttt{~:>} ~{\type{#3}} }

\makeatletter
\newcommand{\predSet}[1]{%
    #1 \pchecknextarg}
\newcommand{\pchecknextarg}{\@ifnextchar\bgroup{\pgobblenextarg}{ }}
\newcommand{\pgobblenextarg}[1]{ ~~ #1 \@ifnextchar\bgroup{\pgobblenextarg}{ }}
\makeatother
\newcommand{\predSets}[2]{#1 \quad #2 }

\makeatletter

\newcommand{\dependSet}[1]{%
\begin{tralign}
#1  \checknextarg}
\newcommand{\checknextarg} {\@ifnextchar\bgroup {\gobblenextarg}{\end{tralign}}} 
\newcommand{\gobblenextarg}[1]{ \\[1.5mm] #1 \@ifnextchar\bgroup{\gobblenextarg} { \end{tralign}} }
\makeatother

\makeatletter
\newcommand{\union}[1]{%
    #1 \uchecknextarg}
\newcommand{\uchecknextarg}{\@ifnextchar\bgroup{\ugobblenextarg}{ }}
\newcommand{\ugobblenextarg}[1]{ \cup #1 \@ifnextchar\bgroup{\ugobblenextarg}{ }}
\makeatother

\makeatletter

\newcommand{\ichecknextarg}{\@ifnextchar\bgroup{\igobblenextarg}{ }}
\newcommand{\igobblenextarg}[1]{ \cap #1 \@ifnextchar\bgroup{\igobblenextarg}{ }}
\makeatother
\makeatletter

\newcommand{\orchecknextarg}{\@ifnextchar\bgroup{\orgobblenextarg}{ }}
\newcommand{\orgobblenextarg}[1]{ \bigvee #1 \@ifnextchar\bgroup{\orgobblenextarg}{ }}
\makeatother

\makeatletter

\newcommand{\andchecknextarg}{\@ifnextchar\bgroup{\andgobblenextarg}{ }}
\newcommand{\andgobblenextarg}[1]{ \land #1 \@ifnextchar\bgroup{\andgobblenextarg}{ }}
\makeatother

\newcommand*\substile[2]{%
  \,\scalebox{0.38}[0.5]{$\sststile[s]{\textstyle#1}{\textstyle#2}$}\,
}

\newcommand{\Proof}{\textsc{Proof.}\ \ }
\newcommand{\Proofsketch}{\textsc{Proof sketch.}\ \ }
\newcommand{\QED}{~\hfill $\square$}

\newcommand{\stream}[1]{\text{\guilsinglleft}\textcolor{mygrey2}{#1}\text{\guilsinglright}}

\newcommand{\nullStream}{\text{\guilsinglleft}\text{\guilsinglright}}
\newcommand{\ckFont}[1]{\textcolor{blue}{\textsf{#1}}}

\newcommand{\ignore}[1]{}

\newcommand{\streamSim}[3]{#1 \Downarrow_{#3} #2}

\newcommand{\expSim}[2]{\streamSim{#1}{#2}{\textsf{e}} }

\newcommand{\ckSim}[2]{\streamSim{#1}{#2}{\textsf{ck}}}
\newcommand{\anySim}[2]{\streamSim{#1}{#2}{\textsf{?}}}

\newcommand{\flatten}[1]{\textcolor{blue}{\flat} (#1) }

\newcommand{\semConst}[3]{\textcolor{blue}{\textsf{const}}~#1~\val{#2}=
{\textcolor{BrickRed}{#3}}}

\newcommand{\whenk}[4]{\textcolor{blue}{\textsf{when}}~\val{#1}~{\textcolor{BrickRed}{#2}}~{\textcolor{BrickRed}{#3}}={\textcolor{BrickRed}{#4}}}

\newcommand{\mapwhenk}[4]{\textcolor{blue}{\widehat{\textsf{when}}}~\val{#1}~{\textcolor{BrickRed}{#2}}~{\textcolor{BrickRed}{#3}}={\textcolor{BrickRed}{#4}}}

\newcommand{\liftunop}[2]{\textcolor{blue}{\widehat{\diamond}} ~
{\textcolor{BrickRed}{#1}} = {\textcolor{BrickRed}{#2} }}

\newcommand{\liftbinop}[3]{{\textcolor{BrickRed}{#1}} \textcolor{blue}{\widehat{\oplus}} {\textcolor{BrickRed}{#2}} = {\textcolor{BrickRed}{#3}} }

\newcommand{\liftnode}[1]{ \textcolor{blue}{\widehat{#1}} }
\newcommand{\htl}[1]{(\textcolor{blue}{\textsf{htl}} ~#1)}
\newcommand{\tl}[1]{(\textcolor{blue}{\textsf{tl}} ~#1)}
\newcommand{\cc}[2]{(#1 \cdot #2)}
\newcommand{\ccnb}[2]{#1 \cdot #2}
\newcommand{\true}{\textcolor{blue}{\textsf{true}}}
\newcommand{\false}{\textcolor{blue}{\textsf{false}}}

\newcommand{\semFby}[3]{\textcolor{blue}{\textsf{fby}_{NL}} ~\val{#1} 
~{\textcolor{BrickRed}{#2}} = {\textcolor{BrickRed}{#3}}}

\newcommand{\semLFby}[3]{{\textcolor{blue}{\textsf{fby}_L}} ~{\textcolor{BrickRed}{#1}} ~{\textcolor{BrickRed}{#2}} = {\textcolor{BrickRed}{#3}}}

\newcommand{\mapsemLFby}[3]{{\textcolor{blue}{\widehat{\textsf{fby}_L}}} ~{\textcolor{BrickRed}{#1}} ~{\textcolor{BrickRed}{#2}} = {\textcolor{BrickRed}{#3}}}

\newcommand{\semLDFby}[4]{\textcolor{blue}{\textsf{fby$_{dl}$}} ~\val{#1} ~{\textcolor{BrickRed}{#2}} ~{\textcolor{BrickRed}{#3}} = {\textcolor{BrickRed}{#4}}}

\newcommand{\semMerge}[4]{\textcolor{blue}{\textsf{merge}}~{\textcolor{BrickRed}{#1}}~{\textcolor{BrickRed}{#2}}~{\textcolor{BrickRed}{#3}}={\textcolor{BrickRed}{#4}}}

\newcommand{\mapsemMerge}[4]{\textcolor{blue}{\widehat{\textsf{merge}}}~{\textcolor{BrickRed}{#1}}~{\textcolor{BrickRed}{#2}}~{\textcolor{BrickRed}{#3}}={\textcolor{BrickRed}{#4}}}

\newcommand{\semIte}[4]{\textcolor{blue}{\textsf{ite}}~{\textcolor{BrickRed}{#1}}~{\textcolor{BrickRed}{#2}}~{\textcolor{BrickRed}{#3}}={\textcolor{BrickRed}{#4}}}

\newcommand{\mapsemIte}[4]{\textcolor{blue}{\widehat{\textsf{ite}}}~{\textcolor{BrickRed}{#1}}~{\textcolor{BrickRed}{#2}}~{\textcolor{BrickRed}{#3}}={\textcolor{BrickRed}{#4}}}

\newcommand{\baseOf}[1]{\textcolor{blue}{\textsf{base-of}}\,#1}

\newcommand{\resClk}[3]{\textcolor{blue}{\textsf{respects-clock}}\,#1\,#2\,#3}

\newcommand{\hist}[1]{H_{*}(#1)}
\newcommand{\Hst}{H_{*}}

\newcommand{\tupstrm}[1]{\textcolor{BrickRed}{\mathbf{#1}}}
\newcommand{\mstrSemExpPred}[4]{#1,#2 ~\textcolor{BrickRed}{\vdash} ~\expSim{#3}{\textcolor{BrickRed}{#4}} }

\newcommand{\mstrSemCkPred}[4]{ #1,#2 ~\textcolor{NavyBlue}{\vdash} ~\ckSim{#3}{#4} }

\newcommand{\mstrSemAnyExpPred}[4]{ #1,#2 ~\textcolor{NavyBlue}{\vdash} ~\anySim{#3}{\textcolor{BrickRed} {#4}} }

\newcommand{\mstrEqnPred}[4]{#1,#2,#3 ~\textcolor{Black}{\vdash}~ #4}

\newcommand{\mstrCallPred}[4]{#1 ~\textcolor{Black}{\substile{s}{}}~ 
{#2}({\textcolor{BrickRed}{#3}}) \Searrow {\textcolor{BrickRed}{#4}} }
  \newcommand{\simpl}[2]{\ckFont{simplify}~#1~\type{#2}}
  \newcommand{\nlsimpl}[2]{\ckFont{simplify$_{NL}$}~#1~\type{#2}}
  \newcommand{\lsimpl}[2]{\textcolor{blue}{\textsf{simplify$_{L}$}}~#1~\type{#2}}

\lstdefinelanguage{Lustre}
{
  morekeywords={
    node,
    when,
    merge,
    fby,
    if,
    then,
    else,
    returns,
    let,
    tel,
    int,
    bool,
    var
  },
  basicstyle=\ttfamily, % Global Code Style
  captionpos=b,
  sensitive=true, % keywords are not case-sensitive
  morecomment=[l]{--}, % l is for line comment
  morecomment=[s]{/*}{*/}, % s is for start and end delimiter
  morestring=[b]", % defines that strings are enclosed in double quotes
  commentstyle=\color{eclipseGreen}, % style of comments
  keywordstyle=\color{eclipsePurple}, % style of keywords
  stringstyle=\color{eclipseBlue}, % style of strings
  mathescape = true,
}

\lstdefinelanguage{Coq}{ 
    mathescape=true,
    texcl=false, 
    morekeywords=[1]{Section, Module, End, Require, Import, Export,
        Variable, Variables, Parameter, Parameters, Axiom, Hypothesis,
        Hypotheses, Notation, Local, Tactic, Reserved, Scope, Open, Close,
        Bind, Delimit, Definition, Let, Ltac, Fixpoint, CoFixpoint, Add,
        Morphism, Relation, Implicit, Arguments, Unset, Contextual,
        Strict, Prenex, Implicits, Inductive, CoInductive, Record,
        Structure, Canonical, Coercion, Context, Class, Global, Instance,
        Program, Infix, Theorem, Lemma, Corollary, Proposition, Fact,
        Remark, Example, Proof, Goal, Save, Qed, Defined, Hint, Resolve,
        Rewrite, View, Search, Show, Print, Printing, All, Eval, Check,
        Projections, inside, outside, Def},
    morekeywords=[2]{forall, exists, exists2, fun, fix, cofix, struct,
        match, with, end, as, in, return, let, if, is, then, else, for, of,
        nosimpl, when},
    morekeywords=[3]{Type, Prop, Set, true, false, option},
    morekeywords=[4]{pose, set, move, case, elim, apply, clear, hnf,
        intro, intros, generalize, rename, pattern, after, destruct,
        induction, using, refine, inversion, injection, rewrite, congr,
        unlock, compute, ring, field, fourier, replace, fold, unfold,
        change, cutrewrite, simpl, have, suff, wlog, suffices, without,
        loss, nat_norm, assert, cut, trivial, revert, bool_congr, nat_congr,
        symmetry, transitivity, auto, split, left, right, autorewrite},
    morekeywords=[5]{by, done, exact, reflexivity, tauto, romega, omega,
        assumption, solve, contradiction, discriminate},
    morekeywords=[6]{do, last, first, try, idtac, repeat},
    morecomment=[s]{(*}{*)},
    showstringspaces=false,
    morestring=[b]",
    morestring=[d],
    tabsize=3,
    extendedchars=false,
    sensitive=true,
    breaklines=false,
    basicstyle=\small,
    captionpos=b,
    columns=[l]flexible,
    identifierstyle={\ttfamily\color{black}},
    keywordstyle=[1]{\ttfamily\color{violet}},
    keywordstyle=[2]{\ttfamily\color{OliveGreen}},
    keywordstyle=[3]{\ttfamily\color{blue}},
    keywordstyle=[4]{\ttfamily\color{blue}},
    keywordstyle=[5]{\ttfamily\color{red}},
    stringstyle=\ttfamily,
    commentstyle={\ttfamily\color{OliveGreen}},
    literate=
    {\\forall}{{\color{BrickRed}{$\forall\;$}}}1
    {\\exists}{{$\exists\;$}}1
    {\\alpha}{{$\alpha\;$}}1
    {\\beta}{{$\beta\;$}}1
    {\\rho}{{$\rho\;$}}1
    {\\theta}{{$\theta\;$}}1
    {\\delta}{{$\delta\;$}}1
    {\\Delta}{{$\Delta\;$}}1
    {<-}{{$\leftarrow\;$}}1
    {=>}{{$\Rightarrow\;$}}1
    {==}{{\code{==}\;}}1
    {==>}{{\code{==>}\;}}1
    {->}{{$\rightarrow\;$}}1
    {<->}{{$\leftrightarrow\;$}}1
    {<==}{{$\leq\;$}}1
    {\#}{{$^\star$}}1 
    {\\o}{{$\circ\;$}}1 
    {\@}{{$\cdot$}}1 
    {\/\\}{{$\wedge\;$}}1
    {\\\/}{{$\vee\;$}}1
    {++}{{\code{++}}}1
    {~}{{\ }}1
    {\@\@}{{$@$}}1
    {\\mapsto}{{$\mapsto\;$}}1
    {\\hline}{{\rule{\linewidth}{0.5pt}}}1
}[keywords,comments,strings]

\begin{document}

\title{Secure Information Flow Typing in \lustre}

\author{Sanjiva Prasad,
%\orcidID{0000-0001-5887-1237}
%\and
R. Madhukar Yerraguntla,
%\orcidID{0000-0001-8219-925X}
%\and
Subodh Sharma
%\orcidID{0000-0003-3069-3744}
}

\begin{abstract}
Synchronous reactive data flow is a paradigm that provides a high-level abstract programming model for embedded and cyber-physical systems, including the locally synchronous components of IoT systems. 
Security in such systems is severely compromised due to low-level programming, ill-defined interfaces and inattention to security classification of data.   
By incorporating a Denning-style lattice-based secure information flow framework into a synchronous reactive data flow language, we provide a framework in which correct-and-secure-by-construction implementations for such systems may be specified and derived.  
In particular, we propose an extension of the \lustre\  programming framework with a \textit{security type system}. 
The novelty of our type system lies in a symbolic formulation of constraints over security type variables, in particular the treatment of node calls, which allows us to reason about secure flow with respect to any security class lattice.

The main theorem is the soundness of our type system with respect to the co-inductive operational semantics of  \lustre, which we prove by showing that well-typed programs exhibit non-interference.
Rather than tackle the full language, we first prove the non-interference result for a well-behaved sub-language called ``Normalised \lustre'' (\nlustre), for which our type system is far simpler.
We then show that Bourke \textit{et al.}'s semantics-preserving ``normalisation'' transformations from \lustre\ to \nlustre\ are security-preserving as well.
This preservation of security types by the normalisation transformations is a property akin to ``subject reduction'' but at the level of compiler transformations.
The main result that well-security-typed \lustre\ programs are non-interfering follows from a reduction to our earlier result of non-interference for \nlustre\  via the semantics-preservation (of Bourke \textit{et al.}) and type preservation results.
\end{abstract}

\maketitle

% 1
\section{Introduction}
\label{SEC:Introduction}
% Introduction and motivation to the entire work.

% Why
\paragraph{Motivation} \ 
The impetus for this work was to address the problems of correctness and security in embedded and cyber-physical systems, especially in the Internet of Things paradigm \cite{Iotstand}.
Several high-profile attacks such as those on CAN systems \cite{Iotbug2,Carshark}, smart lighting \cite{SmartLighting}, and pacemakers \cite{Iotbug1,pacemaker} have exploited vulnerabilities  arising from lacunae such as
(L1) ill-defined interfaces and hidden attack surfaces;
(L2) no secure information flow (SIF) architecture and weak security mechanisms;
(L3) components operating with greater privilege or more capability than necessary.

Our contention is that much of this unfortunate insecurity arises from low-level programming approaches in these domains, which can be avoided by using a high-level programming paradigm.
While Domain-Specific Languages suggest a principled way to build more reliable systems, we contend that for the large subset of locally synchronous systems there already is a quintessential solution -- namely, reactive synchronous data-flow languages. 

The merits to taking this view include:  
(M1) ``Things'', embedded and cyber-physical systems, can be  treated abstractly as generators, consumers and transformers of (clocked) streams of data.
Indeed, this extensional view supports not only abstract ``things'' but composite computations as first-class entities.  
(M2) The data flow model makes explicit all interfaces, connections and data dependencies, and (clocked, named) flows -- thus greatly reducing attack surfaces and unanticipated interactions.
(M3) In synchronous reactive data flow languages, monitoring safety properties is easy and can be achieved using finite-state automata \cite{Rushby12,Halbwachs93}.
Not only can monitors be expressed within the model, but the same framework can be used to specify axioms and assumptions, to constrain behaviour, and specify test cases.
(M4) In particular, \lustre\ \cite{Lustre87,LusMain} is an eminently suitable synchronous reactive data flow language, for which there already exist elegant formal semantics \textit{and} a suite of tools for (a) certified compilation from the high-level model into lower-level imperative languages \cite{SNLustre2017,Lustre2020}, (b) model-checkers \cite{LusMC,Kind2} (c) simulation tools \cite{Lurette}, etc.
Indeed, the simplicity of \lustre\ -- with its underlying deterministic, clocked, structured model -- makes it both attractive and versatile as a programming paradigm:
It can express distributed embedded controllers, numerical computations, and complex Scade 6 \cite{Scade6} safety-critical systems with its support for difference equations. 
Moreover, the gamut of formal structures such as automata, regular expressions, temporal logic, test harnesses, synchronous observers, etc., can all be \textit{efficiently} expressed \textit{within} the \lustre\ model.
The only missing piece in this picture is a security model. 

% What
In this paper, we seek to integrate Denning's lattice-based secure information flow (SIF) framework \cite{Denning:1976} into \lustre\  and propose a type system for SIF. 
In this type system,
(i) each  stream of data is assigned a security type that is mapped to a security class from the security lattice, based on assumptions made about security types of the program variables, and 
(ii) the output streams from a node have security levels at least as high as the security levels of the input streams on which they depend. 
The rules are simple, intuitive and amenable to incorporation into the mechanised certified compilation already developed for \lustre\ \cite{Velus} that integrates into the CompCert effort  \cite{Leroy2009}.

\paragraph{Contributions} \ \ 
The main contributions of this paper are: 
(C1)
the proposal of a \textit{security type system} which ensures SIF in \lustre\ (\autoref{SEC:LSec-types}).
The security types are as simple as possible, which makes possible further refinements.
The main technical achievement lies in formulating appropriate type inference rules for \lustre\ equations and node (function) calls. 
The security types and constraints are stated in a \textit{symbolic} style employing security type variables, thus abstracting the inference rules from any particular security lattice.
The type system is equipped with an equational theory, which is shown to be sound with respect to any security class lattice.
Based on the security typing rules, we propose a \textit{definition of security} for \lustre\ programs. 

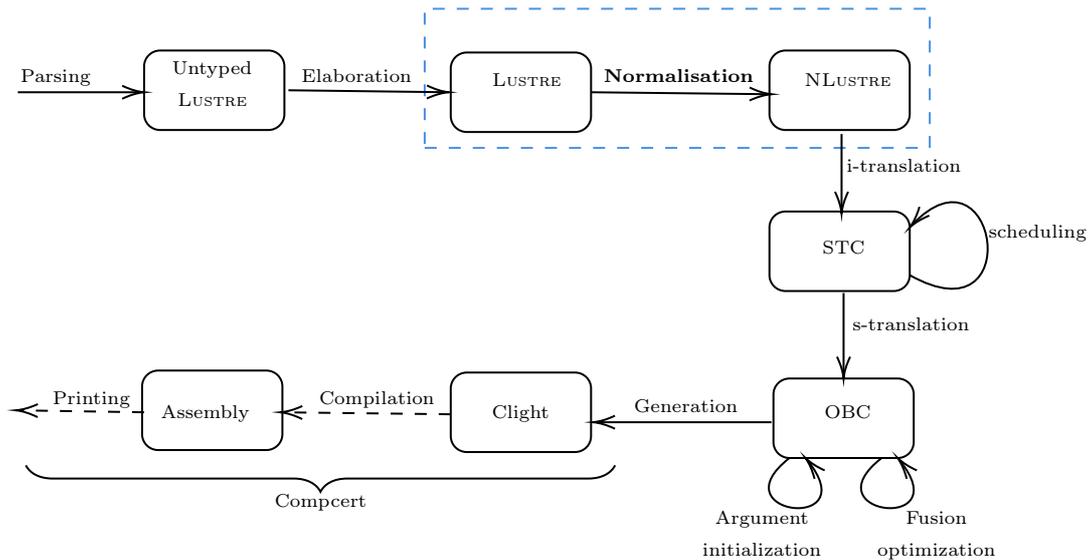
\begin{figure}
    \centering

\tikzset{every picture/.style={line width=0.75pt}} %set default line width to 0.75pt        

\begin{tikzpicture}[x=0.75pt,y=0.75pt,yscale=-1,xscale=1]
%uncomment if require: \path (0,381); %set diagram left start at 0, and has height of 381

%Rounded Rect [id:dp6262754485315697] 
\draw   (137,50) .. controls (137,45.58) and (140.58,42) .. (145,42) -- (199,42) .. controls (203.42,42) and (207,45.58) .. (207,50) -- (207,74) .. controls (207,78.42) and (203.42,82) .. (199,82) -- (145,82) .. controls (140.58,82) and (137,78.42) .. (137,74) -- cycle ;
%Rounded Rect [id:dp2932887415829737] 
\draw   (290,51) .. controls (290,46.58) and (293.58,43) .. (298,43) -- (352,43) .. controls (356.42,43) and (360,46.58) .. (360,51) -- (360,75) .. controls (360,79.42) and (356.42,83) .. (352,83) -- (298,83) .. controls (293.58,83) and (290,79.42) .. (290,75) -- cycle ;
%Rounded Rect [id:dp7774395753124194] 
\draw   (449,50) .. controls (449,45.58) and (452.58,42) .. (457,42) -- (511,42) .. controls (515.42,42) and (519,45.58) .. (519,50) -- (519,74) .. controls (519,78.42) and (515.42,82) .. (511,82) -- (457,82) .. controls (452.58,82) and (449,78.42) .. (449,74) -- cycle ;
%Rounded Rect [id:dp7230175543794733] 
\draw   (449,131) .. controls (449,126.58) and (452.58,123) .. (457,123) -- (511,123) .. controls (515.42,123) and (519,126.58) .. (519,131) -- (519,155) .. controls (519,159.42) and (515.42,163) .. (511,163) -- (457,163) .. controls (452.58,163) and (449,159.42) .. (449,155) -- cycle ;
%Rounded Rect [id:dp7869201381351568] 
\draw   (451,215) .. controls (451,210.58) and (454.58,207) .. (459,207) -- (513,207) .. controls (517.42,207) and (521,210.58) .. (521,215) -- (521,239) .. controls (521,243.42) and (517.42,247) .. (513,247) -- (459,247) .. controls (454.58,247) and (451,243.42) .. (451,239) -- cycle ;
%Rounded Rect [id:dp7898978262944473] 
\draw   (290,212) .. controls (290,207.58) and (293.58,204) .. (298,204) -- (352,204) .. controls (356.42,204) and (360,207.58) .. (360,212) -- (360,236) .. controls (360,240.42) and (356.42,244) .. (352,244) -- (298,244) .. controls (293.58,244) and (290,240.42) .. (290,236) -- cycle ;
%Rounded Rect [id:dp17918845237210834] 
\draw   (136,211) .. controls (136,206.58) and (139.58,203) .. (144,203) -- (198,203) .. controls (202.42,203) and (206,206.58) .. (206,211) -- (206,235) .. controls (206,239.42) and (202.42,243) .. (198,243) -- (144,243) .. controls (139.58,243) and (136,239.42) .. (136,235) -- cycle ;
%Shape: Rectangle [id:dp07808940942282694] 
\draw  [color={rgb, 255:red, 74; green, 144; blue, 226 }  ,draw opacity=1 ][dash pattern={on 4.5pt off 4.5pt}] (277,21) -- (529,21) -- (529,91) -- (277,91) -- cycle ;
%Straight Lines [id:da1430997440048979] 
\draw    (74,63) -- (135,63) ;
\draw [shift={(137,63)}, rotate = 180] [color={rgb, 255:red, 0; green, 0; blue, 0 }  ][line width=0.75]    (10.93,-3.29) .. controls (6.95,-1.4) and (3.31,-0.3) .. (0,0) .. controls (3.31,0.3) and (6.95,1.4) .. (10.93,3.29)   ;
%Straight Lines [id:da4486249704554761] 
\draw    (209,62) -- (287,62.98) ;
\draw [shift={(289,63)}, rotate = 180.72] [color={rgb, 255:red, 0; green, 0; blue, 0 }  ][line width=0.75]    (10.93,-3.29) .. controls (6.95,-1.4) and (3.31,-0.3) .. (0,0) .. controls (3.31,0.3) and (6.95,1.4) .. (10.93,3.29)   ;
%Straight Lines [id:da22681061831722638] 
\draw    (360,63) -- (448,63.98) ;
\draw [shift={(450,64)}, rotate = 180.64] [color={rgb, 255:red, 0; green, 0; blue, 0 }  ][line width=0.75]    (10.93,-3.29) .. controls (6.95,-1.4) and (3.31,-0.3) .. (0,0) .. controls (3.31,0.3) and (6.95,1.4) .. (10.93,3.29)   ;
%Straight Lines [id:da7474091173307768] 
\draw    (485,84) -- (485,122) ;
\draw [shift={(485,124)}, rotate = 270] [color={rgb, 255:red, 0; green, 0; blue, 0 }  ][line width=0.75]    (10.93,-3.29) .. controls (6.95,-1.4) and (3.31,-0.3) .. (0,0) .. controls (3.31,0.3) and (6.95,1.4) .. (10.93,3.29)   ;
%Straight Lines [id:da6300290835751637] 
\draw    (486,164) -- (486,205) ;
\draw [shift={(486,207)}, rotate = 270] [color={rgb, 255:red, 0; green, 0; blue, 0 }  ][line width=0.75]    (10.93,-3.29) .. controls (6.95,-1.4) and (3.31,-0.3) .. (0,0) .. controls (3.31,0.3) and (6.95,1.4) .. (10.93,3.29)   ;
%Straight Lines [id:da8931160541578665] 
\draw    (450,229) -- (363,229) ;
\draw [shift={(361,229)}, rotate = 360] [color={rgb, 255:red, 0; green, 0; blue, 0 }  ][line width=0.75]    (10.93,-3.29) .. controls (6.95,-1.4) and (3.31,-0.3) .. (0,0) .. controls (3.31,0.3) and (6.95,1.4) .. (10.93,3.29)   ;
%Straight Lines [id:da5669693552461903] 
\draw  [dash pattern={on 4.5pt off 4.5pt}]  (290,225) -- (207,224.02) ;
\draw [shift={(205,224)}, rotate = 0.67] [color={rgb, 255:red, 0; green, 0; blue, 0 }  ][line width=0.75]    (10.93,-3.29) .. controls (6.95,-1.4) and (3.31,-0.3) .. (0,0) .. controls (3.31,0.3) and (6.95,1.4) .. (10.93,3.29)   ;
%Straight Lines [id:da9478505922249792] 
\draw  [dash pattern={on 4.5pt off 4.5pt}]  (137,224) -- (75,223.03) ;
\draw [shift={(73,223)}, rotate = 0.9] [color={rgb, 255:red, 0; green, 0; blue, 0 }  ][line width=0.75]    (10.93,-3.29) .. controls (6.95,-1.4) and (3.31,-0.3) .. (0,0) .. controls (3.31,0.3) and (6.95,1.4) .. (10.93,3.29)   ;
% %Shape: Rectangle [id:dp045618772052431145] 
% \draw  [dash pattern={on 4.5pt off 4.5pt}] (71,172) -- (375,172) -- (375,276) -- (71,276) -- cycle ;
%Curve Lines [id:da8370618755996595] 
\draw    (519,155) .. controls (578.4,188.66) and (562.33,85.11) .. (520.28,129.61) ;
\draw [shift={(519,131)}, rotate = 311.86] [color={rgb, 255:red, 0; green, 0; blue, 0 }  ][line width=0.75]    (10.93,-3.29) .. controls (6.95,-1.4) and (3.31,-0.3) .. (0,0) .. controls (3.31,0.3) and (6.95,1.4) .. (10.93,3.29)   ;
%Curve Lines [id:da4867102008830375] 
\draw    (459,247) .. controls (422.37,275.71) and (497.47,284.82) .. (467.93,248.13) ;
\draw [shift={(467,247)}, rotate = 49.9] [color={rgb, 255:red, 0; green, 0; blue, 0 }  ][line width=0.75]    (10.93,-3.29) .. controls (6.95,-1.4) and (3.31,-0.3) .. (0,0) .. controls (3.31,0.3) and (6.95,1.4) .. (10.93,3.29)   ;
%Curve Lines [id:da10933129069986214] 
\draw    (505,247) .. controls (468.37,275.71) and (543.47,284.82) .. (513.93,248.13) ;
\draw [shift={(513,247)}, rotate = 49.9] [color={rgb, 255:red, 0; green, 0; blue, 0 }  ][line width=0.75]    (10.93,-3.29) .. controls (6.95,-1.4) and (3.31,-0.3) .. (0,0) .. controls (3.31,0.3) and (6.95,1.4) .. (10.93,3.29)   ;

% Text Node
\draw (365,50) node [anchor=north west][inner sep=0.75pt]   [align=left] {{\scriptsize \tbf{Normalisation}}};
% Text Node
\draw (214,50) node [anchor=north west][inner sep=0.75pt]   [align=left] {{\scriptsize Elaboration}};
% Text Node
\draw (74,50) node [anchor=north west][inner sep=0.75pt]   [align=left] {{\scriptsize Parsing}};
% % Text Node
% \draw (194,280) node [anchor=north west][inner sep=0.75pt]   [align=left] {{\scriptsize Compcert}};
% % Text Node
\draw (486,95) node [anchor=north west][inner sep=0.75pt]   [align=left] {{\scriptsize i-translation}};
% Text Node
\draw (489,175) node [anchor=north west][inner sep=0.75pt]   [align=left] {{\scriptsize s-translation}};
% Text Node
\draw (380,216) node [anchor=north west][inner sep=0.75pt]   [align=left] {{\scriptsize Generation}};
% Text Node
\draw (223,212) node [anchor=north west][inner sep=0.75pt]   [align=left] {{\scriptsize Compilation}};
% Text Node
\draw (90,212) node [anchor=north west][inner sep=0.75pt]   [align=left] {{\scriptsize Printing}};
% Text Node
\draw (149,46) node [anchor=north west][inner sep=0.75pt]   [align=left] {\begin{minipage}[lt]{30.59pt}\setlength\topsep{0pt}
\begin{center}
{\scriptsize Untyped}\\{\scriptsize \lustre}
\end{center}

\end{minipage}};
% Text Node
\draw (309,52) node [anchor=north west][inner sep=0.75pt]   [align=left] {{\scriptsize \tbf{\lustre}}};
% Text Node
\draw (465,52) node [anchor=north west][inner sep=0.75pt]   [align=left] {{\scriptsize \tbf{\nlustre}}};
% Text Node
\draw (474,136) node [anchor=north west][inner sep=0.75pt]   [align=left] {{\scriptsize STC}};
% Text Node
\draw (475,219) node [anchor=north west][inner sep=0.75pt]   [align=left] {{\scriptsize OBC}};
% Text Node
\draw (309,219) node [anchor=north west][inner sep=0.75pt]   [align=left] {{\scriptsize Clight}};
% Text Node
\draw (144,219) node [anchor=north west][inner sep=0.75pt]   [align=left] {{\scriptsize Assembly}};
% Text Node
\draw (557,128) node [anchor=north west][inner sep=0.75pt]   [align=left] {{\scriptsize scheduling}};
% Text Node
\draw (414,272) node [anchor=north west][inner sep=0.75pt]   [align=left] {\begin{minipage}[lt]{44.4pt}\setlength\topsep{0pt}
\begin{center}
{\scriptsize Argument}\\{\scriptsize initialization}
\end{center}

\end{minipage}};
 \draw [decoration={
        brace,
        raise=-7pt,
        amplitude=10pt,
        mirror
    }, decorate](78,260) -- (372,260) node [midway,below] {{\scriptsize Compcert}};
% Text Node
\draw (501,272) node [anchor=north west][inner sep=0.75pt]   [align=left] {\begin{minipage}[lt]{44.64pt}\setlength\topsep{0pt}
\begin{center}
{\scriptsize Fusion}\\{\scriptsize optimization}
\end{center}

\end{minipage}};

\end{tikzpicture}
\caption{Lustre compilation flow: This paper focuses on security in the phases in the blue dotted rectangle. \textit{q.v.} Fig 1 from \cite{emsoft2021} }
\label{FIG:compilation}
\end{figure}

The main result in this paper 
(\autoref{THM:Derived-Non-interference}) is (C2)
proving that our security type system is sound with respect to the \textit{co-inductive} Stream semantics for \lustre; we do so
by establishing \textit{non-interference} \cite{GoguenM82a} for well-typed programs. 
Rather than directly proving this result for the full language \lustre, we follow an approach that is common in programming language semantics and compiler correctness: 
We first prove (C2a) the non-interference result (\autoref{THM:Non-interference}) for a well-behaved sub-language called ``Normalised \lustre'' (\nlustre) which has a simpler structure and from which subsequent compilation is easier.
\nlustre\ is the language taken from \cite{Lustre2020} (though without the \ttt{reset} operator introduced there) and into which there is a semantics-preserving translation from the full language \cite{Bourke-TECS2021}.
We  show that Bourke \textit{et al.}'s semantics-preserving ``normalisation'' transformations from \lustre\ to \nlustre\ are (C2b) security-preserving as well (\autoref{THM:SecPreserve}).
This preservation of security types by the normalisation transformations is a property akin to ``subject reduction'' but at the level of compiler transformations.
The main result that well-security-typed \lustre\ programs are non-interfering (\autoref{THM:Derived-Non-interference}) follows from a \textit{reduction} to our earlier result of non-interference for \nlustre\  (\autoref{THM:Non-interference}) via the semantics-preservation results of Bourke \textit{et al.} \cite{Bourke-jfla2021,Bourke-TECS2021} and our type preservation result (\autoref{THM:SecPreserve}).
\autoref{FIG:compilation} situates the results within the framework of \lustre's compilation flow.

A supplement, in the Appendices, is a consolidated (Coq-independent) specification  of \lustre's stream semantics,  consistent with the V\'{e}lus formalisation \cite{Velus}.

\paragraph{Novelty}\ \ 
Security type systems have so far been proposed for imperative \cite{Denning:1976,Denning:1977} and for functional languages with imperative features \cite{Volpano96,Heintze98,Barthe04}.
We believe that ours is the first presentation of a SIF type system for a synchronous, reactive \textit{data-flow language}, together with its \textit{soundness} with respect to the operational semantics, 
While our approach to showing that securely-typed programs exhibit non-interference broadly follows that of Volpano \textit{et al} \cite{Volpano96}, we believe that the adaptation to a data-flow setting is both novel and inventive.
In particular, instead of notions of \textit{confinement checking} used to specify security in imperative paradigms, we generate and solve \textit{constraints} for equations and programs.
Thus we go beyond just \textit{checking} that a program is secure to \textit{inferring} constraints that suffice to ensure security, abstracting the SIF analysis from reasoning with respect to a fixed security lattice.

% Motivation for symbolic type inference system
\paragraph{Security type synthesis}\ \
Our treatment wherein security type constraints are accumulated over symbolic variables allows us to {\em synthesize} suitable security types. 
It is noteworthy that the resulting type inference system is divorced from a specific security lattice, but in fact can accommodate on-demand analysis of the various modes under which a system may operate. 
Consider an example (abstracted from a circuit design in which one of us was involved) where one has to implement one-time programmable SoC memory allowing for a {\em secret key} to be written once on the memory. 
The design may operate in one of the many modes:
\begin{enumerate}
\item {\em Write-mode} where the secret-key is burned into the memory,
\item {\em Lock-mode} where one can only verify whether an input
  matches the etched secret key, and
\item {\em Debug-mode} where cyclic redundancy check (CRC) value is
  generated
\end{enumerate}
There are various combinations of input and output security levels to the parameters (key, input, CRC), some of which are secure and others insecure.
For instance, once the memory is in Lock-mode, any access to CRC value
must have a suitably high privilege security level. If a {\em user}
wishes to access the memory in Debug-mode while the secret-key is at
the {\em root} privilege level, then this exhibits a case of insecure
combination of security levels.
By generating type constraints parametric in the types of the secret key, input, and CRC, one can reason over different scenarios \textit{without} having to repeatedly analyse the program or to re-compute the system of constraints. 
The same inferred constraints could be used to determine, \textit{e.g.}, whether or not to escalate a user's security in debug mode when the secret key has been already etched on the memory.

\paragraph{\bf Structure of the paper} \ \ 
In \autoref{SEC:Lustre}, we briefly overview the language \lustre\ (\S\ref{SUB:Lustre-overview}), and present its syntax (\S\ref{SUB:Lustre-syntax}) as well as that of its core sub-language \nlustre.
After a small example (\S\ref{SUB:Example}), we highlight the main features of its Stream semantics (\S\ref{SUB:LustreStreamSem}; see also \ref{APP:LustreStreamSem} for complete details).
In \autoref{SEC:LSec-types}, after recapitulating the notion of 
Denning's lattice-based model for secure information flow (SIF), we motivate the need for SIF types, which are then presented with their equational theory in \S\ref{SUB:SecTypes}. 
The basic type inference system is then presented in a syntax-oriented manner (\S\ref{SUB:SecTypeRules}). 
Based on these rules, we propose a definition of security for \lustre\ programs.
In \S\ref{SUB:NLTypes-COQ}, we provide a glimpse into our formalisation in Coq of the equational theory of our SIF types.
Soundness of the type system (for \nlustre) follows in \autoref{SEC:NLustreNonInterference}, where we  show that securely-typed \lustre\ programs exhibit \textit{non-interference} with respect to the Stream semantics (\autoref{THM:Non-interference}).
Bourke \textit{et al}'s \textit{translation} from \lustre\ to \nlustre\  \cite{Bourke-jfla2021,Bourke-TECS2021} in presented in \autoref{SEC:Normalisation}.
The main result here is the preservation of security types during this translation (\autoref{THM:SecPreserve}).
We illustrate this result with an example (\S\ref{SUB:AnalysisExample}).
The main result (\autoref{THM:Derived-Non-interference}) that well-typed \lustre\ programs exhibit non-interference follows in \autoref{SEC:noninterference}.
The paper concludes with a discussion of the related work (\autoref{SEC:related}) and directions for future work in \autoref{SEC:Conclusion}.
\ref{APP:LustreStreamSem} presents a consolidated specification of the Stream semantics of \lustre\ that is consistent with the V\'{e}lus/CompCert/Coq encoding  on a github repository \cite{Velus} mentioned in \cite{Lustre2020}.
Auxiliary definitions are presented in \ref{APP:FVDef} and \ref{APP:AuxPred}. 

This paper combines the results of \cite{PYS-memocode2020}, in which the type system for \nlustre\ and its semantic soundness were established, with the extension of the type system to the full \lustre\ language and the preservation of types during normalisation which was presented in \cite{PY-ictac2021}, and integrates them into a single self-contained account. 
The motivations and discussions have been expanded, and fortified with worked-out examples.  
Additional formalisation and proof details have been added, as well as directions for future work. 

% 2
\section{\lustre}
 \label{SEC:Lustre}
 
 \subsection{A brief overview of \lustre}
\label{SUB:Lustre-overview}
\lustre\ \cite{LusMain,LusClock,Lustre2020} is a synchronous data-flow language used for modelling, simulating, and verifying a wide range of reactive programs including embedded controllers, safety-critical systems, communication protocols, railway signal networks, etc. 
In \lustre, a reactive system is represented as a data-flow network with \textit{clocked} data streams flowing between operators and \textit{nodes}, \textit{i.e.},  a synchronous analogue to Kahn Process networks \cite{Kahn74}.

The main characteristics of the language are:
\begin{enumerate}
    \item \tti{Declarative Style}: A \lustre\ program consists of a set of definitions of named \textit{nodes}, each parameterised by tuples of clocked \textit{input} and \textit{output} flows.
    Node definitions are unique, and may appear in any order.
    Each node comprises a set of \textit{equations}, which  may be possibly mutually recursive provided they are ``well-clocked'' \cite{LusClock,LusMain}.
    The order in which equations are written has no effect on the semantics of the program.     
    Each equation in a node \textit{uniquely defines} a \textit{local} variable or an \textit{output} flow in terms of flow \textit{expressions}.
    The flow expressions mention only input, output or local variables (nodes do not have free variables), and may involve \textit{node calls}.
    Nodes cannot make recursive calls or have cyclic dependencies; therefore, the dependency order on nodes forms a DAG.
    Equations thus exhibit \tti{referential transparency}, referred to as \lustre's \tbf{Substitution Principle}.
    \item \tti{Deterministic Behaviour:} Program behaviours in \lustre\ are completely determined by sequences of clocked occurrences of events.  
    \lustre's \tbf{Definition Principle} states that the context does not determine the meaning of an expression.
    \item \tti{Synchronous Semantics:} Each variable and expression defines a data \textit{stream}, indexed with respect to a \textit{clock}.
    A clock is a stream of Boolean values.
    A flow takes its $n^\tti{th}$ value on the $n^\tti{th}$ clock tick, \textit{i.e.}, when the clock has value \true.
    A clock is either a \texttt{base} clock or one \textit{derived} from another clock when a variable takes a specific (boolean) value (\texttt{on} $x=k$, where $k \in \{ \ttt{T},\ttt{F} \}$).
    \item \tti{Temporal Operators} \ttt{when} (sampling), %\ttt{whenot}, 
    \ttt{merge} (interpolation) and \ttt{fby} (delay) are used to express complex clock-changing and clock-dependent behaviours.
    Following a static analysis to determine clock dependencies,
    all expressions and equations can be \textit{annotated} with a clock.
\end{enumerate} 
\lustre\ has a carefully designed system of static analyses including type checking, clock checking \cite{LusClock} and cyclic dependency checks \cite{LusMain}, the details of which are beyond the scope of this paper, which ensure the well-formedness and bounded space execution of programs.
\lustre\ has seen a steady development of its suite of tools over three decades, commencing with its introduction \cite{Lustre87} through to formally certified compiler developments \cite{SNLustre2017,Lustre2020}. 

\subsection{Syntax of \lustre\ and \nlustre}
\label{SUB:Lustre-syntax}

 % SYNTAX   
\begin{figure*}[ht]

\begin{minipage}{.45\textwidth}
\small
    \begin{align*}
    e  :={}& \tag{\tbf{expr}} \\
       | {}&~ c \tag{cnst} \\
       | {}&~ x \tag{var} \\
       | {}&~ \unop{e} \tag{unop} \\
       | {}&~ \binop{e}{e} \tag{binop} \\
       | {}&~ \when{\vv{e}}{x=k} \tag{whn}\\
       | {}&~ \lmerge{x}{\vv{e}}{\vv{e}} \tag{mrg}\\
       | {}&~ \ite{e}{\vv{e}}{\vv{e}}  \tag{ite}\\
       | {}&~ \fby{\vv{e}}{\vv{e}}        \tag{fby} \\
       | {}&~ f(\vv{e}) \tag{ncall}\\
  %     | {}&~ \whenot{e}{x} \tag{F-sample}
    % \end{align*}
    % \begin{align*}
   eq  :={}& \tag{\tbf{equation}} \\
      |{}& \vv{x} = \vv{e} \tag{eq} \\
      \\
      \\
    \end{align*}
    \caption{\lustre\ syntax}\label{FIG:LustreSyntax}
\end{minipage}%
\begin{minipage}{.55\textwidth}
   \small
       \begin{align*}
       e  :={}& \tag{\tbf{expr'}} \\
          | {}&~ c \tag{cnst} \\
          | {}&~ x \tag{var} \\
          | {}&~ \unop{e} \tag{unop} \\
          | {}&~ \binop{e}{e} \tag{binop} \\
          | {}&~\when{e}{x=k}  \tag{whn'}\\
      ce  :={}& \tag{\tbf{cntrl expr}} \\
          | {}&~ e \tag{expr'} \\
          | {}&~ \lmerge{x}{ce}{ce} \tag{mrg'} \\
          | {}&~ \ite{e}{ce}{ce} \tag{ite'} \\
      eq  :={}& \tag{\tbf{equation}} \\
          | {}&~ \eqn{x}{ce}{ck} \tag{eq'} \\
          | {}&~ \eqn{x}{\fby{c}{e}}{ck} \tag{fby'} \\
          | {}&~ \eqn{\vv{x}}{\nodecall{f}{e}}{ck} \tag{ncall'} \\
     %     | {}&~ \whenot{e}{x} \tag{F-sample}
       % \end{align*}
       % \begin{align*} 
       \end{align*}
       \caption{\nlustre\ syntax}\label{FIG:NLustreSyntax}
   \end{minipage}

   %  u  :={}&  \tag{\tbf{single flow annot}}\\
   %     | {}&~\etype{\tau}, nck  \\
   %  s  :={}&  \tag{\tbf{synchronized flows annot}}\\
   %     | {}&~\etype{\tau}^{+}, nck  \\  
   \begin{minipage}{.55\textwidth}
      \begin{align*}
      ck  :={}& \tag{\tbf{clock}} \\
       | {}&~ \ttt{base} \tag{base} \\
       | {}&~ \ck{ck}{(x=k)} \tag{on} \\
      \end{align*}
   \end{minipage}%
    \begin{minipage}{.45\textwidth}
      \begin{align*}
      d  :={}& \tag{\tbf{node declr}} \\
      | ~{}& \Node \\
      G:= {}& \vv{d} \tag{\tbf{program}}
      \end{align*}
    \end{minipage}

   %  \begin{align*}
   % G := {}& \tag{\tbf{program}} \\
   %   | {}& ~\vv{d} 
   % \end{align*}
    
    \caption{Common syntax of programs, nodes and clocks}\label{FIG:ClockNodeSyntax}

\end{figure*}

As is common in compilation, the full language \lustre\ can be  translated into a core sub-language \nlustre\ \cite{Bourke-jfla2021}, from which subsequent compilation is easier.
The language \nlustre\ considered in this paper is the core language  taken from \cite{Lustre2020,Auger2013}, but without the \ttt{reset} operator introduced there.
\autoref{FIG:LustreSyntax}--\autoref{FIG:ClockNodeSyntax} 
present the syntax of \lustre\ and \nlustre. 
(Notation: \lustre\ keywords are written in \ttt{teletype} face and coloured in example listings; meta-variables are in $\mathit{italic}$ face.)

\lustre\ expressions (\autoref{FIG:LustreSyntax}) include flows described by constants, variables, unary and binary operations on flows, as well as the flows obtained by sampling when a variable takes a particular boolean value (\texttt{when}), interpolation based on a boolean variable flow (\texttt{merge}), and conditional combinations of flows (\texttt{if\_then\_else}).
Of particular interest are flows involving guarded delays (\texttt{fby}) and \textit{node calls}.

The syntax for clocks, node definitions and programs is given in \autoref{FIG:ClockNodeSyntax}, while the syntax for expressions and equations in \nlustre\ is listed in \autoref{FIG:NLustreSyntax}.

The main differences between \lustre\ and \nlustre\ are 
(i) the former supports \textit{lists} of flows (written $\vv{e}$) for conciseness, whereas in the latter all flows are single streams;
(ii) \nlustre\ requires that conditional and \texttt{merge} ``control'' expressions are not nested below unary and binary operators or sampling;
(iii) node call and delayed flows (\texttt{fby}) are treated as first-class expressions, whereas in \nlustre, they can appear only in the context of equations;
(iv) \lustre\ permits nested node calls, whereas nesting in disallowed in \nlustre;
(v) finally,  the first argument of \texttt{fby} expressions in \nlustre\; must be a \textit{constant}, to enable a well-defined initialisation that can be easily implemented.

The \textit{translation} from \lustre\ to \nlustre\  \cite{Bourke-jfla2021,Bourke-TECS2021} involves \textit{distributing} constructs over the individual components of lists of expressions, and \textit{de-nesting} expressions by introducing fresh local variables (See \autoref{SEC:Normalisation} for details).
% The reader can see an example, adapted from \cite{Bourke-jfla2021}, of a \lustre\ program and its translation into \nlustre\ in  \autoref{FIG:L2NLEx2} (ignoring for the moment the \type{security type annotations} therein).

\subsection{\lustre\ Example}
\label{SUB:Example}
We present a small example of a \lustre\ program.
(For the moment, let us ignore the \textcolor{blue}{\textit{blue}} superscript {security type annotations}.)
\begin{figure*}
   \begin{minipage}{.51\textwidth}
\begin{lstlisting}[language=Lustre,label=LST:EgCounter]
-- a simple counter node with a reset
node Ctr(init$^{\type{\alpha_1}}$, incr$^{\type{\alpha_2}}$:int, rst$^{\type{\alpha_3}}$:bool) 
   returns (n$^{\type{\beta}}$:int); 
var fst$^{\type{\delta_1}}$: bool, pre_n$^{\type{\delta_2}}$: int;
let
  n =$_{base^{\type{\gamma}}}$ if (fst$^{\type{\delta_1}}$ or rst$^{\type{\alpha_3}}$) then init$^{\type{\alpha_1}}$
        else pre_n$^{\type{\delta_2}}$ + incr$^{\type{\alpha_2}}$;        
  fst   =$_{base^{\type{\gamma}}}$ true$^{\type{\bot}}$ fby false$^{\type{\bot}}$;  
  pre_n =$_{base^{\type{\gamma}}}$ 0$^{\type{\bot}}$ fby n$^{\type{\beta}}$;       
tel
    \end{lstlisting}
    \caption{Counter}\label{EG:Ctr-listing}
    \end{minipage}
    \begin{minipage}{.49\textwidth}
    \begin{lstlisting}[language=Lustre,label=LST:SpdMtr]
    -- using the counter node
    node SpdMtr(acc$^{\type{\alpha_4}}$:int) 
       returns (spd$^{\type{\beta_1}}$,pos$^{\type{\beta_2}}$:int);
    let 
      spd =$_{base^{\type{\gamma_2}}}$ Ctr(0$^{\type{\bot}}$,acc$^{\type{\alpha_4}}$,false$^{\type{\bot}}$);
      pos =$_{base^{\type{\gamma_2}}}$ Ctr(3$^{\type{\bot}}$,spd$^{\type{\beta_1}}$,false$^{\type{\bot}}$);
    tel
\end{lstlisting}
    \caption{Speedometer}\label{EG:SpdMtr-listing}
     \end{minipage}   
\end{figure*}
\begin{example}[Counter]\label{EG:Ctr}
\ttt{Ctr}, defined in \autoref{EG:Ctr-listing}, is a node which takes two integer stream parameters \ttt{init} and \ttt{incr} and a boolean parameter \ttt{rst}, representing (respectively) an initial value, the increment and a reset signal stream. 
The output is the integer stream \ttt{n}.
Two local variables are declared:  the boolean stream \ttt{fst}, which is true initially and false thereafter, and the integer stream \ttt{pre\_n}, which latches onto the previous value of \ttt{n}.
The equation for \ttt{n} sets it to the value of \ttt{init} if either \ttt{fst} or \ttt{rst} is true, otherwise adding \ttt{incr} to \ttt{n}'s previous value (\ttt{pre\_n}).
\ttt{pre\_n} is initialised to 0, and thereafter (using \ttt{fby}) trails the value of \ttt{n} by a clock instant. 
All equations here are on the same implicit ``base'' clock, and the calculations may be considered as occurring synchronously. 
\end{example} 

An example run of \ttt{Ctr} is:
\begin{small}
\centering
    \begin{tabular}[t]{|c| c c c c c c c c|} 
    \hline
    \textit{Flow} & {\textit{Values}} & {} & {} & {} & {}  &{} & {} & {}  \\
    \hline
    \hline
    \texttt{init} & \stream{1} & \stream{2} & \stream{1} & \stream{1} & \stream{0} & \stream{2} & \stream{4} & $\ldots$ \\
    \hline
    \texttt{incr} & \stream{1} & \stream{2} & \stream{2} & \stream{3} & \stream{3} & \stream{1} & \stream{2} & $\ldots$ \\
    \hline
    \texttt{rst}& \stream{F} & \stream{F} & \stream{F} & \stream{F} & \stream{T} & \stream{F} & \stream{T} & $\ldots$ \\
    \hline
    \texttt{fst} & \stream{T} & \stream{F} & \stream{F} & \stream{F} & \stream{F} & \stream{F} & \stream{F} & $\ldots$ \\
    \hline
    \texttt{n} & \stream{1} & \stream{3} & \stream{5} & \stream{8} & \stream{0} & \stream{1} & \stream{4} & $\ldots$ \\
    \hline
    \texttt{pre\_n} & \stream{0} & \stream{1} & \stream{3} & \stream{5} & \stream{8} & \stream{0} & \stream{1} & $\ldots$ \\
    
    \hline
   \end{tabular}
\end{small}
\begin{example}[Speedometer]\label{EG:SpdMtr}
In \autoref{EG:SpdMtr-listing}, we next define another node, \ttt{SpdMtr}, using two instances of node \ttt{Ctr}.
\ttt{spd} is calculated by invoking \ttt{Ctr} with suitable initial value $0$ and increment $\ttt{acc}$, while \ttt{pos} is calculated with initial value $3$ and increment \ttt{spd}.
Again, both equations are on the same base clock, and the calculations are synchronous. 
In the exaample code in \autoref{EG:SpdMtr-listing}, the two instances of \ttt{Ctr} are never reset. 
\end{example}
%

% Insert Lustre example and its translation to NLustre

\subsection{Stream Semantics of \lustre}
\label{SUB:LustreStreamSem}

% Name the Rules in a way consistent with the syntax
\ignore{
\begin{figure*}
    $$
    \begin{array}{c}
        \namedJdg{\semConst{c}{bs}{cs}}
        {\mstrSemExpPred{G,\Hst}{bs}{c}{[cs]}}{LScnst} ~~~
        \namedJdg{\Hst(x)={\textcolor{BrickRed}{xs}}}
        {\mstrSemExpPred{G,\Hst}{bs}{x}{[xs]}}{LSvar} \\
        \\
        \namedJdg{\predSet{\mstrSemExpPred{G,\Hst}{bs}{e}{[es]}}
            {\liftunop{es}{os} }}
        {\mstrSemExpPred{G,\Hst}{bs}{\unop{e}}{[os]}}{LSunop} \\
        \\
        \namedJdg{\predSet{\mstrSemExpPred{G,\Hst}{bs}{e_1}{[es_1]}}
            {\mstrSemExpPred{G,\Hst}{bs}{e_2}{[es_2]}}
            {\liftbinop{es_1}{es_2}{os} }}
        {\mstrSemExpPred{G,\Hst}{bs}{\binop{e_1}{e_2}}{[os]}}{LSbinop} \\
        \\
        \namedJdg{\predSet{\forall i ~\mstrSemExpPred{G,\Hst}{bs}{e_i}{\tupstrm{es}_i}}
            {\hist{x}={\textcolor{BrickRed}{xs}}}
            {\forall i:~\mapwhenk{k}{xs}{\tupstrm{es}_i}{\tupstrm{os}_i}}}
        {\mstrSemExpPred{G,\Hst}{bs}{\when{\vv{e_i}}{x=k}}{
        \flatten{\vv{\tupstrm{os}_i}}}}{LSwhn}\\
        \\
        \namedJdg{\dependSet{\predSet{\mstrSemExpPred{G,\Hst}{bs}{e}{[es]}}
        {\forall i:~ ~\mstrSemExpPred{G,\Hst}{bs}{et_i}{\tupstrm{ets}_i}}
        }
        {
           \forall j:~  \mstrSemExpPred{G,\Hst}{bs}{ef_j}{\tupstrm{efs}_j} 
            ~~~~
            \mapsemIte{es}{(\flatten{\vv{\tupstrm{ets}_i}})}{
            (\flatten{\vv{\tupstrm{efs}_j}})}{\tupstrm{os}}
            }}
        {\mstrSemExpPred{G,\Hst}{bs}{\ite{e}{\vv{et_i}}{\vv{ef_j}}}{\tupstrm{os}}}{LSite} \\
        \\
       \namedJdg{\dependSet{\predSet{\hist{x}=xs}{
       \forall i: ~\mstrSemExpPred{G,\Hst}{bs}{et_i}{\tupstrm{ets}_i}
       }}
            {\forall j: ~
            \mstrSemExpPred{G,\Hst}{bs}{ef_j}{\tupstrm{efs}_j}
            ~~~~
             \mapsemMerge{xs}{(\flatten{\vv{\tupstrm{ets}_i}})}{
            (\flatten{\vv{\tupstrm{efs}_j}})}{\tupstrm{os}}}
            }
       {\mstrSemExpPred{G,\Hst}{bs}{\lmerge{x}{\vv{et_i}}{\vv{ef_j}}}{\tupstrm{os}}}{LSmrg} \\
       \\
       \namedJdg{\predSet{\forall i:~ ~\mstrSemExpPred{G,\Hst}{bs}{e0_i}{\tupstrm{e0s}_i}
      ~~~~ \forall j:~ 
       \mstrSemExpPred{G,\Hst}{bs}{e_j}{\tupstrm{es}_j}}
            { ~\mapsemLFby{(\flatten{\vv{\tupstrm{e0s}_i}})}{
            (\flatten{\vv{\tupstrm{es}_j}})}{\tupstrm{os}}}}
        {\mstrSemExpPred{G,\Hst}{bs}{\fby{\vv{e0_i}}{\vv{e_j}}}{\tupstrm{os}}}{LSfby} \\
   \\
        \namedJdg{\forall i \in [1,k] ~ \mstrSemExpPred{G,\Hst}{bs}{e_i}{\tupstrm{es}_i} ~~~
        [\hist{x_1}, \ldots, \hist{x_n}] = 
        \flatten{\vv{\tupstrm{es}_i}}
        }
        {\mstrEqnPred{G}{\Hst}{bs}{\vv{x_j}=\vv{e_i}}}{LSeq} \\
  \\
    \namedJdg{\dependSet{\predSet{\node \in G}
        {\hist{n.\tbf{in}} = \tupstrm{xs}}
        {\baseOf{\tupstrm{xs}} = bs}}
        {\predSet{\hist{n.\tbf{out}} = \tupstrm{ys}}
        {\forall eq \in \vv{eq}:~ \mstrEqnPred{G}{\Hst}{bs}{eq}}
         }
    }{\mstrCallPred{G}{\liftnode{f}}{\tupstrm{xs}}{\tupstrm{ys}}}{LSndef} \\
    \\
           \namedJdg{\predSet{\mstrSemExpPred{G,\Hst}{bs}{\vv{e_i}}{\tupstrm{xs}}}
            {\mstrCallPred{G}{\liftnode{f}}{\tupstrm{xs}}{\tupstrm{ys}}}}
        {\mstrSemExpPred{G,\Hst}{bs}{f(\vv{e_i})}{\tupstrm{ys}}}{LSncall}
    \end{array}
    $$
    \caption{Stream semantics of \lustre}
    \label{FIG:LusStrSemNodeEqn}
\end{figure*}
}

The semantics of \lustre\ and \nlustre\  programs are \textit{synchronous}:
Each variable and expression defines a data stream which pulses with respect to a \textit{clock}.
A clock is a stream of booleans (CompCert/Coq's \cite{Leroy2009,CoqMan} $\true$ and $\false$ in V\'{e}lus).
A flow takes its $n^\tti{th}$ value on the $n^\tti{th}$ clock tick, \textit{i.e.},  some value, written $\stream{v}$, is present at instants when the clock value is \true, and none (written $\nullStream)$ when it is \false.
The \textit{temporal operators} \ttt{when}, \ttt{merge} and \ttt{fby} are used to express the complex clock-changing and clock-dependent behaviours of sampling, interpolation and delay respectively.

%Program behaviours in \lustre\ are completely determined by sequences of clocked occurrences of events.

Formally the stream semantics is defined using predicates over the program graph $G$, a (co-inductive) stream \textit{history} ($\Hst: \tti{Ident} \rightarrow \tti{value}~\tti{Stream}$) that associates value streams to variables, and a clock $bs$ \cite{Lustre2020,PYS-memocode2020,Bourke-jfla2021}.
Semantic operations on (lists of) streams are written in \textcolor{blue}{blue \textsf{sans serif}} typeface.
Streams are written in \textcolor{BrickRed}{red}, with lists of streams usually written in \textbf{\textcolor{BrickRed}{bold face}}.
All these stream operators, defined co-inductively,  enforce the clocking regime, ensuring  the presence of a value when the clock is \ckFont{true}, and absence when \ckFont{false}.
\ref{APP:AuxPred} contains a complete specification of these auxiliary predicates.

The predicate $\mstrSemExpPred{G,\Hst}{bs}{e}{\tupstrm{es}}$ relates an \textit{expression} $e$ to a \textit{list} of streams, written $\tupstrm{es}$.
A list consisting of only a single stream $\textcolor{BrickRed}{es}$ is explicitly denoted as $\textcolor{BrickRed}{[es]}$.
The semantics of \textit{equations} are expressed using the predicate 
$\mstrEqnPred{G}{\Hst}{bs}{\vv{eq_i}}$, which requires \textit{consistency} between the assumed and defined stream histories in $\Hst$ for the program variables, induced by the equations $\vv{eq_i}$.
Finally, the semantics of a \textit{node} named $f$ in program $G$ is given via a predicate $\mstrCallPred{G}{\liftnode{f}}{\tupstrm{xs}}{\tupstrm{ys}}$, which defines a stream history transformer $\liftnode{f}$ that maps the list of streams $\tupstrm{xs}$ to the list of streams $\tupstrm{ys}$.

% Then reference the figure AND 
% explain the rules of the Figure.
We discuss here only some constructs, especially those that are important to the normalisation transformations.
\ref{APP:LustreStreamSem} presents a complete account of the stream semantics for \lustre\ and \nlustre, consistent with the Coq developments in \cite{Velus}.

\ignore{
\autoref{FIG:LusStrSemNodeEqn} presents the stream semantics for \lustre.
While rules for \textit{some} constructs have been variously presented \cite{SNLustre2017,Lustre2020,PYS-memocode2020,Bourke-jfla2021}, our presentation can be considered as a definitive consolidated specification of the operational semantics of \lustre, consistent with the V\'{e}lus compiler encoding \cite{Velus}.
}
\ignore{
Rule (LScnst) states that a constant $c$ denotes a constant stream of the value $\stream{c}$ pulsed according to given clock $bs$.  
This is effected by the semantic operator \ckFont{const}.
}
\[
        \namedJdg{\Hst(x)={\textcolor{BrickRed}{xs}}}
        {\mstrSemExpPred{G,\Hst}{bs}{x}{[xs]}}{LSvar}
\]
Rule (LSvar) associates the expression consisting of a variable $x$ to the stream given by $\Hst(x)$. 
\ignore{
In rule (LSunop), \ckFont{$\hat{\diamond}$} denotes the operation $\diamond$ \textit{lifted} to apply instant-wise to the stream denoted by expression $e$.
Likewise in rule (LSbinop), the  binary operation $\oplus$ is applied paired point-wise to the streams denoted by the two sub-expressions (which should both pulse according to the same clock).
In all these rules, an expression is associated with a \textit{single} stream. 
\[
                \namedJdg{\predSet{\forall i ~\mstrSemExpPred{G,\Hst}{bs}{e_i}{\tupstrm{es}_i}}
            {\hist{x}={\textcolor{BrickRed}{xs}}}
            {\forall i:~\mapwhenk{k}{xs}{\tupstrm{es}_i}{\tupstrm{os}_i}}}
        {\mstrSemExpPred{G,\Hst}{bs}{\when{\vv{e_i}}{x=k}}{
        \flatten{\vv{\tupstrm{os}_i}}}}{LSwhn}
\]
 
The rule (LSwhn) describes \textit{sampling} whenever a variable $x$ takes the boolean value $k$, from the flows arising from a list of expressions $\vv{e_i}$,  returning a list of streams of such sampled values.
The predicate $\widehat{\ckFont{when}}$ \textit{maps} the predicate $\ckFont{when}$ to act on the corresponding components of \textit{lists} of streams, \textit{i.e.},  \[ 
\mapwhenk{k}{xs}{[es_1, \ldots, es_k]}{[os_1, \ldots, os_k]} 
~\textrm{abbreviates} ~ 
\bigwedge_{i \in [1,k]}~ \whenk{k}{xs}{es_i}{os_i}.
\]
(Similarly for the predicates
$\widehat{\ckFont{merge}}$, $\widehat{\ckFont{ite}}$, and $\widehat{\ckFont{fby}_L}$.  
The operation $\flatten{\_}$ flattens a list of lists (of possibly different lengths) into a single list. 
Flattening is required since expression $e_i$ may in general denote a \textit{list} of streams \textcolor{BrickRed}{$\tupstrm{es}_i$}.
}
\ignore{
The expression $\lmerge{x}{\vv{et}_i}{\vv{ef}_j}$ achieves (lists of) streams on a faster clock.
The semantics in rule (LSmrg) assume that for each pair of corresponding component streams from
$\flatten{\tupstrm{ets}_i}$ and
$\flatten{\tupstrm{efs}_j}$, a value is present in the first stream and absent in the second at those instances where $x$ has a true value $\stream{T}$, and complementarily, a value is present in the second stream and absent in the first when $x$ has a false value $\stream{F}$.
Both values must be absent when $x$’s value is absent.
These conditions are enforced by the auxiliary semantic operation \ckFont{merge}. 
In contrast, the conditional expression $\ite{e}{\vv{et}}{\vv{ef}}$
requires that all three argument streams $es$, and the corresponding components from $\flatten{\vv{\tupstrm{ets}_i}}$ and
$\flatten{\vv{\tupstrm{efs}_j}}$ pulse to the same clock.
Again, values are selected from the first or second component streams depending on whether the stream $es$ has the value $\stream{T}$ or $\stream{F}$ at a particular instant.
These conditions are enforced by the auxiliary semantic operation \ckFont{ite}. 
}
\[
       \namedJdg{\dependSet{\predSet{\forall i:~ ~\mstrSemExpPred{G,\Hst}{bs}{e0_i}{\tupstrm{e0s}_i}
      ~~~~ \forall j:~ 
       \mstrSemExpPred{G,\Hst}{bs}{e_j}{\tupstrm{es}_j}}}
       {\predSet{
             ~\mapsemLFby{(\flatten{\vv{\tupstrm{e0s}_i}})}{
            (\flatten{\vv{\tupstrm{es}_j}})}{\tupstrm{os}}}}
            }
        {\mstrSemExpPred{G,\Hst}{bs}{\fby{\vv{e0_i}}{\vv{e_j}}}{\tupstrm{os}}}{LSfby}
\]
A delay operation is implemented by $\fby{e0}{e}$.
The rule (LSfby) is to be read as follows.
Let each expression $e0_i$ denote a list of streams
$\tupstrm{e0s}_i$, and each expression $e_j$ denote a list of streams $\tupstrm{es}_j$.
The predicate $\widehat{\ckFont{fby}_L}$ \textit{maps} the predicate $\ckFont{fby}_L$ to act on the corresponding components of \textit{lists} of streams, \textit{i.e.},  \[ 
\mapsemLFby{\tupstrm{xs}}{\tupstrm{ys}}{\tupstrm{zs}} 
~\textrm{abbreviates} ~ 
\bigwedge_{i \in [1,m]}~ \semLFby{xs_i}{ys_i}{zs_i}
\]
(Similarly for the predicates
$\widehat{\ckFont{when}}$,
$\widehat{\ckFont{merge}}$, and $\widehat{\ckFont{ite}}$.) 
The operation $\flatten{\_}$ flattens a list of lists (of possibly different lengths) into a single list. 
Flattening is required since expression $e_i$ may in general denote a \textit{list} of streams \textcolor{BrickRed}{$\tupstrm{es}_i$}.
The output list of streams $\tupstrm{os}$ consists of streams whose first elements are taken from each stream in $\flatten{\vv{\tupstrm{e0s}_i}}$ with the rest taken from the corresponding component of $\flatten{\vv{\tupstrm{es}_j}}$.
%The output streams are obtained using the semantic operation $\widehat{\ckFont{fby}_L}$.
%
\[
  \namedJdg{\forall i \in [1,..,k] ~ \mstrSemExpPred{G,\Hst}{bs}{e_i}{\tupstrm{es}_i} ~~~
        [\hist{x_1}, \ldots, \hist{x_n}] = 
        \flatten{\vv{\tupstrm{es}_i}}
        }
        {\mstrEqnPred{G}{\Hst}{bs}{\vv{x_j}=\vv{e_i}}}{LSeq}
\]
The rule (LSeq) for equations checks the consistency between the assumed meanings for the defined variables $x_j$ according to the history $\Hst$ with the corresponding components of the tuple of streams $\flatten{\vv{\tupstrm{es}_i}}$ to which a tuple of right-hand side expressions evaluates.

\[
            \namedJdg{\dependSet{\predSet{\node \in G}
        {~~\hist{f.\tbf{in}} = \tupstrm{xs}}
        }
        {\predSet{\hist{f.\tbf{out}} = \tupstrm{ys}}{\baseOf{\tupstrm{xs}} = bs}
        {\forall eq \in \vv{eq}:~ \mstrEqnPred{G}{\Hst}{bs}{eq}}
         }
    }{\mstrCallPred{G}{\liftnode{f}}{\tupstrm{xs}}{\tupstrm{ys}}}{LSndef}
\]
The rule (LSndef) presents the meaning given to the definition of a node named $f \in G$ as a stream list transformer. 
If history $\Hst$ assigns lists of streams to the input and output variables for a node in a manner such that the semantics of the equations $\vv{eq}$ in the node are satisfied, then the semantic function $\liftnode{f}$ transforms input stream list $\tupstrm{xs}$ to output stream list $\tupstrm{ys}$.
The operation \ckFont{base-of} finds an appropriate base clock with respect to which a given list of value streams pulse.
\[
       \namedJdg{\predSet{\mstrSemExpPred{G,\Hst}{bs}{\vv{e}}{\tupstrm{es}}}
       {\mstrCallPred{G}{\liftnode{f}}{\tupstrm{es}}{\tupstrm{os}}}}
        {\mstrSemExpPred{G,\Hst}{bs}{f(\vv{e})}{\tupstrm{os}}}{LSncall}
\]
The rule (LSncall) applies the stream transformer semantic function $\liftnode{f}$ defined in rule (LSndef) to the stream list $\tupstrm{es}$ corresponding to the tuple of arguments $\vv{e}$, and returns the stream list $\tupstrm{os}$.

\ignore{
\paragraph{Clocks and clock-annotated expressions.} \ 
We next present rules for clocks.
Further, we  assume that all (\nlustre) expressions in equations can be clock-annotated, and present the corresponding rules.

\begin{figure*}
$$
   \begin{array}{c}
        \namedJdg{}
        {\mstrSemCkPred{\Hst}{bs}{\ttt{base}}{bs}}{LSbase}  \\ \\
        \namedJdg{
            \dependSet{\predSets{\mstrSemCkPred{\Hst}{bs}{ck}{\cc{\true}{bk}}}
            {\hist{x}=\cc{\stream{k}}{xs}}}
            {\mstrSemCkPred{\htl{\Hst}}{\tl{bs}}{\on{ck}{x}{k}}{bs'}}}
        {\mstrSemCkPred{\Hst}{bs}{\on{ck}{x}{k}}{\cc{\true}{bs'}}}{LSonT} \\

        \namedJdg{
            \dependSet{\predSets{\mstrSemCkPred{\Hst}{bs}{ck}{\cc{\false}{bk}}}
            {\hist{x}=\cc{\nullStream}{xs}}}
            {\mstrSemCkPred{\htl{\Hst}}{\tl{bs}}{\on{ck}{x}{k}}{bs'}}}
        {\mstrSemCkPred{\Hst}{bs}{\on{ck}{x}{k}}{\cc{\false}{bs'}}}{LSonA1} 
        \\
               \namedJdg{
            \dependSet{
             \predSet{\mstrSemCkPred{\Hst}{bs}{ck}{\cc{\true}{bk}}}
                {\hist{x}=\cc{\stream{k}}{xs}}
            }
            {\mstrSemCkPred{\htl{\Hst}}{\tl{bs}}{\onF{ck}{x}{k}}{bs'}}}
        {\mstrSemCkPred{\Hst}{bs}{\onF{ck}{x}{k}}{\cc{\false}{bs'}}}{LSonA2} \\ \\

        \namedJdg{\dependSet{\mstrSemExpPred{\Hst}{bs}{e}{[\ccnb{\nullStream}{es}]}}
        {\mstrSemCkPred{\Hst}{bs}{ck}{\ccnb{\false}{cs}}}}
    {\mstrSemExpPred{\Hst}{bs}{e::ck}{[\ccnb{\nullStream}{es}]}}{NSaeA} ~~
    
        \namedJdg{\dependSet{\mstrSemExpPred{\Hst}{bs}{e}{[\ccnb{\stream{v}}{es}]}}
            {\mstrSemCkPred{\Hst}{bs}{ck}{\ccnb{\true}{cs}}}}
        {\mstrSemExpPred{\Hst}{bs}{e::ck}{[\ccnb{\stream{v}}{es}]}}{NSae} 
    \end{array}
$$   
    \caption{Stream semantics of clocks and annotated expressions}
    \label{FIG:StrSemCk}
\end{figure*}

The predicate 
$\mstrSemCkPred{\Hst}{bs}{ck}{bs'}$ in \autoref{FIG:StrSemCk} defines the meaning of a \nlustre\ clock expression $ck$ with respect to a given history $\Hst$ and a clock $bs$ to be the resultant clock $bs'$.
% Note that a clock is a stream of booleans.  
The \ttt{on} construct lets us define 
coarser clocks derived from  a given clock --- whenever a variable $x$ has the desired value $k$ and the given clock is true. 
The rules (LSonT), (LSonA1), and (LSonA2) 
present the three cases: respectively when variable $x$ has the desired value $k$ and clock is true; the clock is false; and  the program variable $x$ has the complementary value and the clock is true.
The auxiliary operations \ckFont{tl} and \ckFont{htl},  give the tail of a stream, and the tails of streams for each variable according to a given history $\Hst$. 
Rules (NSaeA)-(NSae) describe the semantics of  clock-annotated expressions, where the output stream carries a value exactly when the clock is true.
}

\paragraph{Stream semantics for \nlustre} \ \ 
The semantic relations for \nlustre\ are either identical to (as in constants, variables, unary and binary operations) or else the singleton cases of the rules for \lustre\ (as in \ttt{merge}, \ttt{ite}, \ttt{when}).
The main differences lie in the occurrences of \ttt{fby} (now in a restricted form) and node call, which can only be in the context of (clock-annotated) equations.
\begin{figure*}
   $$
    \begin{array}{c}
       \namedJdg{\predSet{\mstrSemExpPred{\Hst}{bs}{e::ck}{[vs]}}
            {\semFby{c}{vs}{\hist{x}}}}
        {\mstrEqnPred{G}{\Hst}{bs}{\eqn{x}{\fby{c}{e}}{ck}}}{NSfby'} \\
\\  
    \namedJdg{\dependSet{\predSet{\node \in G}
        {~\hist{f.\tbf{in}} = \tupstrm{xs}}
        {~\baseOf{\tupstrm{xs}} = bs}}
        {\predSet{~\resClk{\Hst}{bs}}
        {~~\hist{f.\tbf{out}} = \tupstrm{ys}}
        {~~\forall eq \in \vv{eq}:~ \mstrEqnPred{G}{\Hst}{bs}{eq}}
         }
    }{\mstrCallPred{G}{\liftnode{f}}{\tupstrm{xs}}{\tupstrm{ys}}}{NSndef'} \\
    \\
\ignore{
        \namedJdg{\mstrSemExpPred{\Hst}{bs}{e::ck}{\hist{x}}}
        {\mstrEqnPred{G}{\Hst}{bs}{\eqn{x}{e}{ck}}}{NSeq} 
\\ 
}
   
    \namedJdg{\predSet{\mstrSemExpPred{\Hst}{bs}{\vv{e}}{\tupstrm{es}~~~}}
    {\mstrSemCkPred{\Hst}{bs}{ck}{\baseOf{\tupstrm{es}}}}
    {~~~\mstrCallPred{G}{\liftnode{f}}{\tupstrm{es}}{\vv{\hist{x_i}}}}}
            {\mstrEqnPred{G}{\Hst}{bs}
        {\eqn{\vv{x}}{f(\vv{e})}{ck}  }}{NSncall'}
   
   \end{array}%  
    $$
    
    \caption{Stream semantics of \nlustre\ nodes and equations}
    \label{FIG:StrSemNodeEqn}
\end{figure*}

The (NSfby') rule for \ttt{fby} in an equational context uses the semantic operation \textcolor{blue}{$\textsf{fby}_{NL}$}, which differs from $\ckFont{fby}_L$ in that it requires its first argument to be a constant rather than a stream.
The (NSndef') rule only differs from (LSndef) in that  after clock alignment during \textit{transcription}, we make explicit the requirement of $\Hst$ being in accordance with the base clock $bs$, enforced by \ckFont{respects-clock}.
\ignore{
The (NSeq) rule for simple equations mentions the clock that annotates the defining expression, checking that it is consistent with the assumed history for the defined variable $x$.
}
Finally, the rule rule (NSncall') for node call, now in an equational context, is similar to (LSncall) combined with (LSeq),
with the condition that the base clock of the input flows annotates the equation. 

We end this section with a result that states that only those variables that appear in an expression are relevant to its behaviour.
Since the treatment of \lustre\ expressions and clocks have much in common, as also with ``control expressions'' in \nlustre, for convenience we sometimes speak of  ``general expressions'', written $ge$, and use the generic predicate $\mstrSemAnyExpPred{\Hst}{bs}{ge}{\tupstrm{vs}}$.
\begin{lemma}[Relevant variables for expression evaluation]\label{LEM:relevant-variables}
If $fv(ge) \subseteq X$ and for all $x\in X: 
\Hst(x) = \Hst'(x)$, 
then 
$\mstrSemAnyExpPred{G, \Hst}{bs}{ge}{\tupstrm{vs}}$ iff $\mstrSemAnyExpPred{G, \Hst'}{bs}{ge}{\tupstrm{vs}}$.
\end{lemma}
\Proofsketch
By induction on the structure of $ge$.
\QED

% 3
\section{A Security Type System for {\lustre}}\label{SEC:LSec-types}

\paragraph{Security class lattice}
Denning proposed lattices as the appropriate mathematical model for reasoning about secure information flow \cite{Denning:1976,Denning:1977}.
An information flow model
\(
 \langle \tti{N}, \tti{SC}, \rel, \lub ,\bot \rangle 
\)
consists of a set \tti{N} of all data variables/objects in the system, which are assigned security classes (typically $t$, possibly with subscripts) from  \tti{SC}, which is a (usually finite) lattice, partially ordered by the relation $\rel$, and with $\lub$ being the \tti{least upper-bound} (LUB) operator and $\bot$ the least element of the lattice. 
The intuitive reading of $t_1 \rel t_2$ is that the security class $t_1$ is less secure (\textit{i.e.}, less confidential, or dually, more trusted) than $t_2$, and so a flow from $t_1$ to $t_2$ is permitted.

\paragraph{Information flow leaks}
\label{SUB:SimpleBugs}
% Currently \lustre\ does not have a security framework. 
Suppose we decorated variables in a \lustre\ program with security levels drawn from a Denning-style lattice.
We give two simple instances of insecure expressions which can leak information implicitly.
The conditional expression $\ite{e_0}{e_1}{e_2}$ which, depending on whether $e_0$ is true or false, evaluates expression $e_1$ or else $e_2$ (all expressions are on the same clock), can leak the value of the $e_0$.
Consider the following example, where by observing the public flow named \ttt{c}, we can learn the secret variable \ttt{b}:
\begin{lstlisting}[language=lustre]
-- b secret, c public 
    c = if b then 1 else 0
\end{lstlisting}

Similarly, the expression $\lmerge{x}{e_1}{e_2}$ -- which merges, based on the value of $x$ at each instant, the corresponding value from streams $e_1$ or $e_2$ into a single stream -- also can leak the variable $x$'s values.
This is evident in the following:
\begin{lstlisting}[language=lustre]
-- x secret, c0 public 
    c0 = merge x 1 0
\end{lstlisting}

Our type system aims at preventing such \textit{implicit flows}.
Further, it should be able to correctly combine the security levels of the arguments for all operators, and allow only legal flows in equational definitions of variables, node definitions and node calls.

\subsection{Security Types}\label{SUB:SecTypes}

We define a secure information flow type system, where under security-level type assumptions for program variables, \lustre\ expressions are given a \textit{symbolic security type} $(ST)$, and \lustre\ equations induce a set of \textit{ordering constraints} over security types.

\textbf{Syntax.}  Security type expressions ($\type{\alpha, \beta}$) for \lustre\ are either (i) \textit{type variables}  (written $\type{\delta})$ drawn from a set \textit{STV}, or 
(ii) of the form $\type{\alpha \stlub \beta}$ where $\type{\stlub}$ is interpreted as an associative, commutative and idempotent operation. 
(iii) The identity element of $\type{\stlub}$ is $\type{\bot}$.
While this idempotent abelian monoid structure suffices for \nlustre, node calls in \lustre\; require (iv) \textit{refinement types} $\type{\alpha \{\!| \rho |\!\} }$, where type expression $\type{\alpha}$ is subject to a symbolic constraint $\type{\rho}$.
Constraints on security types, typically $\type{\rho}$,  are (conjunctions of) relations of the form $\type{\alpha \strel \beta}$.
The comparison $\strel$ is defined in terms of the equational theory: $\type{\alpha \strel \beta}$ exactly when
$\type{\alpha \stlub \beta ~=~ \beta}$.
Our proposed security types and their equational theory are presented in Figure \ref{FIG:SecurityTypes}. 
The security types for \nlustre\  and their equational theory \cite{PYS-memocode2020} are highlighted in grey within the diagram.
This congruence on \nlustre\ types (henceforth $\type{\equiv_{NL}}$), which is given in the highlighted second line of Figure \ref{FIG:SecurityTypes}, is significantly simpler since it does not involve refinement types!
(Notation: Security types and constraints are written in \type{blue}. 
In program listings, we will write them as superscripts.)

\begin{figure*}
%   \begin{align*}
        \text{Types: }  
        $\type{\alpha,\beta,\gamma,\theta} ~::= \mathhl{mygrey}{\type{\bot}}  \;|\; \mathhl{mygrey}{\type{\delta} \in \textit{STV}}  \;|\; 
        \mathhl{mygrey}{\type{\alpha \lub \beta}} \;|\; \type{\alpha\{\!| \rho |\!\} }$  ~~
        \text{Constraints:} $\mathhl{mygrey}{\type{\rho} ~::=~ \type{(\theta \rel \alpha)}^*}$ \\ \\
$\mathhl{mygrey}{\type{(\alpha \lub \beta) \lub \theta} ~=~ \type{\alpha \lub (\beta \lub \theta)}, 
~\hfill~
\type{\alpha \lub \alpha} ~=~ \type{\alpha},
~\hfill~
\type{\alpha \lub \beta} ~=~ \type{\beta \lub \alpha},
~\hfill~
\type{\alpha \lub \bot} ~=~ \type{\alpha} ~=~ 
\type{\bot \lub \alpha},}$ \\ \\
$ \type{\alpha\{\!| |\!\}} ~=~ \type{\alpha},     
~~\hfill~~
\type{\alpha_1\{\!| \rho_1 |\!\}} \type{~\stlub}~ \type{\alpha_2\{\!| \rho_2 |\!\}} ~=~ (\type{\alpha_1 \lub \alpha_2}) \type{\{\!|} \union{ \type{\rho_1}} {\type{\rho_2}} \type{|\!\}}, 
~~\hfill~~
\type{\alpha\{\!| \rho_1 |\!\}\{\!| \rho_2 |\!\}} ~=~ \type{\alpha \{\!| } \union{\type{\rho_1}}{\type{\rho_2}} \type{|\!\}}$, \\  \\
$\type{\vv{\alpha_i} \{\!| \rho |\!\}} ~=~ 
\type{ \vv{ \alpha_i  \{\!| \rho |\!\}} }
$,
~~
$\{ \type{\alpha\{\!| \rho_1 |\!\} \rel \beta\{\!| \rho_2 |\!\}} \} ~=~ \union{\{ \type{\alpha \strel \beta} \} } {\union{\type{\rho_1}}{\type{\rho_2}}}$,
~~~
$\mathhl{mygrey}{\type{\vv{\alpha_j}[\theta_i/\delta_i]} ~=~ \type{\vv{\alpha_j[\theta_i/\delta_i]}}}$,
\\ \\
%
%\text{Substitution: }
$\type{\alpha\{\!| \rho |\!\}[\theta_i/\delta_i]} ~=~ \type{\alpha[\theta_i/\delta_i] \{\!| \rho[\theta_i/\delta_i] |\!\}}$,
~~\hfill~~
$\mathhl{mygrey}{(\type{\alpha \strel \beta}) \type{[\theta_i/\delta_i]} ~=~
\type{\alpha[\theta_i/\delta_i] ~\strel~ \beta[\theta_i/\delta_i]}} 
$.

%    \end{align*}
    \caption{Security types, constraints and their properties}
    \label{FIG:SecurityTypes}
\end{figure*}

We write  $\type{\alpha [ \theta_i / \delta_i]}$ for $i=1, \ldots, k$ to denote the (simultaneous) substitution of security types $\type{\theta_i}$ for security type variables $\type{\delta_i}$ in security type $\type{\alpha}$. 
The notation extends to substitutions on tuples ($\type{\vv{\alpha} [ \theta_i / \delta_i]}$) and constraints ($\type{(\alpha \strel \beta) [ \theta_i / \delta_i]}$).

\textbf{Semantics.}  Security types are interpreted with respect to a complete lattice  $\langle \tti{SC}, \rel, \lub, \bot \rangle$ of security levels \cite{Denning:1976}. 
Given a ground instantiation $s: \tti{STV} \rightarrow \tti{SC}$,
security type expressions and tuples are interpreted according to its homomorphic extension:
$s(\type{\bot}) = \bot$,
$s(\type{\alpha \stlub \beta}) = s(\type{\alpha}) \lub s(\type{\beta})$,
$s(\type{\vv{\alpha}}) = \vv{s(\type{\alpha_i})}$,
and constraints are interpreted according to the lattice ordering:
$s(\type{\alpha \strel \beta}) = 
s(\type{\alpha}) \rel s(\type{\beta})$.
The ``refinement types'' are interpreted as:
$s(\type{\alpha\{\!| \rho |\!\}}) = s(\type{\alpha})$ if $s(\type{\rho})$ holds wrt $SC$, \textit{i.e.}, if ``$s$ satisfies $\type{\rho}$'', else is undefined.
% To make sense, refinement type $\type{\alpha \{\!| \rho |\!\}}$ requires the satisfaction of constraint $\type{\rho}$.

\begin{lemma}[Soundness]\label{LEM:TypeEq-sound}
The equational theory induced by the equalities in  \autoref{FIG:SecurityTypes} is sound with respect to any ground instantiation $s$, \textit{i.e.}, 
(i) $\type{\alpha = \beta}$ implies $s(\type{\alpha}) = s(\type{\beta})$, and (ii) $\type{\rho_1} =  \type{\rho_2}$ implies $s(\type{\rho_1})$ is satisfied iff $s(\type{\rho_2})$ is.
\end{lemma}
\Proofsketch
Most of the properties follow from $s$ being a homomorphism, and the bijection between idempotent abelian monoids and join semi-lattices (the monoid operation maps to LUB in the lattice). 
\QED

The following facts are useful since we often want to reason about equality of security types or about constraints independently of any given security lattice.
\begin{lemma}[Confluence]\label{LEM:TypeEq-confluence}
All equations other than those of associativity and commutativity (AC) can be oriented (left-to-right) into rewriting rules.
The rewriting system is confluent modulo AC.
Equal types (respectively, equal constraints) can be rewritten to a common form modulo AC.
\end{lemma}
\Proofsketch
The equational theory $\type{\equiv_{NL}}$ trivially yields a convergent rewriting system modulo AC.
The rules in lines 3 and 4 of Figure \ref{FIG:SecurityTypes} can all be oriented left to right. 
We use Knuth-Bendix-completion \cite{Knuth-Bendix} to introduce rules
$\type{\alpha_1\{\!| \rho_1 |\!\}} \type{~\stlub}~ \type{\alpha_2} \longrightarrow (\type{\alpha_1 \lub \alpha_2}) \type{\{\!|}  \type{\rho_1}  \type{|\!\}}$, when $\type{\alpha_2}$ is not a refinement type.
Type equality and constraints are efficiently decided using 
the theory of strongly coherent rewriting modulo AC \cite{Viry-rewriting}. 
\QED.

% \paragraph{Security Typing Rules.} 
\subsection{Security Typing Rules}\label{SUB:SecTypeRules}
Assume typing environment $\Gamma: \tti{Ident} \rightharpoonup \tti{ST}$, a partial function associating a security type to each free variable \tti{x} in a \lustre\ program phrase.
Expressions and clocks are type-checked using the predicates:
$
%\begin{minipage}{.3\textwidth}
\mexprPred{\Gamma}{e}{\vv{\alpha}}$
and 
%\end{minipage}%
%\begin{minipage}{.3\textwidth}
$\mclkPred{\Gamma}{ck}{\alpha}$ respectively. 
These are read as ``under the context $\Gamma$ mapping variables to security types, $e$ and $ck$ have security types $\type{\vv{\alpha}}$ and $\type{\alpha}$''.
Since the predicates for expressions and clocks (and in \nlustre,  ``control expressions'' too) have much in common, for convenience we use a generalized predicate $\manyexprPred{\Gamma}{ge}{\alpha}$ to represent a parametric analysis over the appropriate syntactic structure $ge$ (this notation is used in stating results of \autoref{SEC:Soundness}).

The types for tupled expressions are (flattened) tuples of the types of the component expressions.
For equations, we use the predicate:
$\meqnPred{\Gamma}{eq}{\rho}$, 
which states that under the context $\Gamma$, equation $eq$ when type-elaborated generates constraints $\type{\rho}$.
Elementary constraints for equations are of the form $\type{\alpha \rel \beta}$, where $\type{\beta}$ is the security type of the defined variable, and $\type{\alpha}$ the security type obtained from that of the defining expression joined with the clock's security type.
Since every flow in \lustre\ is defined \textit{exactly once}, by the Definition Principle, no further security constraints apply.

The security typing rules for \lustre\ are presented in Figures \ref{FIG:LustreClkSecTyping} -- \ref{FIG:LustreEqnType}, plus the rules for node definition and node call.
These rules generalise those in \cite{PYS-memocode2020} to handle expressions representing lists of flows, and nested node calls.
The rules for \nlustre\ expressions other than node call and \ttt{fby} are just the singleton cases. 
Node call and \ttt{fby} are handled by the rule for equations. 
\begin{figure*}
    \[
    \begin{array}{c}
        \namedJdg{\Gamma(\ttt{base}) = \type{\gamma}}{\mclkPred{\Gamma}{\ttt{base}}{\gamma}}{LTbase} ~~~~~~
        \namedJdg{\predSet{\Gamma(x)=\type{\gamma_1}}{\mclkPred{\Gamma}{ck}{\gamma_2}}}
        {\mclkPred{\Gamma}{\ck{ck}{x=k}}{\gamma_1 \stlub \gamma_2}}{LTon} \\
    \end{array}
    \]
    \caption{\lustre\ security typing rules for clocks}
\label{FIG:LustreClkSecTyping}
 \end{figure*}

 \begin{figure*}
    \[
    \begin{array}{c}
        \namedJdg{\predSet{\Gamma(x)=\type{\alpha}}}
        {\mexprPred{\Gamma}{x}{\alpha }}{LTvar} 
        ~~~~~~
        \namedJdg{\predSet{\mexprPred{\Gamma}{e }{\alpha}}}
        {\mexprPred{\Gamma}{{\unop{e}}}{\alpha}}{LTunop}  
~~~~~~
        \namedJdg{
            \predSet{\mexprPred{\Gamma}{e_1}{\alpha_1}} 
            {\mexprPred{\Gamma}{e_2}{\alpha_2}}}
        {\mexprPred{\Gamma}{\binop{e_1}{e_2}}{\alpha_1 \stlub \alpha_2 }} {LTbinop} \\
        \namedJdg{
            \predSet{\type{\theta}=\Gamma(x)}
            {\mexprPred{\Gamma}{\vv{e_t}}{\vv{\alpha}}}
            {\mexprPred{\Gamma}{\vv{e_f}}{\vv{\beta}}}}
    {\mexprPred{\Gamma}{ \lmerge{x}{\vv{e_t}}{\vv{e_f}}} {\vv{(\theta \stlub \alpha_i \stlub \beta_i)_i}}}{LTmrg} 
        ~~~~~
       \namedJdg{}
        {\mexprPred{\Gamma}{c}{\bot}} {LTcnst} \\
    
    \namedJdg{
            \predSet {\mexprPred{\Gamma}{e}{\theta}} 
            {\mexprPred{\Gamma}{\vv{e_t}}{\vv{\alpha}}} 
            {\mexprPred{\Gamma}{\vv{e_f}}{\vv{\beta}}} }
    {\mcexprPred{\Gamma}{\ite{e}{\vv{e_t}}{\vv{e_f}}}{ 
    \vv{(\theta \stlub \alpha_i \stlub \beta_i)_i}}}{LTite} \\
\\
    \namedJdg{
            \predSet{\mexprPred{\Gamma}{\vv{e_0}}{\vv{\alpha}}}
            {\mexprPred{\Gamma}{\vv{e}}{\vv{\beta}}}}
    {\mexprPred{\Gamma}{ \lfby{\vv{e_0}}{\vv{e}}} {\vv{(\alpha_i \stlub \beta_i)_i}}}{LTfby} 
    ~~
     \namedJdg{
          \predSet{\mexprPred{\Gamma}{e_1}{\alpha_1}} 
            \ldots
            {\mexprPred{\Gamma}{e_n}{\alpha_n}}~~
            {\Gamma(x)= \type{\gamma}}}
        {\mexprPred{\Gamma}{\when{\vv{e}}{x = k}}{ 
        \vv{(\alpha_i \stlub \gamma)_i}}}{LTwh}
    \end{array}
\]  
    \caption{\lustre\ Security Typing Rules for Expressions}
\label{FIG:LustreExpSecTyping}
\end{figure*}

%following figure on eqs rules exceeds the borders by 2 chars
 \begin{figure*}
\[
   \begin{array}{c}
    \namedJdg { 
         \predSet {\type{\vv{\beta}} = \Gamma(\vv{x})} 
         {\mexprPred{\Gamma}{\vv{e}}{\vv{\alpha}}}
         {\mclkPred{\Gamma}{ck}{\gamma}}}
     {\meqnPred{\Gamma}{ \vv{x}^{ck} = \vv{e}}{\type{\set{ (\gamma \stlub \alpha_i  \rel \beta_i)_i}}}} {LTeq} 
  \;
   \namedJdg{
        \predSets{\meqnPred{\Gamma}{eq}{\type{\rho}}} %{\mu}}
        {\meqnPred{\Gamma}{eqs}{\type{\rho'}}}}%{\mu'}}
   {\meqnPred{\Gamma}{eq;eqs}{\union{\type{\rho}}{\type{\rho'}}}}{LTeqs}
   \end{array}
 \]
   \caption{\lustre\ security typing rules for equations}
   \label{FIG:LustreEqnType}
\end{figure*}

% Explanation of the rules
In (LTbase), we assume $\Gamma$ maps the base clock $\ttt{base}$ to some security variable ($\type{\gamma}$ by convention).
In (LTon), the security type of the derived clock is the join of the security types of the clock $ck$ and that of the variable $x$.

Constants have security type $\type{\bot}$, irrespective of the context (rule (LTcnst)).
For variables, in rule (LTvar), we look up their security type in the context $\Gamma$.
Unary operations preserve the type of their arguments (rule (LTunop)).
Binary ($\oplus$,\ttt{when} and \ttt{fby}) and ternary (\ttt{if-then-else} and \ttt{merge}) operations on flows generate a flow with a security type that is the join of the types of the operand flows (rules (LTbinop), (LTwhn), (LTmrg), (LTite), and (LTfby).
In operations on \textit{lists of flows}, the security types are computed component-wise.
There is an implicit dependency on the security level of the common clock of the operand flows for these operators.
This dependence on the security level of the clock is made explicit in the rule for equations.
In general, the security type for any constructed expression is the join of those of its components (and of the clock).

\begin{example}[Constraints from equations]
\label{EG:Constraints}
With respect to Example \ref{EG:Ctr}, the constraints generated for the definitions of variables \ttt{n}, \ttt{fst} and \ttt{pre\_n} are the following: \\
$\type{\rho_1} :=  \{ \type{\gamma \stlub \delta_1 \stlub \alpha_3 \stlub \alpha_1 \stlub \delta_2 \stlub \alpha_2 \rel \beta} \}$, \\
$\type{\rho_2} := \{ \type{\gamma \stlub \bot \stlub \bot \rel \delta_1} \}$, \textit{i.e.}, $\{ \type{\gamma \rel \delta_1} \}$, and \\
$\type{\rho_3} := \{ \type{\gamma \stlub \bot \stlub \beta \rel \delta_2} \}$, \textit{i.e.}, $\{ \type{\gamma \stlub \beta \rel \delta_2} \}$ respectively.
\end{example}

\paragraph{Node call}\ 
Node calls assume that we have a security signature for the node definition (described below).
We can then securely type node calls by instantiating the security signature with the types of the actual arguments (and that of the base clock).
Note that the rule (LTncall) creates  refinement types consisting of the output types $\type{\beta_i}$ constrained by $\type{\rho'}$, \textit{i.e.}, the instantiated set of constraints $\type{\rho}$ taken from the node signature:
\[
\namedJdg{%\dependSet 
 % { \securitySignature{f}{(\vv{\type{\alpha}})}{\vv{\type{\beta}}}{\gamma}{\rho}}
 {\predSet
 { \securitySignature{f}{(\vv{\type{\alpha}})}{\vv{\type{\beta}}}{\type{\gamma}}{\type{\rho}}}~~~
 {\mexprPred{\Gamma}{ \vv{e}} {\vv{\type{\alpha'}}}}~~~
 %{~\vv{\type{\beta'}} =\Gamma(\vv{x})~}
 {\Gamma(\ttt{base})=\type{\gamma'}}~~
 %{\mclkPred{\Gamma}{ck}{\type{\gamma'}} }
 {\type{\rho'} = 
 \type{\rho[\gamma'/\gamma][\vv{\alpha'}/\vv{\alpha}]}}}}
 {\mexprPred{\Gamma}{ f(\vv{e})   }{\vv{\beta}\{\!| \type{\rho'} |\!\}}}{LTncall}
\]

\paragraph{Node definition}\ 
A node definition is given a signature
$\securitySignature{f}{(\vv{\type{\alpha}})}{\vv{\type{\beta}}}{\type{\gamma}}{\type{\rho}}$, which is to be read as saying that the node named $f$ relates the security types $\vv{\type{\alpha}}$ of the input variables (and $\type{\gamma}$, that of the base clock) to the types of the output variables $\vv{\type{\beta}}$, via the constraints 
$\type{\rho}$.

Let $\type{\alpha_1,\dots,\alpha_n,\delta_1,\dots\delta_k,\beta_1,\dots\beta_m,\gamma}$ be distinct \textit{fresh type variables}.
Assume these to be the types of the input, local and output variables, and that of the base clock. 
We compute the constraints over these variables induced by the node's equations.
Finally, we eliminate, via substitution using procedure \type{\textsf{simplify}}, the type variables $\type{\delta_i}$ given to the local program variables, since these should not appear in the node's interface.
The security signature of a node definition is thus given as:
\[
\namedJdg{\dependSet{G(f)= n:\{\ttt{in}=\vv{x}, \ttt{out}=\vv{y}, \ttt{var}=\vv{z}, \ttt{eqn} = \vv{eq} \}}%{\node \in G}
    {\Gamma_F := \{ \for{\type{\vv{\alpha}}}{\vv{x}}, \for{\type{\vv{\beta}}}{\vv{y}}, \ttt{base} \mapsto \type{\gamma} \}~~~~
     \Gamma_L := \{\for{\vv{\type{\delta}}}{\vv{z}}\}}
    {\predSet{\meqnPred{\union{\Gamma_F}{\Gamma_L}}{\vv{eq}}{\type{\rho'}}}{(\_, \type{\rho}) = \simpl{(\_,\type{\rho'})}{\vv{\delta}}}}
    } 
    { \securitySignature{f}{(\vv{\type{\alpha}})}{\vv{\type{\beta}}}{\type{\gamma}}{\type{\rho}}}{LTndef}
\]
The node signature (and call) rules can be formulated in this step-wise and modular manner since \lustre\ does not allow recursive node calls and cyclic dependencies.
Further, all variables in a node definition are explicitly accounted for as input and  output parameters or local variables, so no extra contextual information is required. 

\begin{figure*}
\[
\begin{array}{cl}
    \mjdg{}
    {(\type{\vv{\alpha}, \rho}) = \simpl{(\type{\vv{\alpha}, \rho})}{[~]}} 
    % \\ \\
    ~~~~~
    \mjdg{(\type{\vv{\alpha'}, \rho'}) = \simpl{(\type{\vv{\alpha}}[\type{\nu/\delta}],
    \type{\rho}[\type{\nu/\delta}])}{\type{\vv{\delta}}}}
    {(\type{\vv{\alpha'}, \rho'}) = \simpl{(\type{\vv{\alpha}}, \union{\type{\rho}}{\{\type{\nu \rel \delta}\}})}{(\type{\delta} :: \type{\vv{\delta}})}}
     & \mbox{$\type{\delta}$ not in $\type{\nu}$}
\\ \\
    \mjdg{(\type{\vv{\alpha'}, \rho'}) = \simpl{(\type{\vv{\alpha}}[\type{\nu/\delta}], \type{\rho}[\type{\nu/\delta}])}{\type{\vv{\delta}}}}
    {(\type{\vv{\alpha'}, \rho'}) = \simpl{(\type{\vv{\alpha}}, \union{\type{\rho}}{\{\type{\nu \stlub \delta ~\rel~ \delta}\}})}{(\type{\delta} :: \type{\vv{\delta}})}}
    & \mbox{$\type{\delta}$ not in $\type{\nu}$}
\end{array}\]

\caption{Eliminating local variables' security type constraints}\label{FIG:simpl}
\end{figure*}

Observe that in the (LTndef) rule,  $\type{\delta_i}$ are fresh security type variables assigned to the local variables.
Since there will be exactly one defining equation for any local variable $z_i$, note that in constraints $\type{\rho'}$, there will be exactly one constraint in which $\type{\delta_i}$ is on the right, and this is of the form $\type{\nu_i \strel \delta_i}$.
Procedure \type{\textsf{simplify}} (Figure \ref{FIG:simpl}) serially (in some arbitrary but fixed order for the $\type{\delta_i}$) eliminates such type variables via substitution in the types and type constraints. 
Our definition of \ckFont{simplify} here generalises that given for the types of \nlustre\ in \cite{PYS-memocode2020}.

% Example revisited
For Example \ref{EG:Constraints}, 
$\simpl{(\_, \type{\rho_1} \cup \type{\rho_2} \cup \type{\rho_3})}{[\delta_1 ; \delta_2]}$ yields 
$\type{\rho} = \{\type{\gamma \stlub \alpha_1 \stlub \alpha_2 \stlub \alpha_3 \rel \beta}\}$. 
Thus the node signature for \ttt{Ctr} is
\[
\securitySignature{\ttt{Ctr}}{(\alpha_1,\alpha_2,\alpha_3)}{\beta}{\gamma}{\type{\rho}}
\]
For Example \ref{EG:SpdMtr}, the constraints generated for the equations defining \ttt{spd} and \ttt{pos} are \\
$\type{\rho_4} := \{ \type{\gamma_1 \stlub \bot \stlub \alpha_4 \stlub \bot \rel \beta_1} \}$, \textit{i.e.},  
$\{ \type{\gamma_1 \stlub \alpha_4 \rel \beta_1} \}$, and \\
$\type{\rho_5} := \{ \type{\gamma_1 \stlub \bot \stlub \beta_1 \stlub \bot \rel \beta_2} \}$ \textit{i.e.}, 
$\{ \type{\gamma_1 \stlub \beta_1 \rel \beta_2} \}$
respectively.\\
$\type{\rho_4} \cup \type{\rho_5}$ simplifies to $\{ \type{\gamma_1 \stlub \alpha_4 \rel \beta_1},\type{\gamma_1 \stlub \beta_1 \rel \beta_2}\}$.
Since $\type{\gamma_1 \rel \beta_1}$, the latter constraint is equivalent to $\type{\beta_1 \rel \beta_2}$.

\begin{lemma}[Correctness of $\simpl{(\type{\vv{\alpha},\rho})}{\type{\vv{\delta}}}$]\label{LEM:Simplify}
Let $\type{\rho}$ be a set of constraints such that for a security type variable $\type{\delta}$, there is at most one constraint of the form $\type{\mu \rel \delta}$.
%such that variable $\type{\delta}$ appears in $\type{\alpha_2}$.
Let $s$ be a ground instantiation of security type variables wrt a security class lattice $\tti{SC}$ such that $\type{\rho}$ is satisfied by $s$.
\begin{enumerate}
\item If $\type{\rho} = \union{\type{\rho_1}}{\set{\type{\nu \rel \delta}}}$, where variable $\type{\delta}$ is not in $\type{\nu}$,
then $\type{\rho_1}[\type{\nu}/\type{\delta}]$ is satisfied by $s$. (Assume disjoint union.)
\item If $\type{\rho} = \union{\type{\rho_1}}{\set{\type{\nu \stlub \delta \rel \delta}}}$, where variable $\type{\delta}$ is not in $\type{\nu}$,
then $\type{\rho_1}[\type{\nu}/\type{\delta}]$ is satisfied by $s$. (Assume disjoint union.)
\end{enumerate}
\end{lemma}
\Proofsketch 
Note that $\type{\rho_1}$ is satisfied by $s$, and that $\type{\delta}$ appears to the right of $\type{\strel}$ in only one constraint. 
Suppose $\type{\beta_1 \strel \beta_2}$ is a constraint in $\type{\rho_1}$, with variable $\type{\delta}$ appearing in $\type{\beta_1}$.
Since  $\tti{FM} \models s(\type{\nu}) \rel s(\type{\delta})$,
by transitivity and monotonicity of $s$ with respect to $\lub$: $s(\type{\beta_1[\nu/\delta]}) \rel s(\type{\beta_1}) \rel s(\type{\beta_2})$.
\QED

\subsection{Mechanisation in Coq}\label{SUB:NLTypes-COQ}

%%Types:
We are mechanising our proofs using the proof assistant Coq, and  integrating our SIF type system into the V\'{e}lus verified compiler framework \cite{Velus}.
We provide here some snippets related to the formalisation of our type system, focusing here only on parts that can be presented without having to invoke details from the V\'{e}lus development.

\begin{lstlisting}[language=Coq]

Inductive nonCanonST : Type :=
| Bot  : nonCanonST
| Var  : ident -> nonCanonST
| Lub  : nonCanonST -> nonCanonST -> nonCanonST
| Ref  : nonCanonST -> uConstraint -> nonCanonST
with uConstraint: Type := 
| CNil : uConstraint
| Cns  : (nonCanonST * nonCanonST) -> uConstraint -> uConstraint.

(*Canonical types*)
Definition ST := list ident.
Definition constraint := (ST*ST)%type.
\end{lstlisting}

The syntax of security types are represented as an inductive data type \ttt{nonCanonST}, and constraints (type \ttt{uConstraint}) essentially as a list of security type pairs.
We represent the \textit{canonical types} as lists of \ttt{ident} (where identifiers are given some fixed order, \textit{e.g.}, lexical), and a (canonical) constraint as a list of canonical type pairs. 
A routine \ttt{canon} (not shown here) turns the original type syntax into a canonical form using the left-right oriented forms of $\type{\equiv_{L}}$ rewrite rules.

\begin{lstlisting}[language=Coq]
(*Equational Theory*)
Inductive EqL : nonCanonST -> nonCanonST -> Prop :=
| refl_case: \forall \alpha, \alpha $\type{\equiv_{L}}$ \alpha
| lub_assoc: \forall \alpha \beta \theta, (Lub (Lub \alpha \beta) \theta)  $\type{\equiv_{L}}$(Lub \alpha (Lub \beta \theta))
| lub_idem : \forall \alpha, (Lub \alpha  \alpha)  $\type{\equiv_{L}}$ \alpha
| lub_comm   : \forall $\alpha_1$ $\alpha_2$, (Lub $\alpha_1$ $\alpha_2$)  $\type{\equiv_{L}}$ (Lub $\alpha_2$ $\alpha_1$) 
| lub_id : \forall \alpha, (Lub \alpha  \Bot)  $\type{\equiv_{L}}$ \alpha
| ref_base : \forall \alpha, (Ref \alpha CNil)  $\type{\equiv_{L}}$\alpha
| ref_lub  : \forall $\alpha_1$ $\alpha_2$ $\rho_1$ $\rho_2$, 
   (Lub (Ref $\alpha_1$ $\rho_1$) (Ref $\alpha_2$ $\rho_2$))  $\type{\equiv_{L}}$ (Ref (Lub $\alpha_1$ $\alpha_2$) ($\rho_1 \cup \rho_2$))
| ref_ref  : \forall \alpha $\rho_1$ $\rho_2$, (Ref (Ref \alpha $\rho_1$) $\rho_2$)  $\type{\equiv_{L}}$ (Ref \alpha ( $\rho_1 \cup \rho_2$))
(*Knuth Bendix completion rules*)
| ref_lub_assocl: \forall $\alpha_1$ $\alpha_2$ \rho, 
            (Lub $\alpha_1$ (Ref $\alpha_2$ \rho) )  $\type{\equiv_{L}}$ (Ref (Lub $\alpha_1$ $\alpha_2$) \rho)
| ref_lub_assocr: \forall $\alpha_1$ $\alpha_2$ \rho, 
            (Lub (Ref $\alpha_1$ \rho) $\alpha_2$)  $\type{\equiv_{L}}$ (Ref (Lub $\alpha_1$ $\alpha_2$) \rho)
where (a $\type{\equiv_{L}}$ b) = (EqL a b)(at level 40, left associativity).
            
(*Homomorphic extension of mapping '$s$' to canonical types*)

Fixpoint hm_ext ($s$: ident -> Lat.t ) (xs: ST): Lat.t := 
    match xs with
    | nil   => Lat.bottom
    | x::xs' => Lat.join  ($s$ x) (hm_ext $s$ xs')
    end.


Lemma Soundness_EqL1: \forall $\alpha_1$ $\alpha_2$, $\alpha_1$  $\type{\equiv_{L}}$ $\alpha_2$ -> \forall $s$, hm_ext $s$ $\alpha_1$ = hm_ext $s$ $\alpha_2$.

(* $\type{\equiv_{LC}}$ and canon_cons are the equality and rewrite routine (canon) extended to constraints *)
Lemma Soundness_EqL2: \forall $\rho_1$ $\rho_2$, $\rho_1$  $\type{\equiv_{LC}}$ $\rho_2$ -> 
    canon_cons $\rho_1$ = $\rho_1'$ -> 
    canon_cons $\rho_2$ = $\rho_2'$ -> 
    \forall $s$, listSatisfiable $s$ $\rho_1'$ <-> listSatisfiable $s$ $\rho_2'$.

Lemma Confluence_EqL: \forall $\alpha_1$ $\alpha_2$, $\alpha_1$  $\type{\equiv_{L}}$ $\alpha_2$ -> canon $\alpha_1$ = canon $\alpha_2$.

\end{lstlisting}
The function \ttt{hm\_ext} is the homomorphic extension (to canonical types) of the mapping $s$ from \ttt{ident} to lattice elements \ttt{Lat.t}. 
It is straightforward to state and prove the soundness of equational theory and the confluence of the rewrite system (Lemmas \ref{LEM:TypeEq-sound} and \ref{LEM:TypeEq-confluence}.

\begin{lstlisting}[language=Coq]

(*subAll is the substitution routine: subAll cns \delta st $\equiv$ cns[st\{\delta} / \delta].*)
(*subAll eliminates \delta from both cns and st during substitution.*)

(*findST returns a constraint with RHS as the singleton [\delta] from the list cns.*)

 Inductive simplifyRel  : list constraint -> list ident ->  list constraint -> Prop :=
  | baseCase  : \forall cns,
    simplifyRel cns [ ] cns
  | elimCase1 : \forall cns \delta st cns' ds cns'',
    simplifyRel cns' ds cns'' ->
    findST cns \delta = Some st ->
    subAll cns \delta st = cns' ->    
    simplifyRel cns (\delta::ds)  cns''
  | elimCase2 : \forall cns \delta cns' ds,
    findST cns \delta = None ->
    simplifyRel cns ds cns' ->
    simplifyRel cns (\delta::ds) cns'.

(*Specification of simplify is correct*)
Theorem simplifyRel_is_correct :
\forall ($s$: ident -> Lat.t) cns cns' $\delta s$,
    listSatisfiable $s$ cns -> simplifyRel cns $\delta s$ cns' -> listSatisfiable $s$ cns'.
 
\end{lstlisting}

By suitable definitions of substitution, and function to pick a constraint with a single (local) variable to be eliminate, we give a relation specification of \type{\textsf{simplify}}, which can be proved correct quite easily (\textit{q.v.} Lemma \ref{LEM:Simplify}).

%4
\section{Security and Non-Interference for NLustre }
\label{SEC:NLustreNonInterference}

\subsection{Security of Nodes}

We now present the notion of security of nodes.  
Our first result concerns when node calls are secure.

\begin{lemma}[Security of Node Calls in \lustre]\label{LEM:sec-nodecall}
For a call $f(\vv{e})$ to a node with the given security signature
\[ \securitySignature{f}{(\vv{\type{\alpha}})}{\vv{\type{\beta}}}{\type{\gamma}}{\type{\rho}} \]
assume the following
\[
\mexprPred{\Gamma}{\vv{e}}{\vv{\type{\alpha'}}}
~~~\hfill~~~
\mexprPred{\Gamma}{ f(\vv{e})}{\vv{\beta'}}
~~~\hfill~~~
\mclkPred{\Gamma}{ck}{\type{\gamma}}
\]
where $ck$ is the base clock underlying the argument streams $\vv{e}$.
Let $s$ be a ground instantiation of type variables such that for some security classes $\vv{t},w \in \tti{SC}$:
$s(\type{\vv{\alpha'}}) = \vv{t}$ 
and $s(\type{\gamma})=w$. \\[1mm]
Now, if $\type{\rho}$ is satisfied by the ground instantiation $\{ 
\type{\vv{\alpha}} \mapsto  \vv{t},
\type{\vv{\beta}} \mapsto  \vv{u},
\type{\gamma} \mapsto w \}$, 
then the  $s(\type{\vv{\beta'}})$ are defined, and 
$s(\type{\vv{\beta'}}) ~\rel~ s(\type{\vv{\beta} \{\!| \type{\rho} |\!\})}$.
\end{lemma}
Lemma \ref{LEM:sec-nodecall} relates the satisfaction of constraints on security types generated during a node call to satisfaction in  a security lattice via a ground instantiation.
(We rely on the modularity of nodes --- that no recursive calls are permitted, and nodes do not have free variables.)
We do not provide a proof of Lemma \ref{LEM:sec-nodecall} here, since we are currently considering only \nlustre\ programs.
In the context of \nlustre, since node calls only occur in the context of an equation, we consider the (mildly) reformulated version.

\begin{lemma}[Security of Node Calls in \nlustre]\label{LEM:sec-nl-nodecall}
Suppose $f$ is a node named \ttt{f} in the graph $G$ of a \lustre\ program, which has security signature
\[ \securitySignature{f}{(\vv{\type{\alpha}})}{\vv{\type{\beta}}}{\gamma}{\type{\rho}}. 
\]
and also suppose
\[
\meqnPred{\Gamma}{ \eqn{\vv{x}}{f(\vv{e})}{ck}  }{\type{\rho_1}}
\]
Let $s$ be a ground instantiation of type variables such that for some security classes $\vv{t}, \vv{u'},w \in \tti{SC}$:
$s(\vv{\type{\alpha'}}) = \vv{t}$ where 
$\mexprPred{\Gamma}{\vv{e}}{\vv{\type{\alpha'}}}$, 
$s(\vv{\type{\nu}})= \vv{u'}$ where $\Gamma(\vv{x}) = \vv{\type{\nu}}$,
and $s(\type{\gamma})=w$ where
$\mclkPred{\Gamma}{ck}{\type{\gamma}}$. \\[1mm]
Now, if $\type{\rho}$ is satisfied by the ground instantiation $\{ 
\vv{\type{\alpha}} \mapsto  \vv{t},
\vv{\type{\beta}} \mapsto  \vv{u},
\type{\gamma} \mapsto w
\}$ where $\vv{u} \rel \vv{u'}$  (point-wise ordering on the tuples), 
then $\type{\rho_1}$ is satisfied by $s$.
\end{lemma}
\Proofsketch   
Note that in the node signature the constraint $\type{\rho}$ guarantees that the type $\type{\beta_i}$ of each output flow is at least as high as the types $\type{\alpha_j}$ of all input flows on which it depends (and also the clock's level $\type{\gamma}$).
Suppose $\type{\rho}$ is satisfied by the ground instantiation $\{ 
\vv{\type{\alpha}} \mapsto  \vv{t},
\vv{\type{\beta}} \mapsto  \vv{u},
\type{\gamma} \mapsto w
\}$, where $\vv{u} \rel \vv{u'}$.
This ground assignment coincides with $s$ on $\type{\vv{\alpha}}$ and $\type{\gamma}$, and assigns $\vv{u}$ to the output flows.
Now since $\vv{u} \rel \vv{u'}$, then clearly the security classes
$\vv{u'}$ assigned to the defined variables $\vv{x}$ by $s$ are (pointwise) at least as high as those of the outputs flows of the node call $f(\vv{e})$. 
\QED

\begin{definition}[Node Security]
\label{DEF:nodesec}
Let $f$ be a node in the program graph $G$ with security signature
\[ \securitySignature{f}{(\vv{\type{\alpha}})}{\vv{\type{\beta}}}{\gamma}{\type{\rho}}. 
\]
Let $s$ be a ground instantiation that maps the security type variables in the set
$\set{\type{\vars{\alpha}{n}}} \cup \set{\type{\vars{\beta}{m}}} \cup \set{\type{\gamma}}$
to security classes in lattice $\tti{SC}$. \\
Node $f$ is secure with respect to $s$ if 
   (i)  $\type{\rho}$ is satisfied by $s$;
   (ii) For each node $g'$ on which $f$ is directly dependent in the program DAG, each call to $g'$ in $f$ is secure with respect to its appropriate ground instantiation, as given by the conditions in Lemma \ref{LEM:sec-nl-nodecall}.
\end{definition}
This definition captures the intuition of node security in that all the constraints generated for the equations within the node must be satisfied, and that each internal node call should also be secure. 

\subsection{Soundness of the Type System}
\label{SEC:Soundness}

We establish the soundness of the type system by adapting the main idea of Volpano \textit{et al} \cite{Volpano96} to a data-flow setting. 
The novelty of our approach is to dispense with the usual notion of \textit{confinement checking} but instead to generate and solve security type constraints.

% Simple Security Lemma
The Simple Security Lemma for expressions (respectively, control expressions and clock expressions says: ``if, under given security assumptions for the free program variables, the type system gives a general expression $ge$ (expression, control expression, clock expression) a security type $\type{\alpha}$, then all variables which may have been read in evaluating the expression have a security level that is $\type{\alpha}$ or lower''.
\begin{lemma}[Simple Security]
\label{LEM:SimpleSec}
For any general expression $ge$ and security type assumption $\Gamma$, if 
$\manyexprPred{\Gamma}{ge}{\alpha}$, then for all $x \in fv(ge): \Gamma(x) \rel \type{\alpha}$.
\end{lemma}
\Proofsketch By induction on the structure of $ge$.
Constants, variables and \texttt{base} are the base cases.
The result is immediate from the fact that in the rules for $\manyexprPred{\Gamma}{ge}{\alpha}$, the security level of a (generalised) expression is the join of the security levels of the component sub-expressions.
\QED

The notion of non-interference requires limiting observation to streams whose security level is at most a given security level $t$.

\begin{definition}[$(\rel t)$-projected Stream]
    \label{DEF:alphaProj}
    Suppose $t \in \tti{SC}$ is a security class.
    Let $X$ be a set of program variables, $\Gamma$ be security type assumptions for variables in $X$, and $s$ be a ground instantiation,
    \textit{i.e.}, $\Gamma \circ s$ maps variables in $X$ to security classes in $\tti{SC}$.
    Let us define $X_{\rel t} = \{ x \in X ~|~ (\Gamma \circ s)(x) \rel t \}$.
    Let $\Hst$ be a Stream history such that $X \subseteq \tti{dom}(\Hst)$.
    Define $\Hst |_{X_{\rel t}}$ as the projection of $\Hst$ to $X_{\rel t}$, \textit{i.e.}, restricted to those variables that are at security level $t$ or lower:
    \[
    \Hst |_{X_{\rel t}}  (x) = \Hst (x) ~~~~\mbox{for $x \in X_{\rel t}$}.
    \]
\end{definition}

\begin{figure}
\centering
\tikzset{every picture/.style={line width=0.75pt}} %set default line width to 0.75pt        

\begin{tikzpicture}[x=0.75pt,y=0.75pt,yscale=-1,xscale=1]
%uncomment if require: \path (0,380); %set diagram left start at 0, and has height of 380

%Shape: Rectangle [id:dp7891383171685136] 
\draw   (128,112) -- (212,112) -- (212,165) -- (128,165) -- cycle ;
%Shape: Rectangle [id:dp3250735582296461] 
\draw   (347,114) -- (431,114) -- (431,167) -- (347,167) -- cycle ;
%Straight Lines [id:da12873504688425264] 
\draw  [dash pattern={on 4.5pt off 4.5pt}]  (69,141) -- (494,141) ;
%Straight Lines [id:da928020541226335] 
\draw    (99,120) -- (128,120) ;
\draw [shift={(130,120)}, rotate = 180] [color={rgb, 255:red, 0; green, 0; blue, 0 }  ][line width=0.75]    (10.93,-3.29) .. controls (6.95,-1.4) and (3.31,-0.3) .. (0,0) .. controls (3.31,0.3) and (6.95,1.4) .. (10.93,3.29)   ;
%Straight Lines [id:da7863996294983626] 
\draw    (98,154) -- (127,154) ;
\draw [shift={(129,154)}, rotate = 180] [color={rgb, 255:red, 0; green, 0; blue, 0 }  ][line width=0.75]    (10.93,-3.29) .. controls (6.95,-1.4) and (3.31,-0.3) .. (0,0) .. controls (3.31,0.3) and (6.95,1.4) .. (10.93,3.29)   ;
%Straight Lines [id:da6747436901263635] 
\draw    (314,123) -- (343,123) ;
\draw [shift={(345,123)}, rotate = 180] [color={rgb, 255:red, 0; green, 0; blue, 0 }  ][line width=0.75]    (10.93,-3.29) .. controls (6.95,-1.4) and (3.31,-0.3) .. (0,0) .. controls (3.31,0.3) and (6.95,1.4) .. (10.93,3.29)   ;
%Straight Lines [id:da8749926485870415] 
\draw    (313,157) -- (342,157) ;
\draw [shift={(344,157)}, rotate = 180] [color={rgb, 255:red, 0; green, 0; blue, 0 }  ][line width=0.75]    (10.93,-3.29) .. controls (6.95,-1.4) and (3.31,-0.3) .. (0,0) .. controls (3.31,0.3) and (6.95,1.4) .. (10.93,3.29)   ;
%Straight Lines [id:da45301403062537415] 
\draw    (212,122) -- (241,122) ;
\draw [shift={(243,122)}, rotate = 180] [color={rgb, 255:red, 0; green, 0; blue, 0 }  ][line width=0.75]    (10.93,-3.29) .. controls (6.95,-1.4) and (3.31,-0.3) .. (0,0) .. controls (3.31,0.3) and (6.95,1.4) .. (10.93,3.29)   ;
%Straight Lines [id:da8171087493724944] 
\draw    (212,157) -- (241,157) ;
\draw [shift={(243,157)}, rotate = 180] [color={rgb, 255:red, 0; green, 0; blue, 0 }  ][line width=0.75]    (10.93,-3.29) .. controls (6.95,-1.4) and (3.31,-0.3) .. (0,0) .. controls (3.31,0.3) and (6.95,1.4) .. (10.93,3.29)   ;
%Straight Lines [id:da32360803041884556] 
\draw    (431,120) -- (460,120) ;
\draw [shift={(462,120)}, rotate = 180] [color={rgb, 255:red, 0; green, 0; blue, 0 }  ][line width=0.75]    (10.93,-3.29) .. controls (6.95,-1.4) and (3.31,-0.3) .. (0,0) .. controls (3.31,0.3) and (6.95,1.4) .. (10.93,3.29)   ;
%Straight Lines [id:da17418190493586738] 
\draw    (432,159) -- (461,159) ;
\draw [shift={(463,159)}, rotate = 180] [color={rgb, 255:red, 0; green, 0; blue, 0 }  ][line width=0.75]    (10.93,-3.29) .. controls (6.95,-1.4) and (3.31,-0.3) .. (0,0) .. controls (3.31,0.3) and (6.95,1.4) .. (10.93,3.29)   ;
%Curve Lines [id:da3885209032114587] 
\draw [color={rgb, 255:red, 208; green, 2; blue, 27 }  ,draw opacity=1 ] [dash pattern={on 0.84pt off 2.51pt}]  (92.66,100.76) .. controls (121.9,49.76) and (271.9,49.19) .. (299.81,104.44) ;
\draw [shift={(301,107)}, rotate = 247.04] [fill={rgb, 255:red, 208; green, 2; blue, 27 }  ,fill opacity=1 ][line width=0.08]  [draw opacity=0] (8.93,-4.29) -- (0,0) -- (8.93,4.29) -- cycle    ;
\draw [shift={(91,104)}, rotate = 294.44] [fill={rgb, 255:red, 208; green, 2; blue, 27 }  ,fill opacity=1 ][line width=0.08]  [draw opacity=0] (8.93,-4.29) -- (0,0) -- (8.93,4.29) -- cycle    ;
%Curve Lines [id:da13524561969025573] 
\draw [color={rgb, 255:red, 74; green, 144; blue, 226 }  ,draw opacity=1 ] [dash pattern={on 0.84pt off 2.51pt}]  (91.81,174.54) .. controls (123.24,230.68) and (256.14,236.61) .. (298.74,175.86) ;
\draw [shift={(300,174)}, rotate = 123.27] [fill={rgb, 255:red, 74; green, 144; blue, 226 }  ,fill opacity=1 ][line width=0.08]  [draw opacity=0] (8.93,-4.29) -- (0,0) -- (8.93,4.29) -- cycle    ;
\draw [shift={(90,171)}, rotate = 64.98] [fill={rgb, 255:red, 74; green, 144; blue, 226 }  ,fill opacity=1 ][line width=0.08]  [draw opacity=0] (8.93,-4.29) -- (0,0) -- (8.93,4.29) -- cycle    ;
%Curve Lines [id:da06996540034774457] 
\draw [color={rgb, 255:red, 74; green, 144; blue, 226 }  ,draw opacity=1 ] [dash pattern={on 0.84pt off 2.51pt}]  (252.81,177.54) .. controls (284.24,233.68) and (417.14,239.61) .. (459.74,178.86) ;
\draw [shift={(461,177)}, rotate = 123.27] [fill={rgb, 255:red, 74; green, 144; blue, 226 }  ,fill opacity=1 ][line width=0.08]  [draw opacity=0] (8.93,-4.29) -- (0,0) -- (8.93,4.29) -- cycle    ;
\draw [shift={(251,174)}, rotate = 64.98] [fill={rgb, 255:red, 74; green, 144; blue, 226 }  ,fill opacity=1 ][line width=0.08]  [draw opacity=0] (8.93,-4.29) -- (0,0) -- (8.93,4.29) -- cycle    ;

% Text Node
\draw (498,129) node [anchor=north west][inner sep=0.75pt]   [align=left] {t};
% Text Node
\draw (293,150) node [anchor=north west][inner sep=0.75pt]   [align=left] {$\displaystyle x_{L}$};
% Text Node
\draw (79,150) node [anchor=north west][inner sep=0.75pt]   [align=left] {$\displaystyle x_{L}$};
% Text Node
\draw (292,112) node [anchor=north west][inner sep=0.75pt]   [align=left] {$\displaystyle x_{H}$};
% Text Node
\draw (77,112) node [anchor=north west][inner sep=0.75pt]   [align=left] {$\displaystyle x_{H}$};
% Text Node
\draw (245,112) node [anchor=north west][inner sep=0.75pt]   [align=left] {$\displaystyle y_{H}$};
% Text Node
\draw (464,112) node [anchor=north west][inner sep=0.75pt]   [align=left] {$\displaystyle y_{H}$};
% Text Node
\draw (245,150) node [anchor=north west][inner sep=0.75pt]   [align=left] {$\displaystyle y_{L}$};
% Text Node
\draw (464,150) node [anchor=north west][inner sep=0.75pt]   [align=left] {$\displaystyle y_{L}$};
% Text Node
\draw (192,210) node [anchor=north west][inner sep=0.75pt]   [align=left] {\textcolor[rgb]{0.10,0.10,0.80}{=}};
% Text Node
\draw (184,39) node [anchor=north west][inner sep=0.75pt]   [align=left] {$\displaystyle \textcolor[rgb]{0.82,0.01,0.11}{\neq }$};
% Text Node
\draw (353,210) node [anchor=north west][inner sep=0.75pt]   [align=left] {\textcolor[rgb]{0.10,0.10,0.80}{=}};

% Text Node
\draw (162,125) node [anchor=north west][inner sep=0.75pt]   [align=left] {$\displaystyle f$};
% Text Node
\draw (383,125) node [anchor=north west][inner sep=0.75pt]   [align=left] {$\displaystyle f$};

\end{tikzpicture}
\caption{Non-interference:  Two runs of a program $f$. Inputs $x_L$ and
output $y_L$ of security level below $t$ are the same in both runs, whereas inputs on $x_H$ at level above $t$ are different}
\label{FIG:noninterference}
\end{figure}
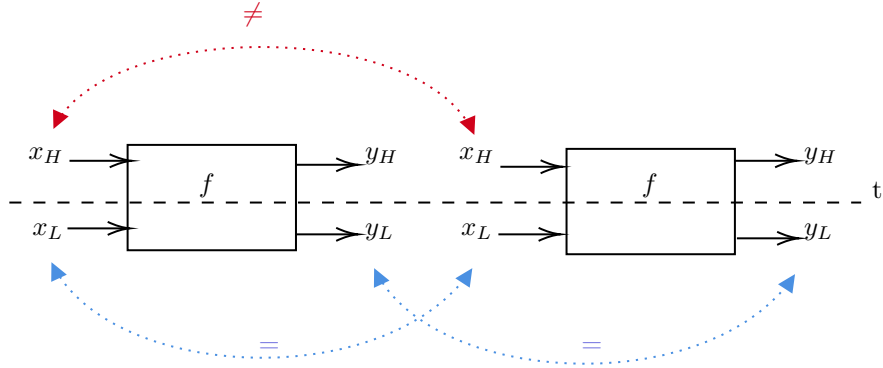

\autoref{FIG:noninterference} conveys the intuition behind the notion of non-interference.
Informally, the idea is that a program displays non-interference if two runs differing only in secret (high) inputs exhibit the same observable behaviour on the public (low) outputs 

\begin{theorem}[Non-interference for \nlustre]
\label{THM:Non-interference}
    Let $f \in G$ be a node with security signature
        %removed references to glb
        \[ \securitySignature{f}{\vv{\type{\alpha}}}{\vv{\type{\beta}}}{\type{\gamma}}{\type{\rho}} \]
    which is secure with respect to ground instantiation $s$ of the type variables. \\
    Let $eqs$ be the set of equations in $f$. 
    Let $X = fv(eqs)-dv(eqs)$, \textit{i.e.}, the input variables in $eqs$. \\
    Let $V = fv(eqs) \cup dv(eqs)$, \textit{i.e.}, the input, output and local variables. \\
    Let $\Gamma$ (and $s$) be such that 
    \(
    \meqnPred{\Gamma}{ eqs }{\rho}%{\type{\mu}}
    \)
    and $\type{\rho}$ is satisfied by $s$.
    Let $t \in \tti{SC}$ be any security level. 
    Let $bs$ be a given (base) clock stream. \\
    Let $\Hst$ and $\Hst'$ be such that
    \begin{enumerate}
        \item for all $eq \in eqs$:  $\mstrEqnPred{G}{\Hst}{bs}{eq}$
        and $\mstrEqnPred{G}{\Hst'}{bs}{eq}$, \textit{i.e.}, both $\Hst$ and $\Hst'$ are consistent Stream histories on each of the equations.
        \item $\Hst |_{X_{\rel t}} = \Hst' |_{X_{\rel t}}$, i.e., 
        $\Hst$ and $\Hst'$ agree on the input variables which are at a security level $t$ or below.
    \end{enumerate}
    Then $\Hst |_{V_{\rel t}} = \Hst' |_{V_{\rel t}}$, i.e, $\Hst$ and $\Hst'$ agree on all variables of the node $f$ that are given a security level $t$ or below.
\end{theorem}

\noindent
\Proof
The proof is by induction on the dependency level of $f \in G$.

For level 0 nodes (leaves in the DAG), the only equations are of the form $\eqn{x}{ce}{ck}$ and $\eqn{x}{\fby{c}{e}}{ck}$.  
We first consider only single equations.
Consider the case when $x \in X_{\rel t}$ (the other case does not matter).
From the rules (CSeqn) and (CSfby), we have $\type{\beta \stlub \gamma \rel \alpha}$, and consequently $s(\type{\beta \stlub \gamma}) \rel s(\type{\alpha}) \rel t$.
By Lemma \ref{LEM:SimpleSec}, $fv(ge) \subseteq  X_{\rel t}$ (otherwise we would contradict $s(\type{\alpha}) \rel t$).
So by Lemma \ref{LEM:relevant-variables}:
$\mstrSemAnyExpPred{\Hst}{bs}{ge}{[vs]}$ iff $\mstrSemAnyExpPred{\Hst'}{bs}{ge}{[vs]}$.
Therefore by the rules in \autoref{FIG:StrSemNodeEqn}, $\Hst(x) = \Hst'(x)$. \\[1mm]
Since the constraints of each equation must be satisfied by $s$, the result extends in a straightforward way to sets of equations.
Thus we have established the result for nodes at dependency level 0.
%%Confinement removed
% We can use the Confinement Lemma \ref{LEM:Confinement} to not consider the cases where $t \rel s(\type{\mu})$.
\\[1mm]
In the general case, we assume that the result holds for all nodes up to a dependency level $k$, and now consider a node at level $k+1$. \\[1mm]
There can now be 3 forms of equations:
$\eqn{x}{ce}{ck}$ and $\eqn{x}{\fby{c}{e}}{ck}$ (as before), and node calls.
For any of the two simple cases of equation, the proof follows the same reasoning given above for level 0 nodes. \\[1mm]
We now consider the case of node call equations $\eqn{\vv{x}}{f'(\vv{e})}{ck}$.
Suppose $x_i \in \set{\vv{x}}$. 
If $(\Gamma \circ s)(x_i) \not\rel t$, there is nothing to show.  
%%Confinement removed
% The Confinement Lemma \ref{LEM:Confinement} says we need \textit{not} consider the cases where $t \sqsubset s(\type{\mu})$. 
%
So we only need to consider the case where 
$s(\type{\nu_i}) \rel t$.\\[1mm]
Since $g$ is secure wrt $s$, by Definition \ref{DEF:nodesec}, each call to a node $g'$ at a dependency level $\leq k$ is secure with respect to the ground instantiation specified in Lemma \ref{LEM:sec-nl-nodecall}. 
Therefore, by invoking the induction hypothesis on $g'$ and the Stream semantics rules in \autoref{FIG:StrSemNodeEqn}, let us consider the  Stream histories $\Hst$ and $\Hst'$ augmented to include the flows on variables of this instance of $g'$.  
Let us call these $\Hst^{+g'}$ and $\Hst'^{+g'}$.
For the corresponding output variable $y'_j$ of security type $\type{\beta''_j}$ in node $g'$ on which $x_i$ depends, since $\type{\beta''_j \rel \nu_i}$ we have:
$\Hst^{+g'}(y'_j) = \Hst'^{+g'}(y'_j)$.
Whence by the rules in \autoref{FIG:StrSemNodeEqn}:  $\Hst(x_i) = \Hst'(x_i)$.
\QED

% 5
\section{Normalisation}
\label{SEC:Normalisation}

%The normalisation pass (\autoref{FIG:Normalisation}, \autoref{FIG:L2NLFbyInit}) and the theorems \autoref{THM:JFLA1} and \autoref{THM:JFLA2} are reproduced from \cite{Bourke-jfla2021}.
%\autoref{THM:SecPreserve} is our contribution.
\begin{figure*}
 %   \begin{align*}
 \[
 \begin{array}{rcl}
% Constant
        \norm{c} &=&  ([c^{\type{\bot}}],[\;]^{\type{\emptyset}}) ~~\hfill \textit{Xcnst} \\
% Variable
        \norm{x^{\type{\alpha}}} &=& 
                 ([x]^{\type{\alpha}}, [\;]^{\type{\emptyset}})
                  ~~\hfill \textit{Xvar} 
                  \\ \\
% Unary op
        \norm{\unop{e}} &=& \textrm{let}~ 
            ([e']^{\type{\alpha}},eqs^{\type{\rho}}) \leftarrow \norm{e}  ~~\hfill \textit{Xunop} \\
        && \textrm{in}~
        ([\unop{e'}]^{\type{\alpha}}, eqs^{\type{\rho}}) 
        \\ \\
% Binary op
        \norm{\binop{e_1}{e_2}} &=& \textrm{let}~
            ([e'_1]^{\type{\alpha_1}}, eqs_1^{\type{\rho_1}}) \leftarrow \norm{e_1} ~\textrm{and}~ ([e'_2]^{\type{\alpha_2}},eqs_2^{\type{\rho_2}}) \leftarrow \norm{e_2} ~~\hfill \textit{Xbinop} \\
        && \textrm{in}~
        ([\binop{e'_1}{e'_2}]^{\type{\alpha_1 \stlub \alpha_2}}, (\union{eqs_1}{eqs_2})^{\union{\type{\rho_1}}{\type{\rho_2}}}) 
        \\ \\
% when
        \norm{\when{\vv{e}}{x^{\type{\gamma}}=k}} &=& \textrm{let}~
        ([e'_1{}^{\type{\alpha_1}}, \ldots, e'_m{}^{\type{\alpha_m}}], eqs^{\type{\rho}})  \leftarrow \norm{\vv{e}} ~~\hfill \textit{Xwhn}\\
        && \textrm{in}~ 
        ([\when{e'_1}{x=k}^{\type{\alpha_1 \stlub \gamma}}, \ldots,  \when{e'_m}{x=k}^{\type{\alpha_m \stlub \gamma}} ], eqs^{\type{\rho}}) 
        \\ \\
% fby
        \norm{\fby{\vv{e_0}}{\vv{e_1}}} &=& 
        \textrm{let}~
        (\vv{e'_0}{}^{\type{\vv{\alpha}}}, eqs_0^{\type{\rho_0}})   \leftarrow \norm{\vv{e_0}} ~\textrm{and}~
     (\vv{e'_1}{}^{\type{\vv{\beta}}}, eqs_1^{\type{\rho_1}})   \leftarrow \norm{\vv{e_1}}  ~~\hfill \textit{Xfby} \\
    && \textrm{in}~ 
    (\vv{x}^{\type{\vv{\delta}}},  % \alpha_1 \stlub \beta_1
        (\{ ( x_i=\fby{e'_{0i}}{e'_{1i}} )_{i=1}^k
                 \} \cup {eqs_0} \cup {eqs_1})^{\type{\rho}}) \\
    &&     \textrm{where}~     \type{\rho} = \{ ( \type{\alpha_i \stlub \beta_i \strel \delta_i} )_{i=1}^k \} \cup
    \type{\rho_0} \cup \type{\rho_1}
                 %}) 
                 \\ \\
% merge 
    \norm{\lmerge{x^{\type{\gamma}}}{\vv{e_1}}{\vv{e_2}}} &=& \textrm{let}~ (\vv{e'_1}^{\type{\vv{\alpha}}},eqs_1^{\type{\rho_1}}) \leftarrow \norm{\vv{e_1}} ~\textrm{and}~  (\vv{e'_2}^{\type{\vv{\beta}}},eqs_2^{\type{\rho_2}}) \leftarrow \norm{\vv{e_2}} ~~\hfill \tti{Xmrg} \\
       && \textrm{in}~
       (\vv{x}^{\type{\vv{\delta}}},  % \alpha_1 \stlub \beta_1
        (\{ ( x_i=\lmerge{x}{e'_{1i}}{e'_{2i}} )_{i=1}^k
                 \} \cup {eqs_1} \cup {eqs_2})^{\type{\rho}}) \\
    &&     \textrm{where}~     \type{\rho} = \{ ( \type{\gamma \stlub \alpha_i \stlub \beta_i \strel \delta_i} )_{i=1}^k \} \cup
    \type{\rho_1} \cup \type{\rho_2} \\ \\
% ite
%writing the ite in one line leads to a overfullbox warning
    \lfloor\ttt{if}~ e~ \ttt{then}~ \vv{e_t}
    \\ 
     ~\ttt{else} ~{\vv{e_f}}\rfloor &=& 
     \textrm{let}~ (e'^{\type{\kappa}},eqs_c^{\type{\rho_c}}) \leftarrow \norm{e} ~\textrm{and}~ (\vv{e'_t}^{\type{\vv{\alpha}}},eqs_t^{\type{\rho_t}}) \leftarrow \norm{\vv{e_t}} ~~\hfill \tti{Xite} \\ 
    &&\textrm{and}~  (\vv{e'_f}^{\type{\vv{\beta}}},eqs_f^{\type{\rho_f}}) \leftarrow \norm{\vv{e_f}} ~ \textrm{in}~ \\
    && 
       (\vv{x}^{\type{\vv{\delta}}},  % \alpha_1 \stlub \beta_1
        (\{ ( x_i=\ite{e'}{e'_{ti}}{e'_{fi}} )_{i=1}^k
                 \} \cup {eqs})^{\type{\rho}}) \\
    && \textrm{where}~ eqs = \union{eqs_c}{eqs_t}{eqs_f} \\
    &&\type{\rho} = \union{( \type{\kappa \stlub \alpha_i \stlub \beta_i \strel \delta_i} )_{i=1}^k }{\type{\rho_c}}{\type{\rho_t}}{\type{\rho_f}}\\ \\
% Node call
        \norm{f(e_1,...,e_n)} &=& \textrm{let}~
        ([e'_1,...,e'_m]^{\type{\vv{\alpha'}}}, eqs^{\type{\rho_1}}) \leftarrow \norm{e_1,...,e_n} 
         ~~\hfill \textit{Xncall}\\
        && \textrm{in}~ ([x_1{}^{\type{\delta_1}},...,x_k{}^{\type{\delta_k}}],\\
        && ~~(\union{\{ (x_1,...,x_k)= f(e'_1,...,e'_m) \}}{eqs})^{\type{\rho_2}}) \\
 && \textrm{where}~ \type{\rho_2} = \type{\rho [\vv{\alpha'}/\vv{\alpha}] [\vv{\delta}/\vv{\beta}][\gamma'/\gamma]
 } \cup \type{\rho_1} \\
 && \textrm{given}~  \securitySignature{f}{(\type{\vv{\alpha}})}{\type{\vv{\beta}}}
 {\type{\gamma}}{\type{\rho}} ~\textrm{and}~ \type{\gamma'} = \Gamma(\texttt{base}) \\ \\
 % tuples
 \norm{[e_1, \ldots, e_k]} &=& \textrm{let~for}~ i \in \{1,\ldots, k\}:  ~~\hfill \textit{Xtup} \\ 
 && ~~~~~~~~~
 ( [e'_{i 1}{}^{\type{\alpha_{i 1}}}, \ldots, 
   e'_{i m_i}{}^{\type{\alpha_{i m_i}}}], eqs_i^{\type{\rho_i}} ) \leftarrow
 \norm{ e_i }  \\
         && \textrm{in}~ 
          ( [ e'_{11}{}^{\type{\alpha_{1 1}}},
          \ldots, 
          e'_{1 m_1}{}^{\type{\alpha_{1 m_1}}},
          \ldots,
         e'_{k 1}{}^{\type{\alpha_{k 1}}} 
          \ldots, 
          e'_{k m_k}{}^{\type{\alpha_{k m_k}}} 
          ], 
          \\ &&
          ~~~~~
         (\bigcup_{i=1..k} eqs_i)^{\cup_i {\type{\rho_i}}}) \\
\\
%clocks
\norm{\ttt{base}} &=& \ttt{base}  ~~\hfill \textit{Xbase}\\ 
\norm{\ck{ck}{x=k}} &=& \ck{\norm{ck}}{x=k} ~~\hfill \textit{Xon}\\
\\
 % Equations
 \norm{\vv{x}^{\vv{\beta}} =_{ck^{\type{\gamma}}} \vv{e}} &=& \textrm{let}~
(\vv{e'}{}^{\type{\vv{\alpha}}}, eqs^{\type{\rho}}) \leftarrow \norm{\vv{e}} 
 ~~\hfill \textit{Xeqs}\\
        && \textrm{in}~ 
    (\{ (\vv{x}_j =_{ck} e'_j)_{j=1}^m \} \cup eqs)^{\{ \type{(\gamma \stlub \alpha_i \strel \beta_i)_{i=1}^k} \} \cup \type{\rho}  }
%   \end{align*}
\end{array}
\]
    \caption{\lustre\ to \nlustre\ normalisation}
    \label{FIG:Normalisation}
\end{figure*}

We now present Bourke \textit{et al.}'s ``normalisation transformations'' , which de-nest and distribute operators over lists (tuples) of expressions, and finally transform \ttt{fby} expressions to a form where the first argument is a constant.

Normalising an $n$-tuple of \lustre\ expressions yields an $m$-tuple of \lustre\ expressions without tupling and nesting, and a set of equations, defining fresh local variables (\autoref{FIG:Normalisation}).
We denote the transformation as 
\[ ([e'_1, \ldots , e'_m]^{\type{\alpha_1, \ldots, \alpha_m}}, eqs^{\type{\rho}}) \leftarrow \lfloor e_1,..., e_n \rfloor \]
where we have additionally decorated the transformations of \cite{Bourke-jfla2021} with security types for each member of the tuple of expressions, and with a set of type constraints for the generated equations.
We show that the normalisation transformations are indeed \textit{typed transformations}.
Our type annotations indicate why security types and constraints of well-security-typed \lustre\ programs are preserved (modulo satisfaction), as in \autoref{THM:SecPreserve}.

% Explain the translation
The rules \textit{(Xcnst)} and \textit{(Xvar)} for constants and variables are straightforward, generating no equations.
The rules  \textit{(Xunop)}-\textit{(Xbinop)} for unary and binary operators are obvious, collecting the equations from the recursive translations of their sub-expression(s) but generating no new equations.
In rule \textit{(Xwhn)}, where the sampling condition is distributed over the members of the tuple, the security type for each component expression in the result of translation is obtained by taking a join of the security type $\type{\alpha_i}$ of the expression $e'_i$ with $\type{\gamma}$, \textit{i.e.}, that of the variable $x$.

Of primary interest are the rules \textit{(Xfby)} for \texttt{fby} and \textit{(Xncall)} for \textit{node call}, where fresh program variables $x_i$ and their defining equations are introduced.
In these cases, we introduce \textit{fresh} security type variables $\type{\delta_i}$ for the $x_i$, and add appropriate type  constraints where the security type expressions for the expressions are bounded above by the type $\type{\delta_i}$.
In \textit{(Xncall)}, where the constraints are obtained from the node signature via substitution, note that this is achieved by substituting the $\type{\delta_i}$ for the security types $\type{\beta_i}$ of the output streams.

The rules \textit{(Xite)} and \textit{(Xmrg)} resemble \textit{(Xfby)} in most respects, collecting the equations (and their corresponding security type constraints) generated for their sub-expressions.  
(In fact, here too, new equations are generated, and thus these ``control expressions'' are not nested below the simpler expressions).

The rules \textit{(Xbase)} and \textit{(Xon)} for clocks also introduce no equations, and are straightforward.
The rules \textit{(Xtup)} for tuples (lists) of expressions and \textit{(Xeqs)} for equations regroup the resulting expressions appropriately.
The translation of node definitions involves translating the equations, and adding the fresh local variables.

\begin{figure*}
        \begin{equation*}
            \norm{x^{\type{\theta}} =_{ck^{\type{\gamma}}} \lfby{e_0^{\type{\alpha}}}{e^{\type{\beta}}}}_{\textit{fby}} =
              \begin{cases}
                 xinit^{\type{\delta_1}} =_{ck^{\type{\gamma}}} \nlfby{\ttt{true}^{\type{\bot}}}{\ttt{false}^{\type{\bot}}} ~\hfill~ 
                 \type{\bot \stlub \gamma ~\strel~ \delta_1} \\
                 px^{\type{\delta_2}} =_{ck^{\type{\gamma}}} \nlfby{c^{\type{\bot}}}{e^{\type{\beta}}} ~\hfill~ 
                 \type{\gamma \stlub \beta ~\strel~ \delta_2}\\
                 x^{\type{\theta}}  =_{ck^{\type{\gamma}}} \ttt{if}~{xinit^{\type{\delta_1}}}\ttt{then}~{e_0^{\type{\alpha}}} ~\hfill~ 
                 \type{\gamma \stlub \delta_1 \stlub \alpha \stlub \delta_2 ~\strel~ \theta}\\
                 ~~~~~~~~~~~\ttt{else} ~{px^{\type{\delta_2}}}
              \end{cases}       
          \end{equation*}

    \caption{Explicit \ttt{fby} initialisation}
    \label{FIG:L2NLFbyInit}
\end{figure*}

For the further transformation involving explicit initialisation of \texttt{fby} with a constant (\autoref{FIG:L2NLFbyInit}), the new equations are constraints are straightforward:  
two local definitions define (i) a flow named $\textit{xinit}$ that is initially true and thereafter false, and (ii) a flow named $\textit{px}$ the delayed expression $e$ with an arbitrarily chosen initial constant $c$ (of appropriate data type).
These are combined via a conditional on $\textit{xinit}$ that chooses the expression $e_0$ in the first instant, and thereafter $\textit{px}$.
Two new security variables $\type{\delta_1}$ and $\type{\delta_2}$ are introduced, and three constraints for the three equations introduced.

\subsection{Subject Reduction and Non-Interference}
\label{SEC:SubjectReduction}

\begin{theorem}[Preservation of security types]
    \label{THM:SecPreserve}
%    Normalisation preserves security of \lustre\ programs.
Let $f \in G$ be a node in \lustre\ program $G$.
If the node signature for $f$ in $G$ is 
$\securitySignature{f}{(\vv{\type{\alpha}})}{\vv{\type{\beta}}}{\type{\gamma}}{\type{\rho}}$, correspondingly in $\norm{G}$ it is $ \securitySignature{f}{(\vv{\type{\alpha}})}{\vv{\type{\beta}}}{\type{\gamma}}{\type{\rho'}}$,
and for any ground instantiation $s$,
$s(\type{\rho})$ implies $s(\type{\rho'})$.
\end{theorem}
\Proofsketch
The proof is on the DAG structure of $G$. 
Here we rely on the topological dependency order on nodes, and in their modularity (i.e., that nodes have no free variables and make no recursive calls), and the correctness of \type{\textsf{simplify}} (Lemma \ref{LEM:Simplify}) in determining node signatures.
Within a node, the proof employs induction on the structure of equations and expressions. 
(The reader can get some intuition about type preservation by inspecting the annotations in \autoref{FIG:Normalisation}).
Lemma \ref{LEM:Simplify} is central to establishing that the type signature of a node does not change in the normalisation transformations of \autoref{SEC:Normalisation}, which introduce equations involving fresh local program variables (and the associated security type constraints involving fresh type variables). 
These fresh type variables are eliminated from the constraints via substitution in \type{\textsf{simplify}}.

For the further explicit initialisation of \texttt{fby} (\autoref{FIG:L2NLFbyInit}), the preservation of security via \type{\textsf{simplify}} is also easy to see.
\QED

% 5.2
\subsection{An Example With Typing Analysis}
\label{SUB:AnalysisExample}

We adapt the examples given in \cite{Bourke-jfla2021} of the translation from \lustre\ to \nlustre, and show how our typing rules and security analysis work at both source and target languages.
In this process, we illustrate the preservation of the security types during the translation.
We annotate the programs with \textcolor{blue}{security types} (as superscripts) on program variables and expressions and write \textcolor{blue}{constraints} over these security types for each equation (as \lustre\ comments), according to the typing rules. 

\begin{figure*}
\footnotesize
    \begin{minipage}{.5\textwidth}
\begin{lstlisting}[language=Lustre]
node cnt_dn
  (res$^{\type{\alpha_1}}$: bool; n$^{\type{\alpha_2}}$: int) 
  returns (cpt$^{\type{\beta}}$: int); 
let
  (cpt$^{ck}$)$^{\type{\beta^\gamma}}$ = if res$^{\type{\alpha_1}}$ then n$^{\type{\alpha_2}}$ 
    else (n$^{\type{\alpha_2}}$ fby (cpt$^{\type{\beta}}$-1));    
  -- $\type{\rho_L} =  \{ \type{ \gamma \lub \alpha_1 \lub \alpha_2 \lub \alpha_2 \lub \beta \lub \bot \strel \beta} \} $ 
tel
\end{lstlisting}
\hrule
\begin{flushleft}
  \footnotesize
    \begin{align*}
{}&\lsimpl{(\type{\beta}, \type{ \rho_L } )}{\{\}} =  \\
{}&  ~~~(\type{\beta}, \{ \type{\gamma \lub \alpha_1 \lub \alpha_2 ~\strel~ \beta } \}) \\
{}&\nlsimpl{ (\type{\beta}, (\type{\rho_1} \cup \type{\rho_2} \cup \type{\rho_3} \cup \type{\rho_4}) )}{\vv{\delta}} \\
{}& = ~~ (\type{\beta}, \{ \type{\gamma \lub \alpha_1 \lub \alpha_2 ~\strel~ \beta } \}) \\
{}&\textit{where} ~ \type{\vv{\delta}} =  ~~\{\{\type{\delta_1},\type{\delta_2},\type{\delta_3}\}\} 
\end{align*}
\end{flushleft}
    \end{minipage}%    
    \begin{minipage}{.5\textwidth}
\begin{lstlisting}[language=Lustre]
node cnt_dn
  (res$^{\type{\alpha_1}}$: bool; n$^{\type{\alpha_2}}$: int) 
 returns (cpt$^{\type{\beta}}$: int); 
 var v14$^{\type{\delta_1}}$, v24$^{\type{\delta_2}}$, v25$^{\type{\delta_3}}$:int; 
let
  v24$^{\type{\delta_2}}$ =$^{\type{\gamma}}$ true fby false;
  -- $\type{\rho_1} = \{ \type{\gamma \lub \bot \lub \bot \rel \delta_2} \}$
  v25$^{\type{\delta_3}}$ =$^{\type{\gamma}}$ 0 fby (cpt$^{\type{\beta}}$ -1);
  -- $\type{\rho_2} = \{ \type{\gamma \lub \beta \lub \bot \rel \delta_3} \}$
  v14$^{\type{\delta_1}}$ =$^{\type{\gamma}}$ if v24$^{\type{\delta_2}}$ then n 
    else v25$^{\type{\delta_3}}$;
  -- $\type{\rho_3}  = \{ \type{\gamma \lub \delta_2 \lub \delta_3 \rel \delta_1} \}$
  cpt$^{\type{\beta}}$ =$^{\type{\gamma}}$ if res$^{\type{\alpha_1}}$ then n$^{\type{\alpha_2}}$ 
    else v14$^{\type{\delta_1}}$;    
  -- $ \type{\rho_4}  = \{ \type{ \gamma \lub \alpha_1 \lub \alpha_2 \lub \delta_1 \strel \beta} \}$ 
tel
\end{lstlisting}
    \end{minipage}

    \noindent\rule{\linewidth}{0.4pt}
    
    \caption{Example of normalisation with security analysis}
    \label{FIG:L2NLEx1}
\end{figure*}
    
The node \ttt{cnt\_dn} defined in \autoref{FIG:L2NLEx1} implements a count-down timer that returns stream \ttt{cpt}, which is initialized with the value of \ttt{n} on $0^{th}$ tick and whenever there is a \ttt{T} on reset \ttt{res},
and which decrements on each clock tick otherwise.
Changing the value of \ttt{n} when the reset \ttt{res} is \ttt{F} doesn't affect the count on \ttt{cpt}.

We assign security types $\type{\alpha_1}$ to
input \ttt{res}, and $\type{\alpha_2}$ to input \ttt{n}.
The output \ttt{cpt} is assigned security type $\type{\beta}$, and the clock $ck$ the type $\type{\gamma}$.
There are no local variables.
Based on the rules (LTvar), (LTbinop), (LTfby) and (LTite), we get constraint $\type{\rho_L}$.
After simplification, the resultant security signature of \ttt{cnt\_dn} is given by:
\[
  \securitySignature{\ttt{cnt\_dn}}
  {(\type{\alpha_1,\alpha_2})}
  {\type{\beta}}
  {\type{\gamma}}
  {\{ \type{\gamma \stlub \alpha_1 \stlub \alpha_2 ~\strel~ \beta} \}}
\]

The normalisation pass de-nests the \ttt{fby} expression and explicitly initializes it into 3 different local streams (\ttt{v14},\ttt{v24},\ttt{v25}).
These have security types
$\type{\delta_1}, \type{\delta_2}, \type{\delta_3}$ respectively.
The local variables generate constraints $\type{\rho_1}, \type{\rho_2}, \type{\rho_3}$ which are eliminated by \ckFont{simplify}.

It can be checked that the resultant signature of \ttt{cnt\_dn} in the translated program is also given by:
\[
  \securitySignature{\ttt{cnt\_dn}}
  {(\type{\alpha_1,\alpha_2})}
  {\type{\beta}}
  {\type{\gamma}}
  {\{ \type{\gamma \stlub \alpha_1 \stlub \alpha_2 ~\strel~ \beta} \}}
\]

% Using cnt_dn
The \ttt{re\_trig} node in Figure \ref{FIG:L2NLEx2} uses the \ttt{cnt\_dn} node of \autoref{FIG:L2NLEx1} to implement a count-down timer that is explicitly triggered whenever there is a rising edge (represented by \ttt{edge})  on \ttt{i}. 
If the count \ttt{v} expires to $0$ before a \ttt{T} on \ttt{i}, the counter isn't allowed restart the count. 
Output \ttt{o} represents an active count in progress. 

The input streams \ttt{i} and \ttt{n} are given security types $\type{\alpha'_1}, \type{\alpha'_2}$ respectively, and the output stream \ttt{o} the security type $\type{\beta'}$.
The local variables \ttt{edge}, \ttt{ck}, and \ttt{v} are given 
security types $\type{\delta'_1}, \type{\delta'_2}, \type{\delta'_3}$ respectively, and further the nested call to \ttt{cnt\_dn} is annotated with $\type{\delta'_6}$.
The typing rules yield the constraints mentioned in the comments.

\begin{figure*}
\footnotesize
    \begin{minipage}{.5\textwidth}
        \begin{lstlisting}[language=Lustre,numbers=left]
node re_trig(i$^{\type{\alpha'_1}}$:bool; n$^{\type{\alpha'_2}}$:int)
  returns (o$^{\type{\beta'}}$ : bool)
  var edge$^{\type{\delta'_1}}$, ck$^{\type{\delta'_2}}$:bool,
   v$^{\type{\delta'_3}}$:int;
let
  (edge$^{base}$)$^{\type{\delta^{'\gamma'}_1}}$ = i$^{\type{\alpha'_1}}$ and 
    (false$^{\type{\bot}}$ fby (not i$^{\type{\alpha'_1}}$)); 
-- $\type{\rho_{1L}}= \{\type{\gamma' \lub \alpha'_1 \lub \bot \lub \alpha'_1 \rel \delta'_1}\}$
  (ck$^{base}$)$^{\type{\delta^{'\gamma'}_2}}$ = edge$^{\type{\delta'_1}}$ or 
    (false$^{\type{\bot}}$ fby o$^{\type{\beta'}}$);
-- $\type{\rho_{2L}}= \{\type{\gamma' \lub \delta'_1 \lub \bot \lub \beta' \rel \delta'_2}\}$
  (v$^{base}$)$^{\type{\delta^{'\gamma'}_3}}$ = merge ck$^{\type{\delta'_2}}$
   (cnt_dn((edge$^{\type{\delta'_1}}$, n$^{\type{\alpha'_2}}$)
    when ck$^{\type{\delta'_2}}$))$^{{\type{\delta'_6\{|\rho'|\}}}^{\type{\delta'_2}}}$ 
   (0 when not ck$^{\type{\delta'_2}}$);
-- $\type{\rho'}= \{ \type{\delta'_2 \lub (\delta'_1 \lub \delta'_2) \lub (\alpha'_2 \lub \delta'_2) \rel\delta'_6}\}$
-- $\type{\rho_{3L}}= \{\type{\gamma' \lub \delta'_2 \lub \delta'_6 \lub \bot \lub \delta'_2 \rel \delta'_3}\} \cup \type{\rho'}$
   (o$^{base}$)$^{\type{\beta^{'\gamma'}}}$= v$^{\type{\delta'_3}}$ > 0$^{\type{\bot}}$;
-- $\type{\rho_{4L}}= \{\type{\gamma' \lub \delta'_3 \lub \bot \rel \beta'}\}$
tel
    \end{lstlisting}
    \end{minipage}%
        %analysis of re_trigger_rising_edge
        \begin{minipage}{.5\textwidth}

\begin{lstlisting}[language=Lustre,numbers=right]
node re_trig(i$^{\type{\alpha'_1}}$:bool; n$^{\type{\alpha'_2}}$:int)
 returns (o$^{\type{\beta'}}$ : bool)
 var edge$^{\type{\delta'_1}}$, ck$^{\type{\delta'_2}}$:bool, v$^{\type{\delta'_3}}$:int,
  v22$^{\type{\delta'_4}}$:bool, v21$^{\type{\delta'_5}}$:bool,
  v24$^{\type{\delta'_6}}$:int when ck;
let
  v22$^{{\type{\delta_4}}}$  =$_{\type{\gamma'}}$ false$^{\type{\bot}}$ fby 
    (not i$^{\type{\alpha'_1}}$);
-- $\type{\rho_1} = \{\type{\gamma' \lub \bot \lub \alpha'_1 \rel \delta'_4}\} $ 
  edge$^{\type{\delta'_1}}$ =$_{\type{\gamma'}}$ i$^{\type{\alpha'_1}}$ and v22$^{\type{\delta'_4}}$;
-- $\type{\rho_2} = \{\type{ \gamma' \lub \alpha'_1 \lub \delta'_4 \rel \delta'_1}\} $ 
  v21$^{\type{\delta'_5}}$ =$_{\type{\gamma'}}$ false$^{\type{\bot}}$ fby o$^{\type{\beta'}}$;
-- $\type{\rho_3} = \{\type{\gamma' \lub \bot \lub \beta' \rel \delta'_5}\}$
  ck$^{\type{\delta'_2}}$  =$_{\type{\gamma'}}$ edge$^{\type{\delta'_1}}$ or v21$^{\type{\delta'_5}}$;
-- $\type{\rho_4} = \{\type{\gamma' \lub \delta'_1 \lub \delta'_5 \rel \delta'_2}\}$
  v24$^{\type{\delta'_6}}$ =$_{\type{\delta'_2}}$ cnt_dn(
    edge$^{\type{\delta'_1}}$ when ck$^{\type{\delta'_2}}$,
   n$^{\type{\alpha'_2}}$ when ck$^{\type{\delta'_2}}$);
-- $\type{\rho_5} = \{\type{\delta'_2 \lub (\delta'_1 \lub \delta'_2) \lub (\alpha'_2 \lub \delta'_2) \rel \delta'_6}\}$
  v$^{\type{\delta'_3}}$ =$_{\type{\gamma'}}$ merge ck$^{\type{\delta'_2}}$ v24$^{\type{\delta'_6}}$ 
    (0$^{\type{\bot}}$ when not ck$^{\type{\delta'_2}}$);
-- $\type{\rho_6}= \{\type{\gamma' \lub \delta'_2 \lub \delta'_6 \lub \bot \lub \delta'_2 \rel \delta'_3}\}$
  o$^{\type{\beta'}}$ =$_{\type{\gamma'}}$ v$^{\type{\delta'_3}}$>0$^{\type{\bot}}$;
-- $\type{\rho_7}= \{\type{\gamma' \lub \delta'_3 \lub \bot \rel \beta'}\}$
tel
\end{lstlisting}
        \end{minipage}
        \noindent\rule{\linewidth}{0.4pt}
        \centering
        \begin{align*}
        \lsimpl{(\type{\beta'}, \{ \union{\type{\rho_{1L}}}{\type{\rho_{2L}}}{\type{\rho_{3L}}}{\type{\rho_{4L}}}\})}{\{\type{\delta'_1},\type{\delta'_2},\type{\delta'_3}, \type{\delta'_6}\}} = 
        (\type{\beta'},
        \{ \type{\gamma' \lub \alpha'_1 \lub \alpha'_2 \rel \beta'}\}) \\
        \nlsimpl{(\type{\beta'}, \{ \union{\type{\rho_{1}}}{\type{\rho_{2}}}{\type{\rho_{3}}}{\type{\rho_{4}}}{\type{\rho_{5}}}{\type{\rho_{6}}}{\type{\rho_{7}}}\})}{\{\type{\delta'_1},\type{\delta'_2},\type{\delta'_3}, \type{\delta'_4}, \type{\delta'_5}, \type{\delta'_6}\}} \\
        \hfill = (\type{\beta'}, \{ \type{\gamma' \lub \alpha'_1 \lub \alpha'_2 \rel \beta'}\})
      \end{align*}
        \caption{Example: Security analysis and normalisation.  \ttt{when  ck} and \ttt{when not ck}] abbreviate \ttt{when ck = T} and \ttt{when ck = F} respectively.}
        \label{FIG:L2NLEx2}
\end{figure*}

Using \type{\textsf{simplify}} to
eliminate the security types $\type{\delta'_1}, \type{\delta'_2}, \type{\delta'_3}$,  and $\type{\delta'_6}$, 
of the local variables \ttt{edge}, \ttt{ck}, \ttt{v} and nested call to \ttt{cnt\_dn} respectively from the constraints (in lines 8,11,16-17,19 on the left), we
get the constraint
$\{ \type{\gamma' \lub \alpha'_1 \lub \alpha'_2 \strel \beta'}\}$.

After simplification, the resultant security signature of \ttt{re\_trig} is given by:
\[
  \securitySignature{\ttt{re\_trig}}
  {(\type{\alpha'_1,\alpha'_2})}
  {\type{\beta'}}
  {\type{\gamma'}}
  {\{ \type{\gamma' \stlub \alpha'_1 \stlub \alpha'_2 ~\strel~ \beta'} \}}
\]

Normalisation introduces local variables (\ttt{v21,v22,v24}) with security types
$\type{\delta'_4}, \type{\delta'_5},  \type{\delta'_6}$ (see lines 7,12,16 on the right).
(Identical names have been used to show the correspondence, especially between the nested call to \ttt{cnt\_dn} and its equational version in the \nlustre\ translation, both of which are given $\type{\delta'_6}$.)
The $\type{\delta'_i}$ are eliminated by \ckFont{simplify}, and the refinement type $\type{\delta'_6 \{\!| \rho' |\!\}}$ for the node call in the \lustre\ version becomes an explicit constraint $\type{\rho_5}$ (line 19) in \nlustre.
Observe that due to \type{\textsf{simplify}} also eliminating the newly introduced type variables $\type{\delta'_4}, \type{\delta'_5},  \type{\delta'_6}$ annotating the new program variables (\ttt{v21,v22,v24}), the security signature  of \ttt{re\_trig} remains the same across the translation.

% 6
\section{Security and Non-Interference for \lustre}\label{SEC:noninterference}

Having shown that \nlustre\ programs which are well-typed exhibit non-interference, and that the translation of a \lustre\ program to \nlustre\ preserves well-typedness, we proceed to show the main result (\autoref{THM:Derived-Non-interference}) that well-security-typed \lustre\ programs are non-interfering.
Our strategy is problem reduction to our earlier result of non-interference for \nlustre\ (\autoref{THM:Non-interference}), using \autoref{THM:SecPreserve}.

\subsection{Semantics preservation}

The remaining piece for the reduction to work is that the semantics of a program does not change during the transformation from \lustre\ to \nlustre.
We recall here (without proofs) the important results from \cite{Bourke-jfla2021,Bourke-TECS2021}, which establish the preservation of stream semantics by the transformations.

\begin{theorem}[Preservation of semantics. Theorem 2 of \cite{Bourke-jfla2021}]
    \label{THM:JFLA1}
    De-nesting and distribution preserve the semantics of \lustre\ programs. 
    (La passe de d\'{e}simbrication et distributivit\'{e} pr\'{e}serve la s\'{e}mantique des programmes.)
    \[
    \forall G~ f ~\tupstrm{xs}~\tupstrm{ys}:~~~ {\mstrCallPred{G}{\liftnode{f}}{\tupstrm{xs}}{\tupstrm{ys}}} \implies {\mstrCallPred{\norm{G}}{\liftnode{f}}{\tupstrm{xs}}{\tupstrm{ys}}}\]
\end{theorem}

\begin{theorem}[Preservation of semantics. Theorem 3 of \cite{Bourke-jfla2021}]
    \label{THM:JFLA2}
    The explicit initialisations of $\ttt{fby}$ preserve the semantics of the programs. 
    (L’explicitation des initialisations pr\'{e}serve la s\'{e}mantique des programmes.)
    \[
    \forall G~ f ~\tupstrm{xs}~\tupstrm{ys}:~~~ {\mstrCallPred{G}{\liftnode{f}}{\tupstrm{xs}}{\tupstrm{ys}}} \implies {\mstrCallPred{\norm{G}_{fby}}{\liftnode{f}}{\tupstrm{xs}}{\tupstrm{ys}}}\]
\end{theorem}

\subsection{Non-Interference for \lustre\ via reduction}\label{SUB:Lustre-Soundness}

\begin{figure}

\centering
\tikzset{every picture/.style={line width=0.75pt}} %set default line width to 0.75pt        

\begin{tikzpicture}[x=0.75pt,y=0.75pt,yscale=-1,xscale=1]

%Shape: Rectangle [id:dp4910268745468237] 
\draw   (107,61) -- (177,61) -- (177,101) -- (107,101) -- cycle ;
%Straight Lines [id:da6606601637719587] 
\draw    (273,81) -- (368,81) ;
\draw [shift={(370,81)}, rotate = 180] [color={rgb, 255:red, 0; green, 0; blue, 0 }  ][line width=0.75]    (10.93,-3.29) .. controls (6.95,-1.4) and (3.31,-0.3) .. (0,0) .. controls (3.31,0.3) and (6.95,1.4) .. (10.93,3.29)   ;
%Straight Lines [id:da8607145718174867] 
\draw    (80,71) -- (104,71) ;
\draw [shift={(106,71)}, rotate = 180] [color={rgb, 255:red, 0; green, 0; blue, 0 }  ][line width=0.75]    (10.93,-3.29) .. controls (6.95,-1.4) and (3.31,-0.3) .. (0,0) .. controls (3.31,0.3) and (6.95,1.4) .. (10.93,3.29)   ;
%Straight Lines [id:da7853412035746739] 
\draw    (81,95) -- (105,95) ;
\draw [shift={(107,95)}, rotate = 180] [color={rgb, 255:red, 0; green, 0; blue, 0 }  ][line width=0.75]    (10.93,-3.29) .. controls (6.95,-1.4) and (3.31,-0.3) .. (0,0) .. controls (3.31,0.3) and (6.95,1.4) .. (10.93,3.29)   ;
%Straight Lines [id:da4964355058115607] 
\draw    (177,72) -- (201,72) ;
\draw [shift={(203,72)}, rotate = 180] [color={rgb, 255:red, 0; green, 0; blue, 0 }  ][line width=0.75]    (10.93,-3.29) .. controls (6.95,-1.4) and (3.31,-0.3) .. (0,0) .. controls (3.31,0.3) and (6.95,1.4) .. (10.93,3.29)   ;
%Straight Lines [id:da7071446257817252] 
\draw    (176,95) -- (200,95) ;
\draw [shift={(202,95)}, rotate = 180] [color={rgb, 255:red, 0; green, 0; blue, 0 }  ][line width=0.75]    (10.93,-3.29) .. controls (6.95,-1.4) and (3.31,-0.3) .. (0,0) .. controls (3.31,0.3) and (6.95,1.4) .. (10.93,3.29)   ;
%Straight Lines [id:da19733579761346542] 
\draw  [dash pattern={on 0.84pt off 2.51pt}]  (120,32) -- (119.07,59) ;
\draw [shift={(119,61)}, rotate = 271.97] [color={rgb, 255:red, 0; green, 0; blue, 0 }  ][line width=0.75]    (10.93,-3.29) .. controls (6.95,-1.4) and (3.31,-0.3) .. (0,0) .. controls (3.31,0.3) and (6.95,1.4) .. (10.93,3.29)   ;
%Shape: Rectangle [id:dp4625001654976626] 
\draw   (459,62) -- (529,62) -- (529,102) -- (459,102) -- cycle ;
%Straight Lines [id:da6719937877020274] 
\draw    (432,72) -- (456,72) ;
\draw [shift={(458,72)}, rotate = 180] [color={rgb, 255:red, 0; green, 0; blue, 0 }  ][line width=0.75]    (10.93,-3.29) .. controls (6.95,-1.4) and (3.31,-0.3) .. (0,0) .. controls (3.31,0.3) and (6.95,1.4) .. (10.93,3.29)   ;
%Straight Lines [id:da16715772407290463] 
\draw    (433,96) -- (457,96) ;
\draw [shift={(459,96)}, rotate = 180] [color={rgb, 255:red, 0; green, 0; blue, 0 }  ][line width=0.75]    (10.93,-3.29) .. controls (6.95,-1.4) and (3.31,-0.3) .. (0,0) .. controls (3.31,0.3) and (6.95,1.4) .. (10.93,3.29)   ;
%Straight Lines [id:da09105767248290131] 
\draw    (529,73) -- (553,73) ;
\draw [shift={(555,73)}, rotate = 180] [color={rgb, 255:red, 0; green, 0; blue, 0 }  ][line width=0.75]    (10.93,-3.29) .. controls (6.95,-1.4) and (3.31,-0.3) .. (0,0) .. controls (3.31,0.3) and (6.95,1.4) .. (10.93,3.29)   ;
%Straight Lines [id:da685426113432854] 
\draw    (528,96) -- (552,96) ;
\draw [shift={(554,96)}, rotate = 180] [color={rgb, 255:red, 0; green, 0; blue, 0 }  ][line width=0.75]    (10.93,-3.29) .. controls (6.95,-1.4) and (3.31,-0.3) .. (0,0) .. controls (3.31,0.3) and (6.95,1.4) .. (10.93,3.29)   ;
%Straight Lines [id:da41708499113369646] 
\draw  [dash pattern={on 0.84pt off 2.51pt}]  (472,33) -- (471.07,60) ;
\draw [shift={(471,62)}, rotate = 271.97] [color={rgb, 255:red, 0; green, 0; blue, 0 }  ][line width=0.75]    (10.93,-3.29) .. controls (6.95,-1.4) and (3.31,-0.3) .. (0,0) .. controls (3.31,0.3) and (6.95,1.4) .. (10.93,3.29)   ;
%Straight Lines [id:da12967254466820022] 
\draw  [dash pattern={on 4.5pt off 4.5pt}]  (250,181) -- (443,181) ;
\draw [shift={(445,181)}, rotate = 179.74] [color={rgb, 255:red, 0; green, 0; blue, 0 }  ][line width=0.75]    (10.93,-3.29) .. controls (6.95,-1.4) and (3.31,-0.3) .. (0,0) .. controls (3.31,0.3) and (6.95,1.4) .. (10.93,3.29)   ;
% Text Node
\draw (286,66) node [anchor=north west][inner sep=0.75pt]   [align=left] {{\footnotesize Normalisation}};

% Text Node
\draw (130,72) node [anchor=north west][inner sep=0.75pt]   [align=left] {$\displaystyle f_{L}$};
% Text Node
\draw (27,62) node [anchor=north west][inner sep=0.75pt]   [align=left] {$\displaystyle x_{1} :\ \type{\alpha _{1}}$};
% Text Node
\draw (27,87) node [anchor=north west][inner sep=0.75pt]   [align=left] {$\displaystyle x_{2} :\ \type{\alpha _{2}}$};
% Text Node
\draw (205,87) node [anchor=north west][inner sep=0.75pt]   [align=left] {$\displaystyle y_{2} :\ \type{\beta _{2}}$};
% Text Node
\draw (204,62) node [anchor=north west][inner sep=0.75pt]   [align=left] {$\displaystyle y_{1} :\ \type{\beta _{1}}$};
% Text Node
\draw (124,25) node [anchor=north west][inner sep=0.75pt]   [align=left] {$\displaystyle ck:\ \type{\gamma} $};
% Text Node
\draw (482,72) node [anchor=north west][inner sep=0.75pt]   [align=left] {$\displaystyle f_{NL}$};
% Text Node
\draw (379,62) node [anchor=north west][inner sep=0.75pt]   [align=left] {$\displaystyle x_{1} :\ \type{\alpha _{1}}$};
% Text Node
\draw (379,87) node [anchor=north west][inner sep=0.75pt]   [align=left] {$\displaystyle x_{2} :\ \type{\alpha _{2}}$};
% Text Node
\draw (557,87) node [anchor=north west][inner sep=0.75pt]   [align=left] {$\displaystyle y_{2} :\ \type{\beta _{2}}$};
% Text Node
\draw (556,62) node [anchor=north west][inner sep=0.75pt]   [align=left] {$\displaystyle y_{1} :\ \type{\beta _{1}}$};
% Text Node
\draw (476,26) node [anchor=north west][inner sep=0.75pt]   [align=left] {$\displaystyle ck:\ \type{\gamma} $};

\draw (95,162) node [anchor=north west][inner sep=0.75pt]   [align=left] {$\displaystyle\securitySignature{f}{(\vv{\type{\alpha}})}{\vv{\type{\beta}}}{\type{\gamma}}{\type{\rho}}$};
% Text Node
\draw (449,164) node [anchor=north west][inner sep=0.75pt]   [align=left] {$\displaystyle \securitySignature{f}{(\vv{\type{\alpha}})}{\vv{\type{\beta}}}{\type{\gamma}}{\type{\rho'}}$};
% Text Node
\draw (292,160) node [anchor=north west][inner sep=0.75pt]   [align=left] {$\displaystyle \type{\rho} \Longrightarrow \type{\rho'}$};
% Text Node
\draw (276,187) node [anchor=north west][inner sep=0.75pt]   [align=left] {{\footnotesize \textcolor[rgb]{0.29,0.56,0.89}{Type Preservation}}};
% Text Node
\draw (259,94) node [anchor=north west][inner sep=0.75pt]   [align=left] {{\footnotesize \textcolor[rgb]{0.25,0.46,0.02}{Semantics Preservation}}};
% Text Node
\draw (425,210) node [anchor=north west][inner sep=0.75pt]   [align=left] {Non-interfering if $\displaystyle s( \type{\rho'})$ holds};

%Shape: Rectangle [id:dp7514025093237444] 
\draw  [color={rgb, 255:red, 74; green, 144; blue, 226 }  ,draw opacity=1 ][dash pattern={on 0.84pt off 2.51pt}] (420,208) -- (608,208) -- (608,232) -- (420,232) -- cycle ;

\end{tikzpicture}

\caption{Non-interference for \lustre\ by reduction, via preservation of semantics and types}
\label{FIG:subjectReduction}
\end{figure}
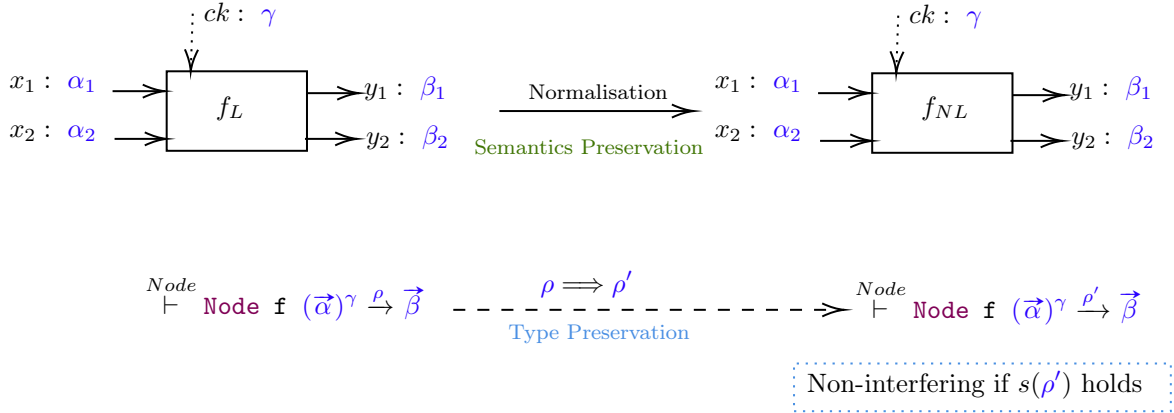

We now prove the main result of the soundness of the type system with respect to \lustre's Stream semantics, via a reduction.
\autoref{FIG:subjectReduction} diagrams our approach.
The upper part of the diagram depicts that the normalisation transformations from \lustre\ to \nlustre\ preserve semantics,  while the lower part of the diagram depicts the type preservation relationship between the  signatures of the source and translated program nodes.
The satisfaction of signature constraints $\type{\rho}$ by an assignment $s$ implies the satisfaction of $\type{\rho'}$ by $s$, and the satisfaction of $\type{\rho'}$ by $s$ indicates that the translated program node exhibits non-interference.

\begin{theorem}[Non-interference for \lustre]
\label{THM:Derived-Non-interference}
If program $G$ is well-security-typed in \lustre, then it exhibits non-interference with respect to \lustre's stream semantics.
\end{theorem}

\Proofsketch \ 
Let $G$ be a \lustre\ program that is well-security-typed (in \lustre's security type system).
This means that each node $f \in G$ is well-security-typed.
By induction on the DAG structure of $G$, using  \autoref{THM:SecPreserve}, each node in $\norm{G}$ is well-security-typed (in \nlustre's security type system).
By \autoref{THM:Non-interference}, $\norm{G}$ exhibits non-interference.
(That is, for any ground instantiation of security type variables, for any given security level $t$, two executions of $\norm{G}$  which agree on the inputs of streams assigned a security level at most $t$ exhibit the same behaviour on all output streams of security level at most $t$.)
By Theorems \ref{THM:JFLA1} and \ref{THM:JFLA2}, $\norm{G}$ and $G$ have the same \textit{extensional} semantics for each node.
Therefore, $G$ exhibits non-interference
(since two executions of $G$  which agree on the inputs of streams assigned a security level at most $t$ will also exhibit the same behaviour on all output streams of security level at most $t$.
\QED

% 7

\section{Related Work}
\label{SEC:related}
\noindent{\bf Security type systems:} 
Denning's seminal paper \cite{Denning:1976} proposed lattices as the appropriate structure for information flow analyses. 
The subsequent paper \cite{Denning:1977} presented static analysis frameworks for certifying secure information flow. 
A gamut of secure flow analyses were based on these foundations.

Only much later did Volpano \textit{et al.} \cite{Volpano96} provide a security type system with a semantic soundness result by showing that security-typed programs exhibit non-interference \cite{GoguenM82a}.
Type systems remain a powerful way of analysing program behaviour, particularly secure information flow.
%Types or typed annotations that specify rules regarding the usage of typed variables offer powerful and general ways by which program behaviors can be analysed, {\em vis-\`a-vis} IFC.  
For instance, the {\sc Jif} compiler \cite{jif} for the {\sc JFlow} language \cite{myers_jflow_1999} (based on Java) not only checks for IFC leaks but also deals with {\em declassification}, using the {\em Decentralised model} of data ownership \cite{MyersLiskov1997}. 
Matos \textit{et al.} proposed a \textit{synchronous reactive} extension of Volpano's imperative framework.  
Their language is at a lower level than \lustre, and has explicit synchronization primitives for broadcast signals, suspension, preemption and scheduling of concurrent threads.
While they employ the notion of \textit{reactive bisimulation} to deal with concurrency, the techniques employed there closely follow Volpano's formulation of the type system (which use \textit{var} and \textit{cmd} ``phrase'' types) and a reduction semantics (necessitating a subject reduction theorem to deal with type preservation during program execution).
These operational formulations of these approaches rely on comparing program state on termination of execution of a program phrase, which is not appropriate for \lustre, where the behaviour of programs involves infinite streams.
In contrast, we are able to leverage the declarative elegance and simplicity of \lustre\ to present a far simpler type system and its soundness proof in terms of \lustre's co-inductive stream semantics. \\[1ex]
\noindent{\bf Semantics and logics:}
Non-interference \cite{GoguenM82a} is considered a standard semantic notion for security although other notions of semantic correctness have been proposed, \textit{e.g.}, \cite{Boudol08}.
Non-interference is a typical \textit{hyperproperty} \cite{Clarkson2010}, \textit{i.e.}, a set of sets of program traces.  
Clarkson \textit{et al.} \cite{Rabe2014} have presented temporal logics {\em HyperLTL, HyperCTL$^{*}$}, for verification of hyperproperties. \\[1ex]
\noindent{\bf Beyond type systems:}
Zanotti \cite{Zanotti2002} proposed an abstract interpretation framework similar to the earlier work from Volpano \textit{et al.}, but strictly more general in its applicability.
Hunt and Sands \cite{HuntSands2006} extended type-based IFC checking with flow-sensitivity.
In {\sc Paragon}, a Java-based language proposed by Broberg \textit{et al.} \cite{Broberg2013}, one can additionally handle the runtime tracking of typestate.
However, in general, type-based techniques can exhibit imprecision as they lack flow and context sensitivity and do not systematically handle unstructured control flow and exceptions in programs.
Hammer and Snelting \cite{hammer_flow-sensitive_2009} proposed the usage of program dependence graphs (PDGs) to offer a flow-, context- and object-sensitive analysis to detect IFC leaks. 
Livshits \textit{et al.} \cite{Livshits2009} use data propagation graphs to automatically infer explicit information flow leaks. \\[1ex]
\noindent{\bf Runtime techniques:} 
Dynamic analyses provide greater precision, particularly in systems which rely on dynamic typing, when static dependency graph or type-based approaches are not adequate.  
Shroff \textit{et al.} \cite{Shroff2007} have proposed dynamic tracking of dependencies to analyse noninterference.  
Austin and Flanagan \cite{AustinFlanagan2009,AustinFlanagan2010} have proposed dynamic checks to efficiently detect implicit flows based on the {\em no-sensitive-upgrade} semantics and the {\em permissive-upgrade} semantics.  
Their subsequent work \cite{AustinFlanagan2017} addressed limitations in the semantics due to which executions where implicit flows cannot be tracked are prematurely terminated. \\[1ex]
\noindent{\bf IFC analyses in hardware and systems:}
We refer to but do not further discuss here work addressing IFC analyses in hardware systems, \textit{e.g.} \cite{TiwariISCA2011,TiwariASPLOS2014,zhang_hardware_2015,ferraiuolo2018hyperflow}, 
in programming languages \cite{sabelfeld2003language,Pottier2003-FlowCaml,liu2017fabric,roy2009laminar}, in operating systems,   \cite{Krohn2007-aa,zeldovich2006-osdi,cheng2012aeolus,efstathopoulos2005asbestos,roy2009laminar}, and in databases \cite{schultz2013ifdb}.
In the context of embedded systems, it will be interesting to see how our higher-level \lustre-based approach compares with  lower-level secure hardware description languages such as SecVerilog \cite{zhang_hardware_2015} and ChiselFlow \cite{ferraiuolo2018hyperflow} in which fine-grained security policies can be expressed.

% \noindent{\bf Hardware analysis:}
% Tiwari \textit{et al.} \cite{TiwariISCA2011} investigate dynamic IFC tracking at the level of logic gates, but could suffer from  overheads in area or power or issues in scaling the technique to large designs. 
% In subsequent work \cite{TiwariASPLOS2014} adds further logic for tracking information flows. 
% Zhang et al. \cite{zhang_hardware_2015} propose a HDL, SecVerilog, which makes it possible to statically analyse control of timing channels of processor designs. 
% \lustre\ has an added advantage over many HDLs that it comes with clean formal semantics, which enables correct-by-construction compilation of systems.

% 8
\section{Conclusions and Future Work}
\label{SEC:Conclusion}

We have presented a simple security type system for a synchronous reactive data-flow language, and shown its semantic soundness with respect to the language's stream semantics in the form of non-interference. 
We took an approach common in programming language research (employed, \textit{e.g.}, by Landin \cite{Landin66}), namely focusing on a core sub-language for which we presented a security type system and its soundness results, and then extended these to the full language, exploiting the fact that the compiler transformations are both semantics-preserving and security-type-preserving. 
The type system for the full language employed a version of refinement types under a sub-typing constraints regime. 
By formulating the type system in a symbolic form rather than in terms of a fixed security lattice, we are able to infer conditions under which an assignment of security levels to variables can ensure security.

Most prior work on type systems for security (based on the flow models of \cite{Denning:1976,Denning:1977}) and their corresponding soundness proofs have been for imperative programming languages and (higher-order) functional-imperative languages such as ML with imperative features. 
In those frameworks, the focus of the analysis has mainly been on control flow effects (including termination), and in the latter case of higher-order languages, on the security type constructions necessitated by records and higher-order function spaces (\textit{e.g.}, the phrase types in \cite{Volpano96}), and the sub-typing relations induced on them.
In contrast, declarative data-flow languages such as \lustre\ pose a different set of issues, since their semantics involve reactive transformations between \textit{infinite} input and output streams.
While our approach to showing that securely-typed programs exhibit non-interference broadly follows the paradigmatic approach of Volpano \textit{et al}  \cite{Volpano96}, we believe that the adaptation to a synchronous data flow setting is both novel and inventive.
The traditional issues of control flow effects and termination cannot be used in the same way, and indeed the results such as the \textit{confinement lemma}, so central to Volpano \textit{et al}'s formulation, become unimportant in the stream semantics model.
Non-interference requires a novel re-interpretation to handle possibly recursively defined flows, and to cater to the infinite stream semantics. 
On the other hand, declarative data-flow languages usually have much simpler data types, and only first-order functions.
Moreover, definitions in these ``single assignment'' frameworks allow us to formulate \textit{type inference rules} and \textit{symbolic sub-type constraints} over type variables.
Finally, the simple and elegant  semantics of \lustre, particularly that all variables have unique definitions and that node calls are not recursive, greatly simplifies our formulation of the type system, the notion of security and the non-interference proof.

While \lustre's type system for data values is quite unremarkable, our security type system is not.
It is therefore satisfying to note that the transformation that was coincidentally dubbed ``normalisation'' happens to satisfy a ``subject reduction'' property, albeit within a compilation phase and for SIF types (and so without a Curry-Howard interpretation).
We believe that there may be other issues of interest in secure information flow in declarative frameworks that differ from those studied in frameworks involving data objects with mutable state.

A difficult aspect encountered during the transcription phase  \cite{Bourke-TECS2021,Bourke-jfla2021} concerns alignment of clocks in the presence of complex clock dependencies. 
We clarify that our type system, being static, only considers \type{security levels of clocks}, not actual clock behaviour, and therefore is free from such complications.
Further, the clocks induce no timing side-channels since the typing rules enforce, \textit{a fortiori}, that the security type of any (clocked) expression is at least as high as that of its clock.

\paragraph{Future work} \ \ 
We are currently developing mechanised proofs of our results, so that these can be integrated into the V\'{e}lus verified compiler framework \cite{Velus}, which dovetails into the CompCert approach of certified compilers  \cite{Leroy2009}.
This formalisation will realise our objective of developing correct-and-secure-by-construction implementations of a variety of embedded and reactive systems.
Accordingly, we have been careful to align our formulations to correspond closely with the V\'{e}lus project.

The next step on our agenda is the following compiler phase, namely the translation from \nlustre\ to the imperative language STC in the V\'{e}lus chain \cite{SNLustre2017,Lustre2020}.
STC programs involve assignments of values to variables, but continue to have equation-style flavour to them.
The only special case arises for the memory elements introduced for the \ttt{fby} construct, where the contents of a memory element are latched to be used the next instant.  

We believe that our type system poses no major problems in this phase.
Our preliminary results indicate that the \textit{instantaneous semantics} for \lustre\ provides the necessary scaffolding in showing that our notions of security and non-interference are mapped by the translation to their counterparts in a more traditional imperative state-transition setting.
Our intuition relies on the following observations about the target STC programs, which considerably simplify the flow analysis, and the preservation of security types:
\begin{itemize}
    \item since in a \nlustre\ node, each defined variable has a unique equation associated with it, the resulting STC program follows a single assignment regime for each variable. 
    An appropriate order of evaluation can easily be determined, which has little or no effect on the security types of  variables that are not dependent on one another.
    The \textit{confinement type} can be inferred easily, and is determined only by the security levels of the variables appearing in its defining expression. 
    \item in each ``instant'',  even in a conditional instruction, the same set of variables are assigned values in different branches.
    \item the initialisation by constants in the delayed (\ttt{fby} flows construct) of a \nlustre\ program ensures that the security type associated with a flow is determined by the type of the remainder of the flow (it can at best go up once, from $\type{\bot}$ to this security level).
\end{itemize}

% Monotone increasing types?

\textbf{Acknowledgements:} This work was initiated under an Indo-Japanese project DST/INT/JST/P-30/2016(G) \textit{Security in the IoT Space}, DST, Govt of India.

\bibliographystyle{acm}
\bibliography{bibliography}

\appendix

\section{Free Variable Definitions}
\label{APP:FVDef}
The definitions of free variables ($fv$) in expressions and equations, and defined variables ($dv$) in equations are given in \autoref{FIG:FVEdef} and \autoref{FIG:FVEQdef}.

\begin{figure}[ht]
\[
\begin{array}{rcl}
    fv(c) &=& \{\}  \\
    fv(x) &=& \{x\}  \\
    fv(\unop{e})&=& fv(e) \\
    fv(\binop{e_1}{e_2})&=& \union{fv(e_1)}{fv(e_2)}\\
    fv(\when{\vv{e}}{x=k})&=&\union{fv(\vv{e})}{\set{x}}\\
    fv(\lmerge{x}{\vv{e_1}}{\vv{e_2}}) &=& 
     \union{\set{x}}{fv(\vv{e_1})}{fv(\vv{e_2})} \\
    fv(\fby{\vv{e_1}}{\vv{e_2}}) &=& 
     \union{fv(\vv{e_1})}{fv(\vv{e_2})} \\
    fv(\ite{e_1}{\vv{e_2}}{\vv{e_3}}) &=& 
    \union{fv(e_1)}{fv(\vv{e_2})}{fv(\vv{e_3})} \\
    fv(f(\vv{e})) &=& \bigcup_{i} fv(\vv{e_i}) \\ \\[1mm]
%    \union{fv(e_1)}{fv(e_2)}{fv(e_3)}\\
    fv(\ttt{base}) &=& \{\ttt{base}\} \\
    fv(\ck{ck}{x=k}) &=& \union{fv(ck)}{\set{x}} \\ \\[1mm]
    fv(\vv{e})  &=& \bigcup_i {fv(e_i)} \\ \\[1mm]
    fv(e::ck) &=& \union{fv(e)}{fv(ck)} 
\end{array}
\]
    \caption{Free variables for expressions}
    \label{FIG:FVEdef}
\end{figure}

\begin{figure}[ht]
\[
\begin{array}{rcl}
    fv(\vv{x}=\vv{e} ) &=& fv(\vv{e}) \setminus \{\vv{x} \} \\
    \\[1mm]
    dv(\vv{x} =\vv{e}) &=& \{\vv{x}\}\\
\end{array}
\]
%old Nlustre equations

\[
\begin{array}{rcl}
    fv(\eqn{x}{ce}{ck}) &=& \union{fv(ck)}{fv(ce)} \setminus \{x\} \\
    fv(\eqn{x}{\fby{c}{e}}{ck}) &=& \union{fv(ck)}{fv(e)} \setminus \{x\} \\
    fv(\eqn{\vv{x}}{f(\vv{e})}{ck}) &=& \union{fv(ck)}{fv(\vv{e})} \setminus \set{\vv{x}} \\ 
    \\[1mm]
 dv(\eqn{x}{ce}{ck}) &=&  \{x\} \\
    dv(\eqn{x}{\fby{c}{e}}{ck}) &=&  \{x\} \\
    dv(\eqn{\vv{x}}{f(\vv{e})}{ck}) &=& \set{\vv{x}} 
\end{array}
\]
    \caption{Free and defined variables for equations}
    \label{FIG:FVEQdef}
\end{figure}

\ignore{
\begin{figure*}[ht]
\begin{minipage}{.3\textwidth}
\begin{align*}
    &fv(x)~ = ~ \{x\}  \\
    &fv(c)~ =~ \{\}  \\
    &fv(\unop{e})~ =~ fv(e) \\
    &fv(\binop{e_1}{e_2})~=~ \union{fv(e_1)}{fv(e_2)}\\
    &fv(\when{e}{x=k})~=~\union{fv(e)}{\set{x}}\\
    &fv(\ttt{base})~=~ \{\ttt{base}\} \\
\end{align*}

\end{minipage}%
\begin{minipage}{.7\textwidth}
\begin{align*}
    &fv(\vv{e})  ~=~ \bigcup_i {fv(e_i)} \\
    &fv(\lmerge{x}{e_1}{e_2}) ~ =~ \\
    &\qquad \union{\set{x}}{fv(e_1)}{fv(e_2)} \\
    &fv(\ite{e_1}{e_2}{e_3})~ =~ \\
    &\qquad \union{fv(e_1)}{fv(e_2)}{fv(e_3)}\\
    &fv(\ck{ck}{x=k}) ~=~ \union{fv(ck)}{\set{x}} \\
\end{align*}

\end{minipage}
\begin{minipage}{.4\textwidth}
\begin{align*}
&dv(\eqn{x}{ce}{ck}) ~=~  \{x\} \\
    &dv(\eqn{x}{\fby{c}{e}}{ck}) ~=~  \{x\} \\
    &dv(\eqn{x}{f(\vv{e})}{ck}) ~=~ \set{\vv{x}} \\
\end{align*}
\end{minipage}%
\begin{minipage}{.7\textwidth}
\begin{align*}
    &fv(\eqn{x}{ce}{ck}) ~=~ \union{fv(ck)}{fv(ce)} \setminus \{x\} \\
    &fv(\eqn{x}{\fby{c}{e}}{ck}) ~=~ \union{fv(ck)}{fv(e)} \setminus \{x\} \\
    &fv(\eqn{x}{f(\vv{e})}{ck}) ~=~ \union{fv(ck)}{fv(\vv{e})} \setminus \set{\vv{x}} \\
\end{align*}
\end{minipage}%
    \caption{Definition of $fv$ and $dv$ functions}
    \label{FIG:FVdef}
\end{figure*}
}

\section{Auxiliary Predicate Definitions}
\label{APP:AuxPred}
The definitions of the auxiliary semantic stream predicates \ckFont{when}, \ckFont{const}, \ckFont{merge}, \ckFont{ite} are given in \autoref{FIG:Def1}. All predicates except $\textcolor{blue}{\textsf{fby}_L}$ and $\textcolor{blue}{\textsf{fby}_{NL}}$ (defined in \autoref{FIG:LusFbyPred}) are reused to define semantics of both \lustre\ and \nlustre.% and \autoref{FIG:Def2}.

\begin{figure*}[ht]
\[
    \begin{array}{c}
        \namedJdg{\semConst{bs'}{c}{cs'}}
        {\semConst{\cc{\true}{bs'}}{c}{\ccnb{\stream{c}}{cs'}}}{DcnstT} ~~~~
        \namedJdg{\semConst{bs'}{c}{cs'}}
            {\semConst{\cc{\false}{bs'}}{c}{\ccnb{\nullStream}{cs'}}}{DcnstF} 
        \\ \\
        \namedJdg{\predSet{\liftunop{es'}{os'}}{\val{v'}= \unop{v}}}
            {\liftunop{\cc{\stream{v}}{es'}}{
                \ccnb{\stream{v'}}{os'}}}{Dunop}  
        ~~~~
        \namedJdg{\liftunop{es'}{os'}}
                {\liftunop{\cc{\nullStream}{es'}}{
                    \ccnb{\nullStream}{os'}}}{DunopA} 
                \\ \\
\namedJdg{es = \ccnb{v}{es'}}
    {\tl{es} = es'}{Dtl}    
~~~~
%            \\ \\
        \namedJdg{
            \predSet{\liftbinop{es_1'}{es_2'}{os'}}{\binop{v_1}{v_2}=\val{v}}}
        {\liftbinop{\cc{\stream{v_1}}{es_1'}}{\cc{\stream{v_2}}{es_2'}}{
            \ccnb{\stream{v}}{os'}}}{Dbinop}
            \\ \\
    \namedJdg{x \in \tti{dom}(\Hst)}
    {\htl{\Hst}(x) =  \tl{\hist{x}}}{Dhtl}
    ~~~~    

        \namedJdg{\liftbinop{es_1'}{es_2'}{os'}}
        {\liftbinop{\cc{\nullStream}{es_1'}}{\cc{\nullStream}{es_2'}}{
            \ccnb{\nullStream}{os'}}}{DbinopA} 
   \\ \\
   \namedJdg{\resClk{\Hst}{bs}{vs}}
    {\resClk{\Hst}{\cc{\false}{bs}}{\cc{\nullStream}{vs}}}{DresA} 
    ~~~
     \namedJdg{\baseOf{vs}= ~bs}
    {\baseOf{\cc{\val{v}}{vs}} = ~\ccnb{\true}{bs} }{Dbase1}  
     \\ \\
    \namedJdg{\resClk{\Hst}{bs}{vs}}
    {\resClk{\Hst}{\cc{\true}{bs}}{\cc{\stream{v}}{vs}}}{Dres} 
    ~~~~
    \namedJdg{\baseOf{vs}= ~bs}
    {\baseOf{\cc{\nullStream}{vs}} = ~\ccnb{\false}{bs} }{Dbase2}     
    \end{array}
\]
    \caption{Definitions of auxiliary predicates-1}
    \label{FIG:Def1}
\end{figure*}

\begin{figure*}
    \[
    \begin{array}{c}
        \namedJdg{\semMerge{xs'}{ts'}{fs'}{os'}}
        {\semMerge{
            \cc{\stream{T}}{xs'}}
            {\cc{\stream{v_t}}{ts'}}
            {\cc{\nullStream}{fs'}}
            {\ccnb{\stream{v_t}}{os'}}}{DmrgT} 
            \\   \\
        \namedJdg{\semMerge{xs'}{ts'}{fs'}{os'}}
        {\semMerge{
            \cc{\stream{F}}{xs'}}
            {\cc{\nullStream}{ts'}}
            {\cc{\stream{v_f}}{fs'}}
            {\ccnb{\stream{v_f}}{os'}}}{DmrgF} 
            \\ \\
        \namedJdg{\semMerge{xs'}{ts'}{fs'}{os'}}
        {\semMerge{
            \cc{\nullStream}{xs'}}
            {\cc{\nullStream}{ts'}}
            {\cc{\nullStream}{fs'}}
            {\ccnb{\nullStream}{os'}}}{DmrgA} 
            \\ \\
        \namedJdg{\semIte{es'}{ts'}{fs'}{os'}}
        {\semIte{
            \cc{\stream{T}}{es'}}
            {\cc{\stream{v_t}}{ts'}}
            {\cc{\stream{v_f}}{fs'}}
            {\ccnb{\stream{v_t}}{os'}}}{DiteT}  
            \\ \\
            \namedJdg{\semIte{es'}{ts'}{fs'}{os'}}
            {\semIte{
                \cc{ \stream{F}}{es'}}
                {\cc{\stream{v_t}}{ts'}}
                {\cc{\stream{v_f}}{fs'}}
                {\ccnb{\stream{v_f}}{os'}}}{DiteF} 
                \\ \\
            \namedJdg{\semIte{es'}{ts'}{fs'}{os'}}
                 {\semIte{  \cc{\nullStream}{es'}}
                    {\cc{\nullStream}{ts'}}
                    {\cc{\nullStream}{fs'}}
            {\ccnb{\nullStream}{os'}}}{DiteA} \\ 
            \\
                            \namedJdg{\whenk{k}{xs'}{es'}{os'}}
            {\whenk{k}{\cc{\stream{\neg k}}{xs'}}{\cc{\stream{v}}{es'}}{\ccnb{\nullStream}{os'}}}{DwhA1} 
            \\ \\
                \namedJdg{\semFby{v}{xs}{ys}}
            {\semFby{c}{\cc{\stream{v}}{xs}}{\ccnb{\stream{c}}{ys}}}{Dfby} 
            ~~
            \namedJdg{\whenk{k}{xs'}{es'}{os'}}
            {\whenk{k}{\cc{\stream{k}}{xs'}}{\cc{\stream{v}}{es'}}{\ccnb{\stream{v}}{os'}}}{Dwhk}
            \\ \\
            \namedJdg{\semFby{c}{xs}{ys}}
            {\semFby{c}{\cc{\nullStream}{xs}}{\ccnb{\nullStream}{ys}}}{DfbyA} 
            ~~
            \namedJdg{\whenk{k}{xs'}{es'}{os'}}
            {\whenk{k}{\cc{\nullStream}{xs'}}{\cc{\nullStream}{es'}}{\ccnb{\nullStream}{os'}}}{DwhA2} 
         \\ \\
    \end{array}
    \]
    \caption{Definitions of auxiliary predicates-2}
\end{figure*}

All auxiliary stream operators are defined to behave according to the clocking regime.
For example, the rule (DcnstF) ensures the absence of a value when the clock is \ckFont{false}.
Likewise the unary and binary operators lifted to stream operations \ckFont{$\hat{\diamond}$} and \ckFont{$\hat{\oplus}$} operate only when the argument streams have values present, as in (Dunop) and (Dbinop), and mark absence when the argument streams' values are absent, as shown in (DunopA) and (DbinopA).
The rules (Dtl) and (Dhtl) are obvious.

Note that in the rules (DmrgT) and (DmrgF) for \ckFont{merge}, a value is present on one of the two streams being merged and absent on the other.
When a value is absent on the stream corresponding to the boolean variable, values are absent on all streams (DmrgA).
The rules for \ckFont{ite} require all streams to have values present, \textit{i.e.},  (DiteT) and (DiteF), or all absent, \textit{i.e.}, (DiteA).
We have already discussed the \ckFont{when} operation in some detail earlier.

The $\textcolor{blue}{\textsf{fby}_{NL}}$ operation is a bit subtle, and rule (Dfby) may look non-intuitive.  
However, its formulation corresponds exactly to the V\'{e}lus formalisation,  ensuring that a value from the first argument stream is prepended exactly when a leading value would have been present on the second argument stream. 
The operation \ckFont{base-of} converts a value stream to a clock, \textit{i.e.}, a boolean stream.
The operation \ckFont{respects-clock} is formulated corresponding to the V\'{e}lus definition. \\

The main difference between $\textcolor{blue}{\textsf{fby}_L}$ and $\textcolor{blue}{\textsf{fby}_{NL}}$ is that the former takes a \tbf{stream} while the latter takes a constant value. 
The $\textcolor{blue}{\textsf{fby}_{NL}}$ predicate assigns the currently saved constant as the stream value and delays the current operand stream by storing its current value for the next clock cycle (effectively functioning as an initialized D-flip flop).
$\textcolor{blue}{\textsf{fby}_L}$ on the other hand extracts the $0^{th}$ tick value of the first operand stream and  uses the  predicate $\textcolor{blue}{\textsf{fby}_{dl}}$ for delaying.

\begin{figure*}
    $$
    \begin{array}{c}
      \mjdg{\semLFby{xs}{ys}{os}}
      {\semLFby{(\ccnb{\nullStream}{xs})}{(\ccnb{\nullStream}{ys})}{\ccnb{\nullStream}{os}}}  ~~~~
      \mjdg{\semLDFby{y}{xs}{ys}{os}}
      {\semLFby{(\ccnb{\stream{x}}{xs})}{(\ccnb{\stream{y}}{ys})}{ \ccnb{\stream{x}}{os}}}\\
      \\
      \mjdg{\semLDFby{v}{xs}{ys}{os}}
      {\semLDFby{v}{(\ccnb{\nullStream}{xs})}{(\ccnb{\nullStream}{ys})}{\ccnb{\nullStream}{os}}} ~~~~
      \mjdg{\semLDFby{y}{xs}{ys}{os}}
      {\semLDFby{v}{(\ccnb{\stream{x}}{xs})}{(\ccnb{\stream{y}}{ys})} {\ccnb{\stream{v}}{os}}}
    \end{array}
    $$
    \caption{\lustre's \ckFont{fby} semantic predicates}
    \label{FIG:LusFbyPred}
\end{figure*}

\section{Stream Semantics}
\label{APP:LustreStreamSem}

We present here a specification of core \lustre's co-inductive stream semantics, with some commentary and intuition.
This consolidates various earlier presentations of rules \cite{SNLustre2017,Lustre2020,PYS-memocode2020,Bourke-jfla2021}, and can be seen as an abstract Coq-independent specification of the semantics encoded in  the V\'{e}lus development.

% Name the Rules in a way consistent with the syntax
\begin{figure*}
    $$
    \begin{array}{c}
        \namedJdg{\semConst{c}{bs}{cs}}
        {\mstrSemExpPred{G,\Hst}{bs}{c}{[cs]}}{LScnst} ~~~
        \namedJdg{\Hst(x)={\textcolor{BrickRed}{xs}}}
        {\mstrSemExpPred{G,\Hst}{bs}{x}{[xs]}}{LSvar} \\
        \\
        \namedJdg{\predSet{\mstrSemExpPred{G,\Hst}{bs}{e}{[es]}}
            {\liftunop{es}{os} }}
        {\mstrSemExpPred{G,\Hst}{bs}{\unop{e}}{[os]}}{LSunop} \\
        \\
        \namedJdg{\predSet{\mstrSemExpPred{G,\Hst}{bs}{e_1}{[es_1]}}
            {\mstrSemExpPred{G,\Hst}{bs}{e_2}{[es_2]}}
            {\liftbinop{es_1}{es_2}{os} }}
        {\mstrSemExpPred{G,\Hst}{bs}{\binop{e_1}{e_2}}{[os]}}{LSbinop} \\
        \\
        \namedJdg{\predSet{\forall i ~\mstrSemExpPred{G,\Hst}{bs}{e_i}{\tupstrm{es}_i}}
            {\hist{x}={\textcolor{BrickRed}{xs}}}
            {\forall i:~\mapwhenk{k}{xs}{\tupstrm{es}_i}{\tupstrm{os}_i}}}
        {\mstrSemExpPred{G,\Hst}{bs}{\when{\vv{e_i}}{x=k}}{
        \flatten{\vv{\tupstrm{os}_i}}}}{LSwhn}\\
        \\
        \namedJdg{\dependSet{\predSet{\mstrSemExpPred{G,\Hst}{bs}{e}{[es]}}
        {\forall i:~ ~\mstrSemExpPred{G,\Hst}{bs}{et_i}{\tupstrm{ets}_i}}
        }
        {
           \forall j:~  \mstrSemExpPred{G,\Hst}{bs}{ef_j}{\tupstrm{efs}_j} 
            ~~~~
            \mapsemIte{es}{(\flatten{\vv{\tupstrm{ets}_i}})}{
            (\flatten{\vv{\tupstrm{efs}_j}})}{\tupstrm{os}}
            }}
        {\mstrSemExpPred{G,\Hst}{bs}{\ite{e}{\vv{et_i}}{\vv{ef_j}}}{\tupstrm{os}}}{LSite} \\
        \\
       \namedJdg{\dependSet{\predSet{\hist{x}=xs}{
       \forall i: ~\mstrSemExpPred{G,\Hst}{bs}{et_i}{\tupstrm{ets}_i}
       }}
            {\forall j: ~
            \mstrSemExpPred{G,\Hst}{bs}{ef_j}{\tupstrm{efs}_j}
            ~~~~
             \mapsemMerge{xs}{(\flatten{\vv{\tupstrm{ets}_i}})}{
            (\flatten{\vv{\tupstrm{efs}_j}})}{\tupstrm{os}}}
            }
       {\mstrSemExpPred{G,\Hst}{bs}{\lmerge{x}{\vv{et_i}}{\vv{ef_j}}}{\tupstrm{os}}}{LSmrg} \\
       \\
       \namedJdg{\dependSet{\predSet{\forall i:~ ~\mstrSemExpPred{G,\Hst}{bs}{e0_i}{\tupstrm{e0s}_i}
      ~~~~ \forall j:~ 
       \mstrSemExpPred{G,\Hst}{bs}{e_j}{\tupstrm{es}_j}}}
            { ~\mapsemLFby{(\flatten{\vv{\tupstrm{e0s}_i}})}{
            (\flatten{\vv{\tupstrm{es}_j}})}{\tupstrm{os}}}}
        {\mstrSemExpPred{G,\Hst}{bs}{\fby{\vv{e0_i}}{\vv{e_j}}}{\tupstrm{os}}}{LSfby} \\
   \\
        \namedJdg{\forall i \in [1,k] ~ \mstrSemExpPred{G,\Hst}{bs}{e_i}{\tupstrm{es}_i} ~~~
        [\hist{x_1}, \ldots, \hist{x_n}] = 
        \flatten{\vv{\tupstrm{es}_i}}
        }
        {\mstrEqnPred{G}{\Hst}{bs}{\vv{x_j}=\vv{e_i}}}{LSeq} \\
  \\
    \namedJdg{\dependSet{\predSet{\node \in G}
        {\hist{n.\tbf{in}} = \tupstrm{xs}}
        {\baseOf{\tupstrm{xs}} = bs}}
        {\predSet{\hist{n.\tbf{out}} = \tupstrm{ys}}
        {\forall eq \in \vv{eq}:~ \mstrEqnPred{G}{\Hst}{bs}{eq}}
         }
    }{\mstrCallPred{G}{\liftnode{f}}{\tupstrm{xs}}{\tupstrm{ys}}}{LSndef} \\
    \\
           \namedJdg{\predSet{\mstrSemExpPred{G,\Hst}{bs}{\vv{e}}{\tupstrm{xs}}}
            {\mstrCallPred{G}{\liftnode{f}}{\tupstrm{xs}}{\tupstrm{ys}}}}
        {\mstrSemExpPred{G,\Hst}{bs}{f(\vv{e})}{\tupstrm{ys}}}{LSncall} \\ \\
       \namedJdg{ \predSet{\forall i: \mstrSemExpPred{G,\Hst}{bs}{e_i}{\tupstrm{es}_i}}}
        {\mstrSemExpPred{G,\Hst}{bs}{\vv{e_i}}{\flatten{\vv{\tupstrm{es}_i}}}}{LStup}
    \end{array}
    $$
    \caption{Stream semantics of \lustre}
    \label{FIG:APPLusStrSemNodeEqn}
\end{figure*}

The semantics of \lustre\ and \nlustre\  programs are \textit{synchronous}:
Each variable and expression defines a data stream which pulses with respect to a \textit{clock}.
A clock is a stream of booleans (CompCert/Coq's $\true$ and $\false$ in Velus).
A flow takes its $n^\tti{th}$ value on the $n^\tti{th}$ clock tick, \textit{i.e.},  some value, written $\stream{v}$, is present at instants when the clock value is \true, and none (written $\nullStream)$ when it is \false.
The \textit{temporal operators} \ttt{when}, \ttt{merge} and \ttt{fby} are used to express the complex clock-changing and clock-dependent behaviours of sampling, interpolation and delay respectively.

%Program behaviours in \lustre\ are completely determined by sequences of clocked occurrences of events.

Formally the stream semantics is defined using predicates over the program graph $G$, a (co-inductive) stream \textit{history} ($\Hst: \tti{Ident} \rightarrow \tti{value}~\tti{Stream}$) that associates value streams to variables, and a clock $bs$ \cite{Lustre2020,PYS-memocode2020,Bourke-jfla2021}.
Semantic operations on (lists of) streams are written in \textcolor{blue}{blue \textsf{sans serif}} typeface.
Streams are written in \textcolor{BrickRed}{red}, with lists of streams usually written in \textbf{\textcolor{BrickRed}{bold face}}.
All these stream operators, defined co-inductively,  enforce the clocking regime, ensuring  the presence of a value when the clock is \ckFont{true}, and absence when \ckFont{false}.

The predicate $\mstrSemExpPred{G,\Hst}{bs}{e}{\tupstrm{es}}$ relates an \textit{expression} $e$ to a \textit{list} of streams, written $\tupstrm{es}$.
A list consisting of only a single stream $\textcolor{BrickRed}{es}$ is explicitly denoted as $\textcolor{BrickRed}{[es]}$.
The semantics of \textit{equations} are expressed using the predicate 
$\mstrEqnPred{G}{\Hst}{bs}{\vv{eq_i}}$, which requires \textit{consistency} between the assumed and defined stream histories in $\Hst$ for the program variables, as induced by the equations.
Finally, the semantics of \textit{nodes} is given as a stream history transformer predicate
$\mstrCallPred{G}{\liftnode{f}}{\tupstrm{xs}}{\tupstrm{ys}}.$

% Then reference the figure AND 
% explain the rules of the Figure.
\autoref{FIG:APPLusStrSemNodeEqn} presents the stream semantics for \lustre.
While rules for \textit{some} constructs have been variously presented \cite{SNLustre2017,Lustre2020,PYS-memocode2020,Bourke-jfla2021}, our presentation can be considered as a definitive consolidated specification of the operational semantics of \lustre, consistent with the V\'{e}lus compiler encoding \cite{Velus}.

\begin{itemize}
\item Rule (LScnst) states that a constant $c$ denotes a constant stream of the value $\stream{c}$ pulsed according to given clock $bs$.  
This is effected by the semantic operator \ckFont{const}.
\item
Rule (LSvar) associates a variable $x$ to the stream given by $\Hst(x)$.
\item
In rule (LSunop), \ckFont{$\hat{\diamond}$} denotes the operation $\diamond$ \textit{lifted} to apply instant-wise to the stream denoted by expression $e$.
\item
Likewise in rule (LSbinop), the  binary operation $\oplus$ is applied paired point-wise to the streams denoted by the two sub-expressions (which should both pulse according to the same clock).
In all these rules, an expression is associated with a \textit{single} stream. 

\item
The rule (LSwhn) describes \textit{sampling} whenever a variable $x$ takes the boolean value $k$, from the flows arising from a list of expressions $\vv{e_i}$,  returning a list of streams of such sampled values.
The predicate $\widehat{\ckFont{when}}$ \textit{maps} the predicate $\ckFont{when}$ to act on the corresponding components of \textit{lists} of streams, \textit{i.e.},  \[ 
\mapwhenk{k}{xs}{[es_1, \ldots, es_k]}{[os_1, \ldots, os_k]} 
~\textrm{abbreviates} ~ 
\bigwedge_{i \in [1,k]}~ \whenk{k}{xs}{es_i}{os_i}.
\]
%Similarly for the predicates $\widehat{\ckFont{merge}}$, $\widehat{\ckFont{ite}}$, and $\widehat{\ckFont{fby}_L}$.  
The operation $\flatten{\_}$ flattens a list of lists (of possibly different lengths) into a single list. 
Flattening is required since expression $e_i$ may in general denote a \textit{list} of streams \textcolor{BrickRed}{$\tupstrm{es}_i$}.

\item
The expression $\lmerge{x}{\vv{et}_i}{\vv{ef}_j}$ produces (lists of) streams on a faster clock.
The semantics in rule (LSmrg) assume that for each pair of corresponding component streams from
$\flatten{\tupstrm{ets}_i}$ and
$\flatten{\tupstrm{efs}_j}$, a value is present in the first stream and absent in the second at those instances where $x$ has a true value $\stream{T}$, and complementarily, a value is present in the second stream and absent in the first when $x$ has a false value $\stream{F}$.
Both values must be absent when $x$’s value is absent.
These conditions are enforced by the auxiliary semantic operation \ckFont{merge}. 
The predicate $\widehat{\ckFont{merge}}$ \textit{maps} the predicate $\ckFont{merge}$ to act on the corresponding components of \textit{lists} of streams, \textit{i.e.},  \[ 
\mapsemMerge{xs}{[ets_1, \ldots, ets_k]}{[efs_1, \ldots, efs_k]}{[os_1, \ldots, os_k]} 
~\textrm{abbreviates} ~ 
\bigwedge_{i \in [1,k]}~ \semMerge{xs}{ets_i}{efs_i}{os_i}.
\]

\item
In contrast, the conditional expression $\ite{e}{\vv{et}}{\vv{ef}}$
requires that all three argument streams $es$, and the corresponding components from $\flatten{\vv{\tupstrm{ets}_i}}$ and
$\flatten{\vv{\tupstrm{efs}_j}}$ pulse to the same clock.
Again, values are selected from the first or second component streams depending on whether the stream $es$ has the value $\stream{T}$ or $\stream{F}$ at a particular instant.
These conditions are enforced by the auxiliary semantic operation \ckFont{ite}. 
The predicate $\widehat{\ckFont{ite}}$ \textit{maps} the predicate $\ckFont{ite}$ to act on the corresponding components of \textit{lists} of streams, \textit{i.e.},  \[ 
\mapsemIte{xs}{[ets_1, \ldots, ets_k]}{[efs_1, \ldots, efs_k]}{[os_1, \ldots, os_k]} 
~\textrm{abbreviates} ~ 
\bigwedge_{i \in [1,k]}~ \semIte{xs}{ets_i}{efs_i}{os_i}.
\]

\item
A delay operation is implemented by $\fby{e0}{e}$.
The rule (LSfby) is to be read as follows.
Let each expression $e0_i$ denote a list of streams
$\tupstrm{e0s_i}$, and each expression $e_j$ denote a list of streams $\tupstrm{es_j}$.
The output list of streams consists of streams whose first elements are taken from the each stream in $\flatten{\vv{\tupstrm{e0s}_i}}$ with the rest taken from the corresponding component of $\flatten{\vv{\tupstrm{es}_j}}$.
These are achieved using the semantic operation \textcolor{blue}{\textsf{fby$_L$}}.
The predicate $\widehat{\ckFont{fby}_L}$ \textit{maps} the predicate $\ckFont{fby}_L$ to act on the corresponding components of \textit{lists} of streams, \textit{i.e.},  \[ 
\mapsemLFby{\tupstrm{xs}}{\tupstrm{ys}}{\tupstrm{zs}} 
~\textrm{abbreviates} ~ 
\bigwedge_{i \in [1,m]}~ \semLFby{xs_i}{ys_i}{zs_i}
\]
\item The rule (LStup) deals with tuples of expressions, where the lists of streams from each expression are flattened into a list of streams.
\end{itemize}

\paragraph{Equations, node definitions and node call} \ 
\begin{itemize}
\item
The rule (LSeq) for equations checks the consistency between the assumed meanings for the defined variables $x_j$ according to the history $\Hst$ with the corresponding components of the list of streams $\flatten{\vv{\tupstrm{es}_i}}$ to which a tuple of right-hand side expressions evaluates.

\item
The rule (LSndef) presents the meaning given to the definition named $f$ of a node as a stream list transformer $\liftnode{f}$. 
If history $\Hst$ assigns lists of streams to the input and output variables for a node in a manner such that the semantics of the equations in the node are satisfied, then the semantic funnction $\liftnode{f}$ transforms input stream list $\tupstrm{xs}$ to output stream list $\tupstrm{ys}$.
The operation \ckFont{base-of} finds an appropriate base clock with respect to which a given list of value streams pulse.

\item
The rule (LSncall) applies the stream transformer semantic function $\liftnode{f}$ to the stream list $\tupstrm{xs}$ corresponding to the tuple of arguments $\vv{e_i}$, and returns the stream list $\tupstrm{ys}$.
\end{itemize}

\paragraph{Clocks and clock-annotated expressions} \ 
We next present rules for clocks.
Further, we  assume that all (\nlustre) expressions in equations can be clock-annotated, and present the corresponding rules.

\begin{figure*}
$$
   \begin{array}{c}
        \namedJdg{}
        {\mstrSemCkPred{\Hst}{bs}{\ttt{base}}{bs}}{LSbase}  \\ \\
        \namedJdg{
            \dependSet{\predSets{\mstrSemCkPred{\Hst}{bs}{ck}{\cc{\true}{bk}}}
            {\hist{x}=\cc{\stream{k}}{xs}}}
            {\mstrSemCkPred{\htl{\Hst}}{\tl{bs}}{\on{ck}{x}{k}}{bs'}}}
        {\mstrSemCkPred{\Hst}{bs}{\on{ck}{x}{k}}{\cc{\true}{bs'}}}{LSonT} \\

        \namedJdg{
            \dependSet{\predSets{\mstrSemCkPred{\Hst}{bs}{ck}{\cc{\false}{bk}}}
            {\hist{x}=\cc{\nullStream}{xs}}}
            {\mstrSemCkPred{\htl{\Hst}}{\tl{bs}}{\on{ck}{x}{k}}{bs'}}}
        {\mstrSemCkPred{\Hst}{bs}{\on{ck}{x}{k}}{\cc{\false}{bs'}}}{LSonA1} 
        \\
               \namedJdg{
            \dependSet{
             \predSet{\mstrSemCkPred{\Hst}{bs}{ck}{\cc{\true}{bk}}}
                {\hist{x}=\cc{\stream{k}}{xs}}
            }
            {\mstrSemCkPred{\htl{\Hst}}{\tl{bs}}{\onF{ck}{x}{k}}{bs'}}}
        {\mstrSemCkPred{\Hst}{bs}{\onF{ck}{x}{k}}{\cc{\false}{bs'}}}{LSonA2} \\ \\

        \namedJdg{\dependSet{\mstrSemExpPred{\Hst}{bs}{e}{[\ccnb{\nullStream}{es}]}}
        {\mstrSemCkPred{\Hst}{bs}{ck}{\ccnb{\false}{cs}}}}
    {\mstrSemExpPred{\Hst}{bs}{e::ck}{[\ccnb{\nullStream}{es}]}}{NSaeA} ~~
    
        \namedJdg{\dependSet{\mstrSemExpPred{\Hst}{bs}{e}{[\ccnb{\stream{v}}{es}]}}
            {\mstrSemCkPred{\Hst}{bs}{ck}{\ccnb{\true}{cs}}}}
        {\mstrSemExpPred{\Hst}{bs}{e::ck}{[\ccnb{\stream{v}}{es}]}}{NSae} 
    \end{array}
$$   
    \caption{Stream semantics of clocks and annotated expressions}
    \label{FIG:StrSemCk}
\end{figure*}

The predicate 
$\mstrSemCkPred{\Hst}{bs}{ck}{bs'}$ in \autoref{FIG:StrSemCk} defines the meaning of a \nlustre\ clock expression $ck$ with respect to a given history $\Hst$ and a clock $bs$ to be the resultant clock $bs'$.
% Note that a clock is a stream of booleans.  
The \ttt{on} construct lets us define 
coarser clocks derived from  a given clock --- whenever a variable $x$ has the desired value $k$ and the given clock is true. 
The rules (LSonT), (LSonA1), and (LSonA2) 
present the three cases: respectively when variable $x$ has the desired value $k$ and clock is true; the clock is false; and  the program variable $x$ has the complementary value and the clock is true.
The auxiliary operations \ckFont{tl} and \ckFont{htl},  give the tail of a stream, and the tails of streams for each variable according to a given history $\Hst$. 
Rules (NSaeA)-(NSae) describe the semantics of  clock-annotated expressions, where the output stream carries a value exactly when the clock is true.

\paragraph{Stream semantics for \nlustre} \ \ 
The semantic relations for \nlustre\ are either identical to (as in constants, variables, unary and binary operations) or else the (simple) singleton cases of the rules given for \lustre\ (as in \ttt{merge}, \ttt{ite}, \ttt{when}).

The significant differences are in treatment of \ttt{fby}, and the occurrence of \ttt{fby} and node call only in the context of equations.

\begin{figure*}
   $$
    \begin{array}{c}
    \namedJdg{\dependSet{\predSet{\node \in G}
        {\hist{n.\tbf{in}} = \tupstrm{xs}}
        {\baseOf{\tupstrm{xs}} = bs}}
        {\predSet{\resClk{\Hst}{bs}}
        {\hist{n.\tbf{out}} = \tupstrm{ys}}
        {\forall eq \in \vv{eq}:~ \mstrEqnPred{G}{\Hst}{bs}{eq}}
         }
    }{\mstrCallPred{G}{\liftnode{f}}{\tupstrm{xs}}{\tupstrm{ys}}}{NSndef'} \\
    \\
        \namedJdg{\mstrSemExpPred{\Hst}{bs}{e::ck}{\hist{x}}}
        {\mstrEqnPred{G}{\Hst}{bs}{\eqn{x}{e}{ck}}}{NSeq} 
\\ 
   \namedJdg{\predSet{\mstrSemExpPred{\Hst}{bs}{e::ck}{[vs]}}
            {\semFby{c}{vs}{\hist{x}}}}
        {\mstrEqnPred{G}{\Hst}{bs}{\eqn{x}{\fby{c}{e}}{ck}}}{NSfby'} \\
\\     
    \namedJdg{\predSet{\mstrSemExpPred{\Hst}{bs}{\vv{e}}{\tupstrm{vs}}}
    {\mstrSemCkPred{\Hst}{bs}{ck}{\baseOf{\tupstrm{vs}}}}
    {\mstrCallPred{G}{\liftnode{f}}{\tupstrm{vs}}{\vv{\hist{x_i}}}}}
            {\mstrEqnPred{G}{\Hst}{bs}
        {\eqn{\vv{x}}{f(\vv{e})}{ck}  }}{NSncall'}
   
   \end{array}%  
    $$
    
    \caption{Stream semantics of \nlustre\ nodes and equations}
    \label{FIG:APPStrSemNodeEqn}
\end{figure*}

The (NSndef') rule only differs from (LSndef) in that 
post-transcription clock alignment, we have an additional requirement of $\Hst$ being in accordance with the base clock $bs$, enforced by \ckFont{respects-clock}.
The (NSeq) rule for simple equations mentions the clock that annotates the defining expression, checking that it is consistent with the assumed history for the defined variable $x$.
The (NSfby') rule for \ttt{fby} in an equational context uses the semantic operation \ckFont{fby}, which differs from $\ckFont{fby}_L$ in that it requires its first argument to be a constant rather than a stream.
Finally, the rule rule (NSncall') for node call, now in an equational context, is similar to (LSncall) except that it constrains the clock modulating the equation to be the base clock of the input flows.

\end{document}